\documentclass[10pt,conference]{IEEEtran}
\usepackage{cite}
\usepackage{amsmath,amssymb,amsfonts}
\usepackage{algorithmic}
\usepackage{graphicx}
\usepackage{textcomp}
\usepackage{xcolor}
\usepackage[hyphens]{url}
\usepackage{fancyhdr}
\usepackage{hyperref}
\usepackage{comment}
\usepackage{xspace}
\usepackage{todonotes}
\usepackage{subcaption}
\usepackage{siunitx}
\usepackage{listings}
\usepackage{multirow}
\usepackage{relsize}
\usepackage{booktabs}
\usepackage{enumitem}
\usepackage{pifont}
\usepackage[table]{xcolor}

\newcommand{\oursys}{SLINFER\xspace}

\title{Towards Resource-Efficient Serverless LLM Inference with SLINFER}

\def\hpcacameraready{} 

\newcommand\hpcaauthors{Chuhao Xu$\dagger$, Zijun Li$\dagger\ddagger$, Quan Chen$\dagger$, Han Zhao$\dagger$, Xueyan Tang$\ddagger$, and Minyi Guo$\dagger$}
\newcommand\hpcaaffiliation{Shanghai Jiao Tong University$\dagger$, Nanyang Technological University$\ddagger$}
\newcommand\hpcaemail{}

\author{
  \ifdefined\hpcacameraready
    \IEEEauthorblockN{\hpcaauthors{}}
      \IEEEauthorblockA{
        \hpcaaffiliation{} \\
        \hpcaemail{}
      }
  \else
    \IEEEauthorblockN{\normalsize{HPCA \hpcayear{} Submission
      \textbf{\#\hpcasubmissionnumber{}} -- Confidential Draft -- Do NOT Distribute!!}
    }
  \fi 
}

\fancypagestyle{camerareadyfirstpage}{%
  \fancyhead{}
  
  \fancyhead[C]{
    \ifdefined\aeopen
    \parbox[][12mm][t]{13.5cm}{\hpcayear{} IEEE International Symposium on High-Performance Computer Architecture (HPCA)}    
    \else
      \ifdefined\aereviewed
      \parbox[][12mm][t]{13.5cm}{\hpcayear{} IEEE International Symposium on High-Performance Computer Architecture (HPCA)}
      \else
      \ifdefined\aereproduced
      \parbox[][12mm][t]{13.5cm}{\hpcayear{} IEEE International Symposium on High-Performance Computer Architecture (HPCA)}
      \else
    \fi 
    \fi 
    \fi 
    \ifdefined\aeopen 
      \includegraphics[width=12mm,height=12mm]{ae-badges/open-research-objects.pdf}
    \fi 
    \ifdefined\aereviewed
      \includegraphics[width=12mm,height=12mm]{ae-badges/research-objects-reviewed.pdf}
    \fi 
    \ifdefined\aereproduced
      \includegraphics[width=12mm,height=12mm]{ae-badges/results-reproduced.pdf}
    \fi
  }
  \fancyfoot[C]{}
}
\begin{document}
\maketitle

\ifdefined\hpcacameraready 
  \thispagestyle{camerareadyfirstpage}
  \pagestyle{empty}
\else
  \thispagestyle{plain}
  \pagestyle{plain}
\fi

\newcommand{\hpcaheight}{0mm}
\ifdefined\eaopen
\renewcommand{\hpcaheight}{12mm}
\fi


\begin{abstract}
The rise of LLMs has driven demand for private serverless deployments, characterized by moderate-sized models and infrequent requests. While existing serverless solutions follow exclusive GPU allocation, we take a step back to explore modern platforms and find that: Emerging CPU architectures with built-in accelerators are capable of serving LLMs but remain underutilized, and both CPUs and GPUs can accommodate multiple LLMs simultaneously.
	
We propose \oursys, a resource-efficient serverless inference scheme tailored for small- to mid-sized LLMs that enables elastic and on-demand sharing across heterogeneous hardware. \oursys tackles three fundamental challenges: (1) precise, fine-grained compute resource allocation at token-level to handle fluctuating computational demands; (2) a coordinated and forward-looking memory scaling mechanism to detect out-of-memory hazards and reduce operational overhead; and (3) a dual approach that consolidates fragmented instances through proactive preemption and reactive bin-packing. Experimental results on 4 32-core CPUs and 4 A100 GPUs show that \oursys improves serving capacity by 47\% - 62\% through sharing, while further leveraging CPUs boosts this to 86\% - 154\%.
	
\end{abstract}

\section{Introduction}

Large Language Models (LLMs) have seen widespread adoption, with many providers (e.g., OpenAI~\cite{chatgpt}, Anthropic~\cite{Claude}).
Meanwhile, driven by the need for customization and privacy, individuals and enterprises are increasingly seeking to deploy private models on the cloud~\cite{google-serverless, aws-sagemaker}, offloading the burden of infrastructure management. 
Consequently, cloud platforms are hosting a large number of LLMs and have turned to serverless approach~\cite{huggingface-serverless,azure-serverless} to maximize serving capacity while meeting service-level objectives (SLOs).

A closer examination of this deployment reveals two key characteristics that closely align with the typical patterns of serverless~\cite{DBLP:conf/usenix/ShahradFGCBCLTR20,DBLP:conf/cloud/JoosenHASDWB23} workloads: 
(1) small- to mid-sized models dominate in popularity—87\% of downloads on HuggingFace are for LLMs no larger than 8B parameters~\cite{huggingface-popular}; and (2) invocation patterns are highly variable and infrequent—For instance, LMSYS hosts diverse HuggingFace LLMs, 56\% of which receive fewer than 5 requests per hour on average~\cite{DBLP:conf/iclr/ZhengC0LZW00LXG24}.

Given the high resource demands and the stringent SLOs, existing serverless LLM inference solutions~\cite{DBLP:conf/osdi/FuXHBUPM24, DBLP:conf/asplos/ZengXGCL25,DBLP:conf/usenix/HuXLHCXLMZWDDRL25} allocate exclusive GPUs to each model in an event-driven manner upon request arrival. 
However, they still struggle to handle the scenario where numerous small-sized LLMs are infrequently invoked. 
For instance, when using ServerlessLLM~\cite{DBLP:conf/osdi/FuXHBUPM24} to host 64 3B- to 13B-sized LLMs on 4 A100-80GB GPUs, 33\% of the requests fail to meet their SLOs due to long queuing, despite the average memory utilization per GPU being only 23\%.
The key issue lies in the scarcity of GPUs relative to the number of models, while the resource over-provisioning makes each model occupy an entire GPU.

Through systematic investigation of modern platforms, we re-examine the deployment characteristics for small- to mid-sized LLMs, revealing two key opportunities.
First, clusters have abundant idle CPUs, and utilizing their built-in accelerators (e.g., Intel Advanced Matrix Extensions, AMX~\cite{AMX,DBLP:conf/isscc/NassifMMPLYMHVK22}) can independently support them while meeting production-grade SLOs.
Second, given the low-frequency, serverless-like workload patterns, individual LLMs usually do not fully saturate the entire CPU/GPU, making it practical to colocate multiple LLMs by provisioning resources on demand.

\begin{figure}
	\centering
	\vspace{-1mm}
	\includegraphics[width=.93\linewidth]{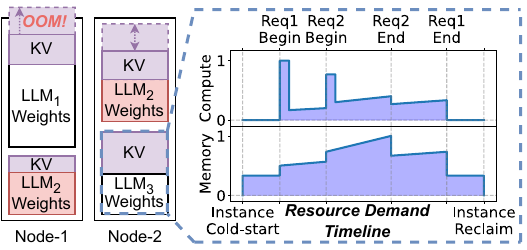}
	\vspace{-1mm}
	\caption{Example of normalized resource demand variation for an instance under multi-LLM sharing. LLM$_2$ is fragmented.}

	\vspace{-1mm}
	\label{fig:challenge}
\end{figure}

We are therefore motivated to design a serverless LLM inference scheme that embraces hardware-agnostic resource allocation and on-demand, elastic sharing across heterogeneous platforms. However, as illustrated in Figure~\ref{fig:challenge}, the dynamic and diverse patterns of per-instance compute and memory demands introduce three fundamental design challenges.

First, computational demand fluctuates sharply during token generation, especially as the first token of each request undergoes the prefill stage~\cite{DBLP:conf/isca/PatelCZSGMB24,DBLP:conf/osdi/AgrawalKPMKGTR24}. 
Since instances continuously receive new requests~\cite{DBLP:conf/osdi/YuJKKC22}, it becomes difficult to allocate just-enough compute resources. Over-provisioning for peak usage leads to wasted resources, while aggressive sharing risks violating SLOs.
Furthermore, instances also go through startup and idle phases where compute demand is negligible.

Second, the memory demand per instance varies with the request load. Dynamically managing memory is non-trivial: each instance requires pre-allocated space for the KV-cache~\cite{DBLP:conf/sosp/KwonLZ0ZY0ZS23}, and we find that resizing incurs noticeable overhead~\cite{DBLP:conf/asplos/ZengXGCL25}. 
More critically, other memory operations such as model weights loading and unloading are also frequent and sensitive. 
When multiple instances share a node, arbitrary memory adjustments can easily trigger out-of-memory (OOM) errors.

Third, in a congested shared environment, the vertical scalability of individual instances is often suppressed, resulting in fragmented deployments of the same model. 
In the example shown in Figure~\ref{fig:challenge}, multiple fragmented instances of LLM$_2$ not only incur redundant memory overhead for model weights but also reduce batching opportunities that could have been leveraged by a single consolidated instance. 
This fragmentation degrades both compute and memory efficiency.

To address these challenges, we closely examine the compute and memory characteristics of LLM inference instances and their implications for resource efficiency.
(1) Since compute demand varies at the granularity of tokens, there is potential to provision compute resources dynamically at the same granularity—provided that we can precisely quantify and budget per-instance demand.
(2) Given the overhead of memory adjustments and the potential OOM risks, it is important to reconsider the trade-off between utilization and operational cost, while coordinating instances to ensure safe and efficient sharing.
(3) Rather than blindly following serverless-style horizontal scaling that leads to fragmented, inefficient instances, identifying or even actively seeking opportunities for vertical scaling can significantly improve efficiency.

Based on above observations, we propose \textbf{\oursys}, a \underline{S}erverless \underline{L}LM \underline{Infer}ence scheme achieving the resource-efficient deployment for small- to mid-sized LLMs.
\oursys abstracts heterogeneous hardware into CPU/GPU nodes, decoupling resource management through compute and memory subsystems.
The compute subsystem, driven by request headroom, efficiently schedules instances via shadow validation and real-time token-level resource provision.
For memory subsystem, it performs watermark-based scaling considering the trade-off, and orchestrates multiple memory adjustments in a controlled and parallel manner to avoid OOM hazards. 
Lastly, to maintain efficiency, \oursys introduces a dual-approach consolidator: prioritizing vertical scaling through proactive preemption while employing a bin-packing strategy to eliminate fragmentation.

The main contributions of this paper are as follows.

\begin{itemize}[leftmargin=6mm]
	\item \textbf{Systematic investigation of LLM serving on heterogeneous resources. }
	The identified CPU/GPU sharing opportunities motivate a resource-efficient design.
	\item \textbf{Solutions for sharing small- to mid-sized LLMs under serverless paradigm.}
	Based on investigation, we construct guidelines considering unique characteristics of LLM inference procedure and serverless workloads.
	\item \textbf{A resource management system with unified hardware abstraction.}
	Based on \oursys, we implement two subsystems that transparently manage hardware while ensuring efficient and precise on-demand resource sharing.
\end{itemize}

We evaluate \oursys with real-world LLM datasets~\cite{DBLP:conf/isca/PatelCZSGMB24} and serverless workloads~\cite{DBLP:conf/usenix/ShahradFGCBCLTR20}.
Experimental results on 4 32-core CPU nodes and 4 A100-80GB GPU nodes demonstrate that \oursys improves serving capacity by 47\% - 62\% through elastic sharing, and leveraging CPU resources further boosts this improvement to 86\% - 154\%.

\section{Related Work}

\textbf{Heterogeneous serverless computing.}
Designing serverless systems with heterogeneous hardware~\cite{DBLP:conf/middleware/PfandzelterDFCE23,DBLP:conf/cloud/RomeroZYK21} offers significant opportunities. 
Molecule~\cite{DBLP:conf/asplos/DuLJXZC22} enables serverless computing to run seamlessly across heterogeneous computers, DSCS-Serverless~\cite{DBLP:conf/asplos/MahapatraGAKWXK24} leverages programmable accelerators to unlock the potential of data centers, IceBreaker~\cite{DBLP:conf/asplos/RoyPT22} improves cold-start by mixing heterogeneous instances, and INFaaS~\cite{DBLP:conf/usenix/Romero0YK21} reduces costs for serving traditional models by automatically selecting the optimal hardware architecture. 
In the context of LLM serving, \oursys also identifies opportunities to leverage heterogeneous hardware effectively.

\textbf{Traditional and serverless model serving systems.}
Before the rise of LLM, traditional model serving systems~\cite{DBLP:conf/nsdi/CrankshawWZFGS17,DBLP:conf/ppopp/FangYZZ21,DBLP:conf/osdi/LeeSCSWI18,DBLP:conf/usenix/ChoiLKPKH22,DBLP:conf/osdi/LiZZL00HCZGS23}—such as Clockwork~\cite{DBLP:conf/osdi/GujaratiKAHKVM20}, Cocktail~\cite{DBLP:conf/nsdi/GunasekaranMTSK22}, and SHEPHERD~\cite{DBLP:conf/nsdi/0025TKS23}—had introduced numerous optimizations in scheduling and resource management. 
Among them, BATCH~\cite{DBLP:conf/sc/AliP0S20}, INFless~\cite{DBLP:conf/asplos/YangZLZLZCL22}, and Dilu~\cite{DBLP:conf/asplos/Lv0LHTZZ25} explored applying serverless paradigms. 
However, traditional models differ significantly from LLMs in their resource demands and execution patterns. 
The latter defines SLOs at token-level and executes in a multi-iteration manner with fluctuating compute/memory demand, necessitating the specialized serving systems.

In response, a wave of LLM-oriented solutions~\cite{DBLP:conf/osdi/YuJKKC22,DBLP:conf/asplos/MiaoOZCWZWZYSSC24,DBLP:conf/asplos/OhKKKLC024,DBLP:conf/sosp/WuLZ0L024,DBLP:conf/asplos/PatelCZGWMB24} has emerged. 
vLLM~\cite{DBLP:conf/sosp/KwonLZ0ZY0ZS23} enhances memory efficiency with paged-attention, Llumnix~\cite{DBLP:conf/osdi/SunHZXZL024} dynamically schedules requests across instances, and SpotServe~\cite{DBLP:conf/asplos/MiaoSDXL0J24} considers preemptible instances. 
A series of approaches~\cite{DBLP:journals/corr/abs-2410-18038,DBLP:conf/isca/PatelCZSGMB24,DBLP:conf/osdi/ZhongLCHZL0024,DBLP:conf/osdi/AgrawalKPMKGTR24} have been proposed to consider the differences between prefill and decode stages. 
They primarily focus on high-load scenarios with a single LLM. 
Meanwhile, MuxServe~\cite{DBLP:conf/icml/DuanLDLZLS024} adopts static GPU sharing for multi-LLM serving but relies on predictable workloads, which does not hold in serverless settings with highly dynamic and bursty workloads.
Finally, for serverless LLM serving, ServerlessLLM~\cite{DBLP:conf/osdi/FuXHBUPM24}, Medusa~\cite{DBLP:conf/asplos/ZengXGCL25}, and ParaServe~\cite{lou2025swiftserverlessllmcold} improve cold-start but still allocate dedicated GPUs to each LLM. 
\oursys focuses on resource sharing through elastic allocation and is orthogonal to them.


\textbf{CPU-assisted LLM inference.}
Given the scarcity of GPUs, many works~\cite{DBLP:journals/corr/abs-2411-09317,DBLP:conf/sosp/SongMX024,DBLP:journals/corr/abs-2401-11240} explore leveraging CPUs to assist LLM inference. 
Early systems such as PowerInfer~\cite{DBLP:conf/sosp/SongMX024} offload infrequently accessed model parameters to the CPU.
NEO~\cite{jiang2025neo} and FastDecode~\cite{DBLP:journals/corr/abs-2403-11421} further offload KV-cache along with the associated attention computations to the CPU, thereby alleviating GPU memory pressure.
In these designs, CPUs primarily serve as auxiliary resources, handling lightweight or memory-bound tasks, while GPUs remain the dominant compute devices.

Recently, the emergence of CPUs equipped with matrix acceleration units (e.g., Intel AMX~\cite{AMX,DBLP:conf/isscc/NassifMMPLYMHVK22}) has reshaped this landscape.
LIA~\cite{DBLP:conf/isca/KimWX0YK25} demonstrates that AMX-enabled Intel CPUs can deliver matrix multiplication throughput comparable to certain low-end GPUs.
Building on this capability, systems such as FlexInfer~\cite{na2025flexinfer} further offload parts of model layers to CPUs.
However, these works still rely on GPUs as the base of execution and require tight CPU–GPU coupling.
In contrast, \oursys explores the CPU’s potential for independent serving, improving deployment density through unified heterogeneous resource management.

\section{Background and Motivation}

\subsection{LLM Inference Process}
\label{sec:llm_inference_process}
In LLM inference, users submit requests containing input tokens, which the inference engine processes iteratively~\cite{DBLP:conf/asplos/MiaoOZCWZWZYSSC24,DBLP:conf/sosp/WuLZ0L024,DBLP:conf/osdi/AgrawalKPMKGTR24,DBLP:conf/osdi/SunHZXZL024}. Each iteration generates one output token, which is streamed back to user in real time.

A request undergoes two stages~\cite{DBLP:conf/asplos/PatelCZGWMB24,DBLP:conf/isca/PatelCZSGMB24,DBLP:conf/osdi/ZhongLCHZL0024}. 
The prefill stage occurs during the first iteration, where the engine builds the key-value (KV) cache~\cite{DBLP:conf/sosp/KwonLZ0ZY0ZS23,TensorRT} and generates the first output token. 
In the decode stage, the engine appends to the KV-cache and generates one token per iteration. 
To improve concurrency, inference engines adopt continuous batching~\cite{DBLP:conf/osdi/YuJKKC22} to dynamically incorporate new requests into ongoing batches.

Interactive LLM serving systems should follow strict Service Level Objectives (SLOs). 
Two key metrics are Time-to-First-Token (TTFT) and Time-per-Output-Token (TPOT). 
TTFT is typically constrained to a few seconds~\cite{DBLP:conf/osdi/ZhongLCHZL0024} and grows with input length, while TPOT should keep up with human reading speed, which is around 250 tokens/min~\cite{DBLP:conf/osdi/AgrawalKPMKGTR24}.

Once LLM serving meets above SLOs, it can operate as a reliable productivity tool like ChatGPT~\cite{chatgpt}.
Ongoing contributions from open-source communities~\cite{touvron2023llama2openfoundation,gemmateam2024gemmaopenmodelsbased} have further expanded the accessibility and diversity of LLMs.

\subsection{Small- to Mid-Sized LLMs and Private Deployments}

\begin{figure}
	\centering
	\begin{minipage}{0.48\linewidth}
		\centering
		\includegraphics[width=.74\linewidth]{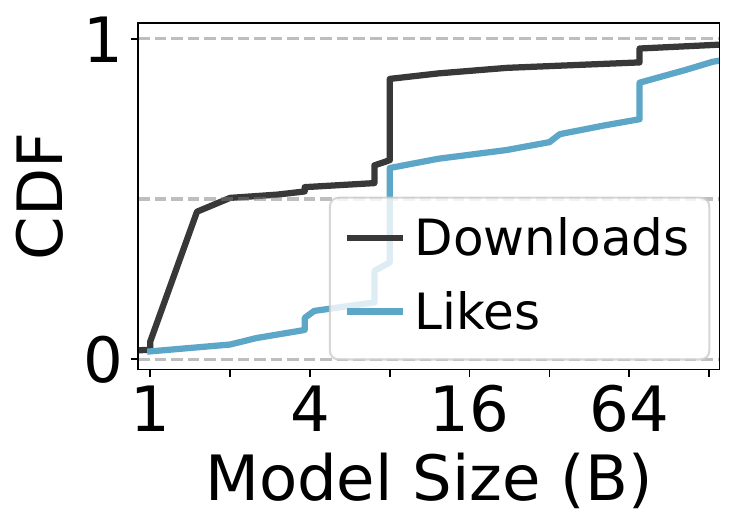}
				\vspace{-1mm}
		\caption{Popularity of LLMs' size from HuggingFace~\cite{huggingface-popular}.}
        \vspace{-1mm}
		\label{fig:model_popularity}
	\end{minipage}
	\hfill
	\begin{minipage}{0.48\linewidth}
		\centering
		\includegraphics[width=.84\linewidth]{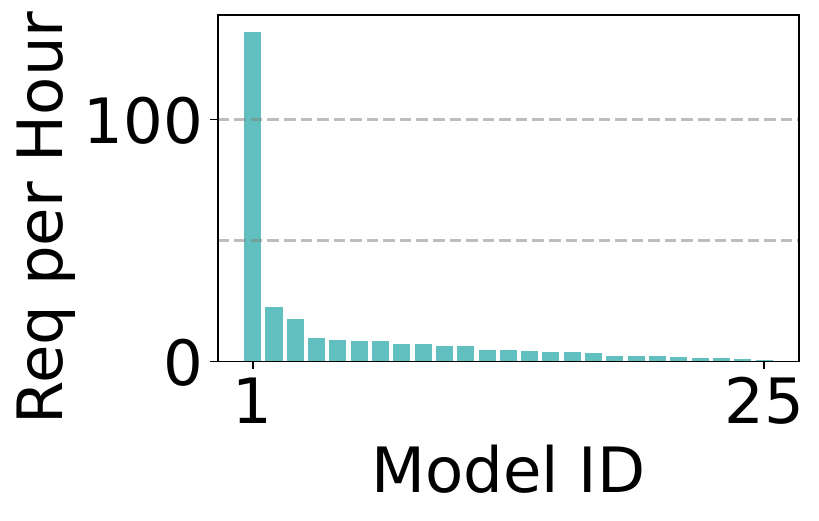}
				\vspace{-1mm}
		\caption{Invocation frequencies of 25 LLMs in LMSYS~\cite{DBLP:conf/iclr/ZhengC0LZW00LXG24}.}
        \vspace{-1mm}
		\label{fig:model_frequency}
	\end{minipage}
\end{figure}

In practice, small- to mid-sized LLMs (e.g., 7B and 13B) have proven effective in addressing most application scenarios, while offering significantly lower operational costs~\cite{DBLP:journals/corr/abs-2408-00118,Phi-2}.
Meanwhile, increasing demands for customization, coupled with privacy concerns, have driven many users to adopt private deployments.
For instance, the developers have created over 1,100 customized variants of Llama-2-7B alone~\cite{llama-2-7b}.
A closer examination of this trend reveals two key characteristics:

\begin{itemize}[leftmargin=6mm]
	\item First, small- to mid-sized models dominate private deployments, as Figure \ref{fig:model_popularity} shows. HuggingFace data~\cite{huggingface-popular} indicates that models with fewer than 8 billion parameters constitute 60\% of user preferences and 87\% of total downloads, reflecting practical concerns about cost efficiency.

	\item Second, invocations are infrequent and highly variable~\cite{DBLP:conf/iclr/ZhengC0LZW00LXG24,DBLP:conf/osdi/FuXHBUPM24}, as Figure \ref{fig:model_frequency} shows. 
	In the most popular multi-LLM dataset, \textit{LMSYS-Chat-1M}~\cite{DBLP:conf/iclr/ZhengC0LZW00LXG24}, most models receive only a handful of requests per hour on average.
	This stems from private deployments serving limited user base, unlike the high-throughput public APIs\cite{chatgpt,Claude}.

\end{itemize}

Given the growing demand for private LLM deployments, cloud providers have introduced one-stop hosting solutions~\cite{huggingface-serverless,google-serverless,azure-serverless}, where users simply upload their models while offloading the complexity of infrastructure management.

\subsection{Problems with Existing Serverless LLM Solutions}

\begin{figure}
	\centering
	\begin{minipage}{0.45\linewidth}
		\centering
		\includegraphics[width=.82\linewidth]{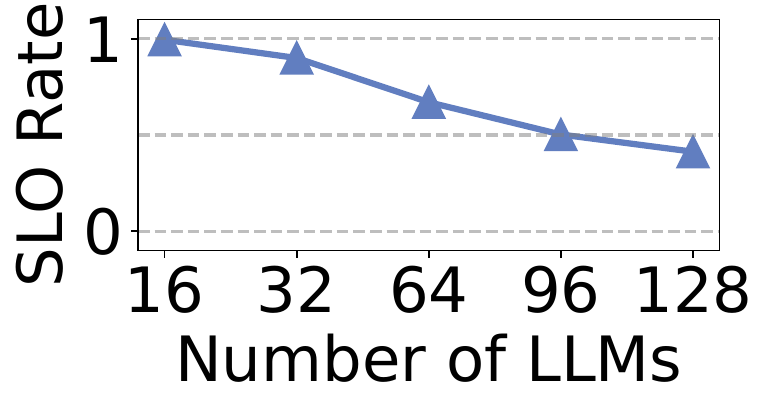}
		\caption{ServerlessLLM's serving capacity across different workload levels.}
		\vspace{-1mm}
		\label{fig:sllm_moti_SLO}
	\end{minipage}
	\hspace{2mm}
	\begin{minipage}{0.45\linewidth}
		\centering
		\includegraphics[width=.81\linewidth]{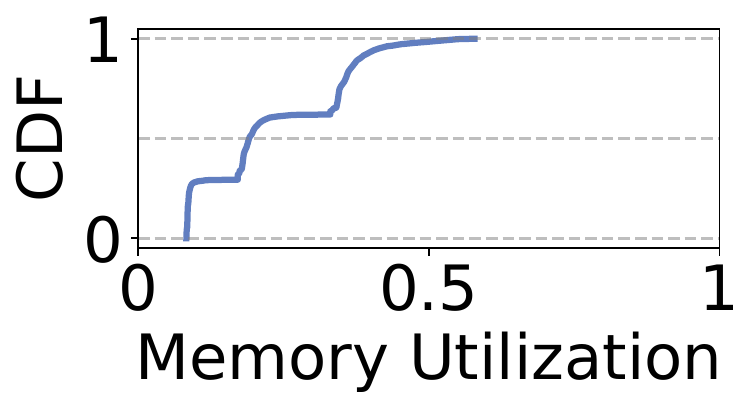}
		\caption{GPU memory utilization when serving 128 LLMs with ServerlessLLM.}
		\vspace{-1mm}
		\label{fig:sllm_moti_memory}
	\end{minipage}
\end{figure}

To improve serving capacity in private deployments, researchers have begun exploring serverless architecture for orchestrating and managing multiple LLMs on the cloud. 
Representative systems such as ServerlessLLM~\cite{DBLP:conf/osdi/FuXHBUPM24}, Medusa~\cite{DBLP:conf/asplos/ZengXGCL25}, and DeepServe~\cite{DBLP:conf/usenix/HuXLHCXLMZWDDRL25} host multiple LLMs within a cluster and dynamically allocate GPUs to each model on demand.
Upon receiving a request, the system launches a new model instance on an available GPU if none is currently running. If no GPU is idle, the request is queued for available resources.

However, we observe that existing solutions still struggle to handle large numbers of low-traffic, small- to mid-sized LLMs. 
Taking ServerlessLLM as a typical example: It enables fast model loading and utilizes vLLM~\cite{DBLP:conf/sosp/KwonLZ0ZY0ZS23} as the internal inference engine. 
We use it to host a mix of 3B, 7B, and 13B LLMs on four A100-80GB GPUs, following the same setup in \S~\ref{sec:experimental_setup}. 
As shown in Figure~\ref{fig:sllm_moti_SLO}, it performs well at small scales. But as the number of LLMs increases, the SLO attainment rate drops sharply as requests heavily queue for limited GPUs.

This situation arises because existing serverless solutions over-provisioning GPU resources for each model: When being allocated the entire GPU memory, each instance utilizes only 23\% of it on average, as shown in Figure~\ref{fig:sllm_moti_memory}.
Moreover, the CPUs are mostly idle, as the computations happens on GPUs.

These observations motivate us to take a step back and reassess the evolving architectures and practical workload scales of small- to mid-sized LLMs. 
Instead of being constrained by scarce GPUs, alternative hardware like CPUs might offer viable solutions. 
Moreover, these heterogeneous resources could potentially enable efficient multi-model sharing, rather than being exclusively allocated.
To this end, we next conduct a systematic investigation of heterogeneous architectures to explore the sharing opportunities in serverless LLM serving.

\begin{figure*}
	\centering
	\begin{minipage}{0.20\textwidth}
		\centering
		\includegraphics[width=\linewidth]{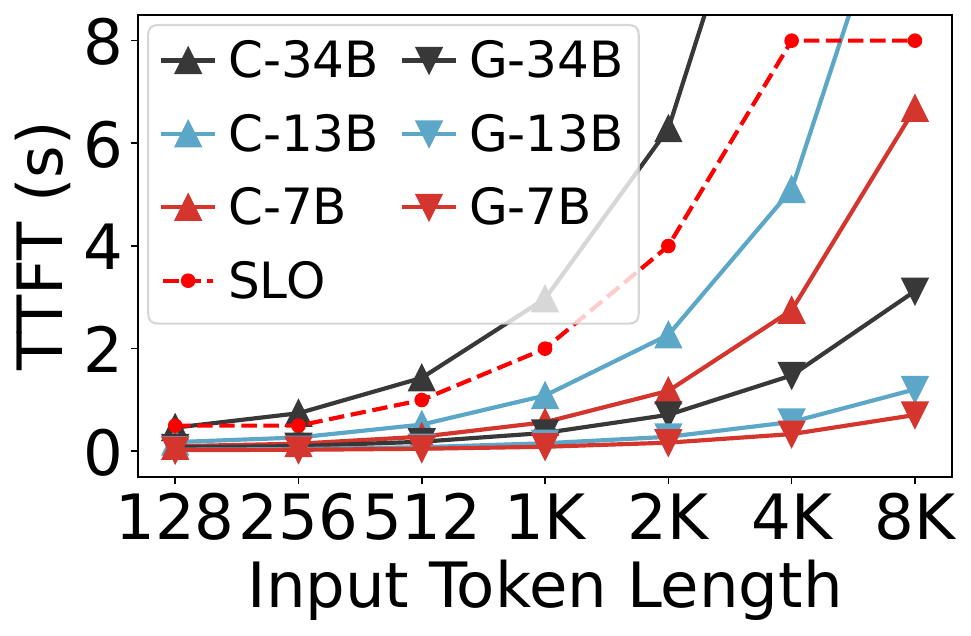}
		\caption{The TTFT metric of diverse models using CPU/GPU.}
		\label{fig:cpu_ttft_profile}
		\vspace{-1mm}
	\end{minipage}
	\hspace{1.5mm}
	\begin{minipage}{0.225\textwidth}
		\centering
		\includegraphics[width=.98\linewidth]{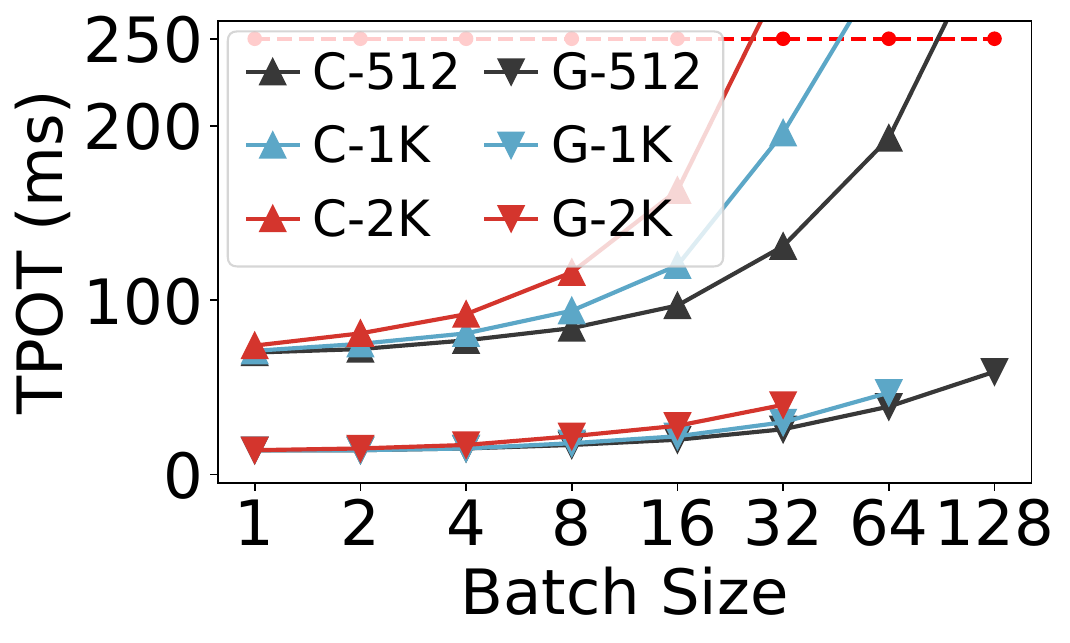}
		\caption{The TPOT metric under different token length of Llama-2-7B.}
		\label{fig:cpu_tpot_profile_7b}
		\vspace{-1mm}
	\end{minipage}
	\hspace{1.5mm}
	\begin{minipage}{0.225\textwidth}
		\centering
		\includegraphics[width=.98\linewidth]{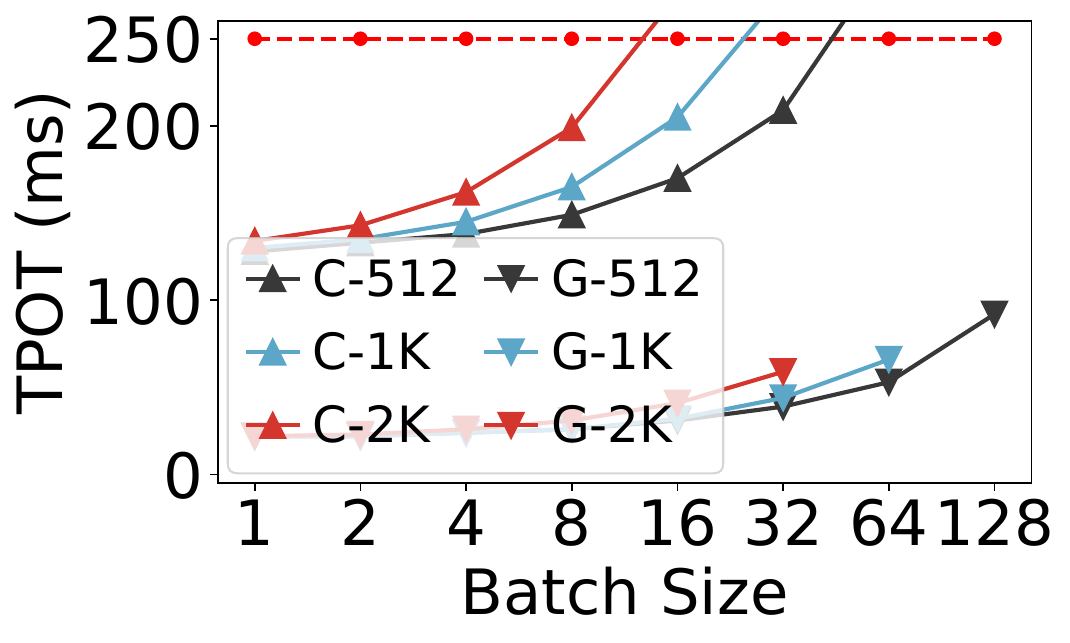}
		\caption{The TPOT metric under different token length of Llama-2-13B.}
		\label{fig:cpu_tpot_profile_13b}
		\vspace{-1mm}
	\end{minipage}
	\hspace{1.5mm}
	\begin{minipage}{0.275\textwidth}
		\centering
		\includegraphics[width=.98\linewidth]{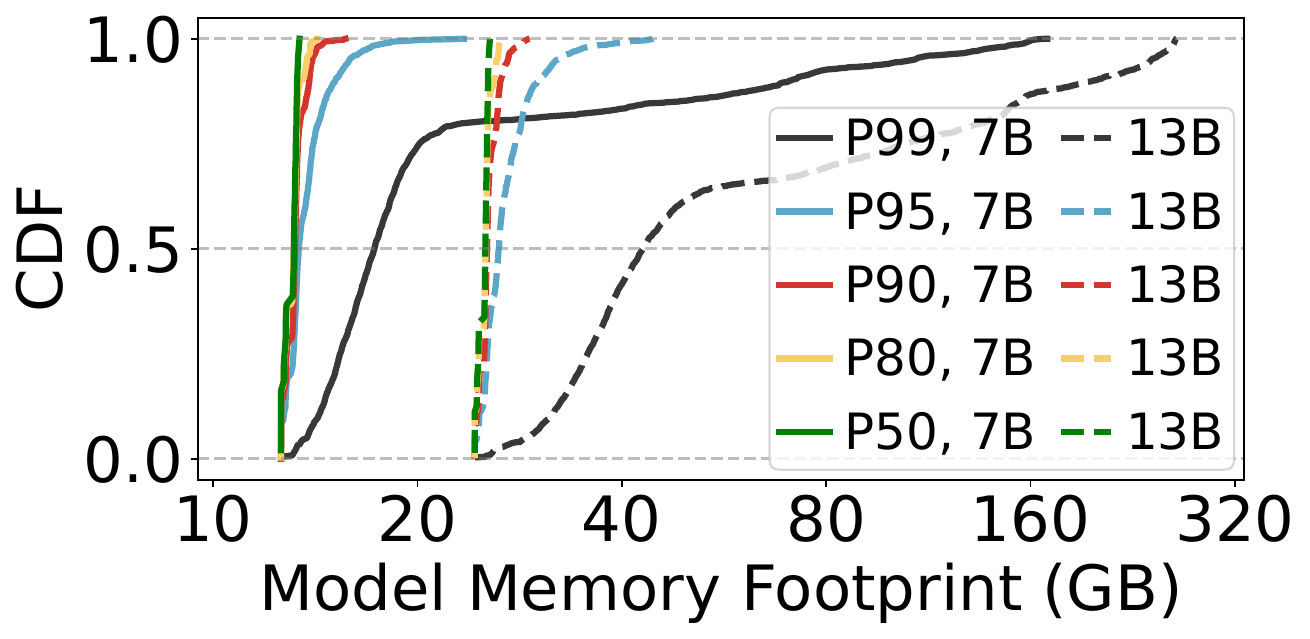}
		\caption{The memory footprint of different models under real-world workloads.}
		\label{fig:moti_model_memory_footprint}
		\vspace{-1mm}
	\end{minipage}
\end{figure*}

\begin{figure}
	\centering
	\begin{minipage}{0.55\linewidth}
		\centering
		\includegraphics[width=.90\linewidth]{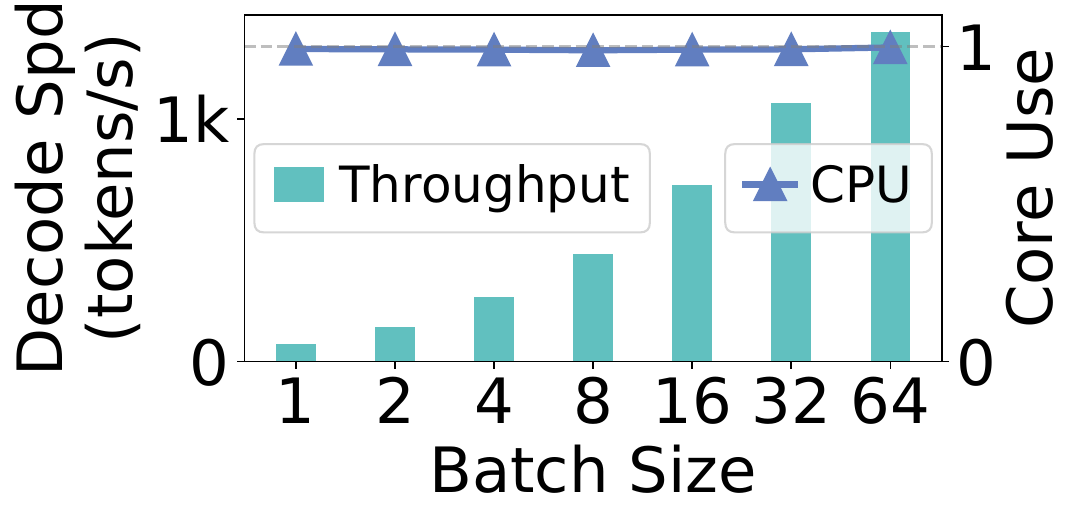}
		\caption{vLLM's GPU decode throughput and CPU core usage under different batch sizes.}
		\vspace{-1mm}
		\label{fig:vllm_cpu_usage}
	\end{minipage}
	\hspace{2mm}
	\begin{minipage}{0.40\linewidth}
		\centering
		\includegraphics[width=.9\linewidth]{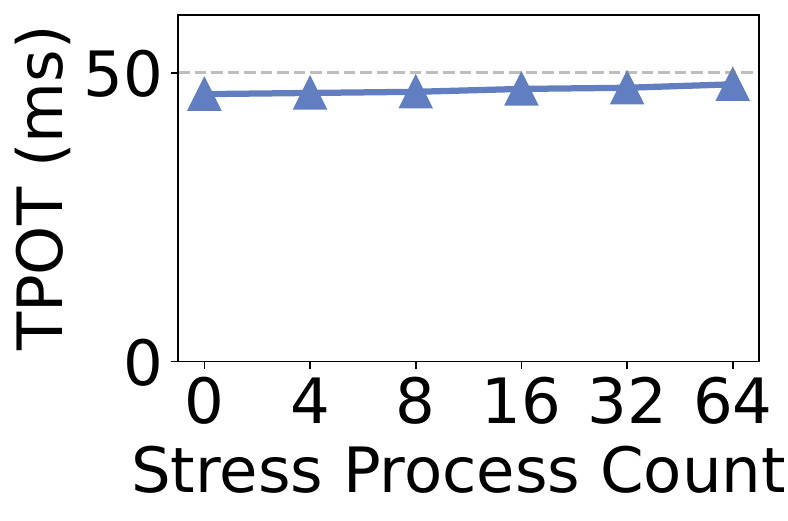}
		\caption{vLLM's TPOT slowdown under background CPU stress.}
		\vspace{-1mm}
		\label{fig:vllm_cpu_pressure}
	\end{minipage}
\end{figure}

\section{Heterogeneous Resource Sharing}
\label{sec:resource_sharing}

Modern data centers are inherently heterogeneous~\cite{DBLP:conf/osdi/JiangZLYCG20,DBLP:conf/osdi/NarayananSKPZ20,DBLP:journals/tpds/ChenLOZL18}. Even a GPU cluster is equipped with CPU nodes for preprocessing tasks. 
Given the reported low CPU utilization on GPU nodes~\cite{DBLP:conf/osdi/JiangZLYCG20,DBLP:conf/usenix/JeonVPQXY19} and the emerging CPU architecture~\cite{DBLP:conf/isscc/NassifMMPLYMHVK22}, it is worth exploring the potential of idle CPU resources. 

However, due to the fundamental architectural differences, CPUs typically offer limited parallelism and are well-known to be compute-bound for LLM inference~\cite{DBLP:conf/sosp/SongMX024,DBLP:conf/IEEEpact/Park024}. 
On the other hand, GPUs are often memory-bound due to their limited memory capacity~\cite{DBLP:conf/osdi/SunHZXZL024,DBLP:journals/corr/abs-2403-11421}. 
Therefore, it is crucial to evaluate the computation latencies in CPUs, and the memory footprints in GPUs to further assess the sharing potential.

\subsection{CPU Sharing Opportunity}

\subsubsection{Spare CPU Resources}
\label{sec:spare_cpu_resource}
We measured the CPU utilization of state-of-the-art inference engine vLLM~\cite{DBLP:conf/sosp/KwonLZ0ZY0ZS23} when serving Llama-2-7B model on an A100 GPU with a 32-core CPU.

In Figure \ref{fig:vllm_cpu_usage}, vLLM's throughput increases with batch size, but never consumes more than one CPU core. 
To further evaluate vLLM's CPU sensitivity, we launched background CPU stress processes while running it with a batch size of 64. 
As shown in Figure \ref{fig:vllm_cpu_pressure}, even with 64 stress processes competing for 32 CPU cores, vLLM suffers only a 4\% performance loss.
Given that GPU nodes typically feature dozens or even hundreds of CPU cores~\cite{DBLP:conf/nsdi/WengXYWWHLZLD22}, substantial CPU resources are waiting to be utilized under LLM inference scenarios.

\subsubsection{CPU Computational Capability}
\label{sec:cpu_capability}
Despite the presence of spare CPU resources, their feasibility for LLM inference remains uncertain due to stringent SLOs and high compute loads.
However, it is worth noting that recent CPU architectures have integrated specialized components to accelerate the AI workloads.
Starting with 4th-Gen Intel Xeon, Intel introduced Advanced Matrix Extensions (AMX)~\cite{AMX,DBLP:conf/isscc/NassifMMPLYMHVK22}, a dedicated hardware block designed for matrix operations. 

Although using AMX has been shown to provide acceleration~\cite{DBLP:conf/iiswc/NaJA0KK24}, detailed latency data under SLO constraints remains underexplored.
To benchmark it, we replace vLLM's GPU backend with OpenVINO~\cite{openvino}, the state-of-the-art for CPU inference.
We use a AMX-equipped 32-core Intel Xeon 6462C CPU, testing three LLMs of varying sizes (Llama-2-7B, Llama-2-13B, and CodeLlama-34B) under different token lengths and batch sizes. 
Following previous works~\cite{DBLP:conf/osdi/ZhongLCHZL0024,DBLP:conf/osdi/AgrawalKPMKGTR24}, we set TTFT SLO to $\min(\max(0.5, \text{input\_length} / 512),8)$ s and TPOT SLO to 0.25 s.

Figure \ref{fig:cpu_ttft_profile} presents the TTFT data.
The label ``C-7B'' denotes using CPU with Llama-2-7B.
We compare the results with an A100 GPU and the SLO. 
CPUs can meet the SLOs of 7B and 13B LLMs under short inputs, which cover most usage scenarios—e.g., 97.9\% of conversation and 85.9\% of coding inputs in the Azure LLM trace are under 4K tokens~\cite{DBLP:conf/isca/PatelCZSGMB24}.

We further examine the TPOT data of the 7B and 13B LLMs, which characterizes the per-token latency during decode, as shown in Figure \ref{fig:cpu_tpot_profile_7b} and Figure \ref{fig:cpu_tpot_profile_13b}. The label ``C-512'' denotes using CPU with a token length of 512. 
We find that the CPU not only meets TPOT SLO with ease but can also utilize batching to improve throughput, similar to GPU. 
For example, serving 7B LLM on CPU with a token length of 1K, the TPOT for a 4-batch increases by only 14\% compared to a 1-batch.
We also find that the TPOT also correlates with token length. 
For instance, serving 13B LLM on CPU with a 32-batch results in a 2X increase in TPOT when the length increases from 512 to 2K, with the latter violating the SLO.

\begin{table}[t]
	\small
	\centering
	\caption{Llama-2-7B's performance under 3rd- (32-core@2.7GHz) and 4th-Gen (32-core@3.3GHz) Xeon CPUs. ``bs'' denotes ``batch size''. Red cells indicate SLO violations.}
	\label{tab:cpu-comparison}
	\renewcommand{\arraystretch}{1.2}
	\setlength{\tabcolsep}{2.8pt}
	\begin{tabular}{lccccccc}
		\toprule
		\multirow{2.5}{*}{CPU} & 
		\multicolumn{3}{c}{TTFT (ms)} & 
		\multicolumn{4}{c}{TPOT (ms)} \\
		\cmidrule(lr){2-4} \cmidrule(lr){5-8}
		& 256 & 1K & 4K & 1bs-1K & 32bs-1K & 1bs-4K & 32bs-4K \\
		\midrule
		3rd Gen & \cellcolor{red!20}1003 & \cellcolor{red!20}4113 & \cellcolor{red!20}18612 & 100 & \cellcolor{red!20}338 & 110 & \cellcolor{red!20}697 \\
		4th Gen & 149 & 567 & 2748 & 71 & 196 & 80 & \cellcolor{red!20}459 \\
		Speedup & 6.7$\times$ & 7.3$\times$ & 6.8$\times$ & 1.4$\times$ & 1.7$\times$ & 1.4$\times$ & 1.5$\times$ \\
		\bottomrule
	\end{tabular}
\end{table}

\textbf{Limitations and Applicable Scenarios.}
Overall, although CPUs offer enhanced capability, they have several limitations:
(1) \textit{Dependence on newer hardware.}
Older CPUs without specialized matrix acceleration block are generally unsuitable~\cite{DBLP:conf/isca/KimWX0YK25}. As shown in Table~\ref{tab:cpu-comparison}, a 32-core 3rd Gen Xeon 8369B (without AMX) running Llama-2-7B with 1K inputs results in a TTFT of 4.1 s—far exceeding the SLOs.
(2) \textit{Sensitivity to model size and workload.}
CPUs can only handle small LLMs ($\le$13B), short inputs ($\le$5.6K for a 13B model), and limited batch sizes.
(3) \textit{Inability under tight SLOs.}
Under a 100 ms TPOT SLO, only 7B or smaller LLMs are feasible, with batch sizes limited to 9 for 1K-length and 3 for 4K-length. At 50 ms, even 7B LLMs become infeasible.
Nevertheless, in serverless scenarios with many small- to mid-sized LLMs and infrequent requests, AMX-equipped CPUs present opportunities for resource sharing under moderate SLOs.

\subsection{GPU Sharing Opportunity}
\label{sec:gpu_opportunity}

An LLM instance's memory footprint primarily consists of model weights and KV-cache. 
While the weights are fixed, the KV-cache is dynamic with request concurrency and token length. 
To capture realistic memory usage, we sample token lengths from Azure LLM Trace~\cite{DBLP:conf/isca/PatelCZSGMB24}. 
Since it lacks multi-LLM invocation patterns, following ServerlessLLM~\cite{DBLP:conf/osdi/FuXHBUPM24}, we fire requests based on Azure Serverless Trace~\cite{DBLP:conf/usenix/ShahradFGCBCLTR20}.

Figure \ref{fig:moti_model_memory_footprint} shows the memory usage of the 7B and 13B model under real-world workloads on 4 A100-80GB GPUs. 
The label ``P99, 7B'' represents mapping the Llama-2-7B model to the top 1\% most frequently invoked function in the Azure Trace.
Since each instance occupies 1 GPU, a footprint exceeding 80GB implies that multiple instances are created.

\begin{figure}
	\centering
	\begin{minipage}{0.38\linewidth}
		\centering
		\includegraphics[width=\linewidth]{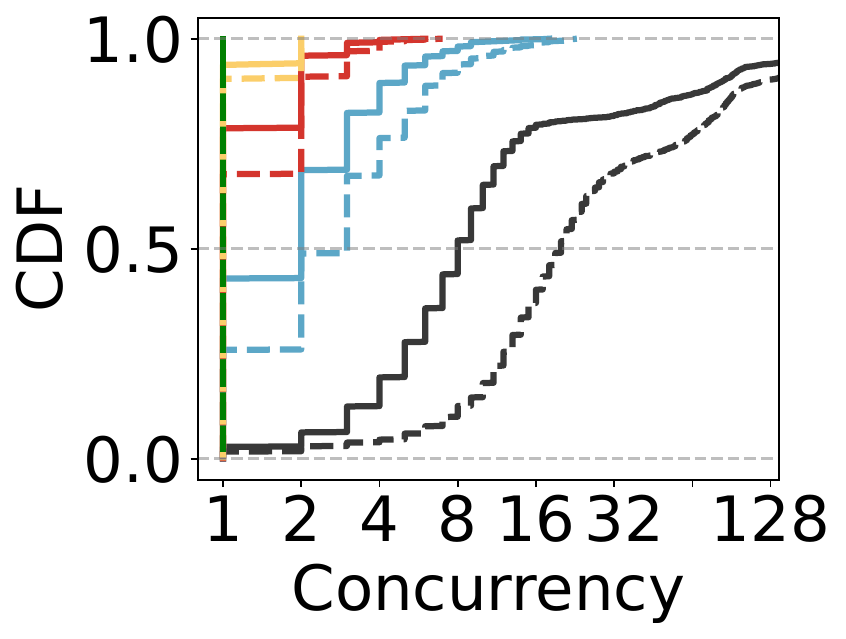}
		\caption{CDF of workload concurrency. Legend same as Fig.~\ref{fig:moti_model_memory_footprint}.}
		\vspace{-1mm}
		\label{fig:moti_batch_cdf}
	\end{minipage}
	\hspace{1mm}
	\begin{minipage}{0.58\linewidth}
		\centering
		\footnotesize
		
		\begin{tabular}{l@{\hskip 4pt}c@{\hskip 6pt}c@{\hskip 6pt}c@{\hskip 6pt}c}
			\toprule
			\textbf{Scenarios} & \textbf{4$\times\frac{\bf1}{\bf4}$} & \textbf{3$\times\frac{\bf1}{\bf3}$} & \textbf{2$\times\frac{\bf1}{\bf2}$} & \textbf{1} \\
			\midrule
			C-7B-2K & -  & $3\!\times\!2$  & $2\!\times\!9$ & 27 \\
			C-7B-4K & -  & $3\!\times\!1$  & $2\!\times\!4$  & 15 \\
			G-7B-2K & $4\!\times\!6$  & $3\!\times\!12$  & $2\!\times\!26$ & 66 \\
			G-7B-4K & $4\!\times\!3$  & $3\!\times\!6$  & $2\!\times\!13$  & 32 \\
			G-13B-2K & -  & -  & $2\!\times\!7$ & 33 \\
			G-13B-4K & -  & -  & $2\!\times\!3$  & 16 \\
			\bottomrule
		\end{tabular}
		\captionof{table}{\label{tab:moti_batch_efficiency}Aggregated concurrency limits of instances under varying resource specifications.}
	\end{minipage}
\end{figure}

For 7B and 13B LLMs, they need at least 14GB and 26GB of memory, respectively, corresponding to the model weights, regardless of the workload. 
Under the top 1\% workload, memory footprint can peak at 169GB (7B) and 263GB (13B), due to bursts of over 128 concurrent requests (shown in Figure~\ref{fig:moti_batch_cdf}), necessitating exclusive use of GPUs. 
However, even under the top 1\%, more than 50\% of the time, memory footprint remains below 17GB (7B) and 43GB (13B).

\textbf{Takeaway.}
One model's memory footprint remains low in most cases.
Given that GPUs like A100 feature 80GB of memory, LLMs can be co-located under serverless workloads.

\subsection{Summary and Challenges}

The above explorations reveal opportunities for resource sharing on both CPUs and GPUs. 
However, straightforward approaches—such as statically assigning a fraction of resources to each model instance—yield negligible improvement in serving capacity. 
This stems from the inability of small instances to effectively absorb bursty traffic, as large batches typically require full hardware access.
For example, as shown in Table~\ref{tab:moti_batch_efficiency}, partitioning a GPU into three smaller instances when serving 7B LLMs achieves only about half the aggregate concurrency limit of a single large instance.
Yet in serverless workloads, most requests originate from a few hot functions exhibiting bursty behavior~\cite{DBLP:conf/usenix/ShahradFGCBCLTR20}. 
As shown in Figure~\ref{fig:moti_batch_cdf}, the top 1\% experiences concurrency levels ranging from 1 to over 128, and alone contributes to 26\% of the total requests.
This coexistence of burstiness, low frequency, and variability makes static partitioning fundamentally inefficient, which we further evaluate in Sections~\ref{sec:eval_e2e}, ~\ref{sec:eval_mixed_deployment}, and ~\ref{sec:gpu_efficiency}.

Given the workload characteristics of small- to mid-sized LLMs, elastic and dynamic sharing based on each instance’s real-time demand presents a promising approach to maximizing serving capacity.
To realize such sharing, we closely examine the compute and memory behaviors of LLM instances and encounter three design challenges (recall Figure~\ref{fig:challenge}).


\textbf{Challenge-1: Timely and precise compute resource allocation.}
The compute demand of an instance fluctuates sharply at token level.
In Section \ref{sec:cpu_capability}, Llama-2-7B running on a 32-core CPU takes 567 ms to generate the first token for a 1024-token input request, while subsequent tokens requires significantly less time (e.g., 71 ms). 
In addition, the token length and batching behavior introduces further variability.
Unlike traditional setups with dedicated resources, \textit{multi-model sharing under serverless scenarios requires the system to precisely budget and allocate compute resources on a per-token basis, dynamically adjusting to fluctuating demands across concurrent instances to consistently meet SLOs.}

\textbf{Challenge-2: Efficient and safe memory sharing.}
A model’s memory demand is highly bursty—its peak can reach up to 12× in Figure~\ref{fig:moti_model_memory_footprint}.
While dynamic memory resizing is essential for efficiency, we observe that such resizing incurs non-trivial overhead: under widely-used paged attention mechanism~\cite{DBLP:conf/sosp/KwonLZ0ZY0ZS23}, changing the KV-cache requires allocating new matrices~\cite{DBLP:conf/asplos/ZengXGCL25,DBLP:conf/asplos/PrabhuNMRP25} and migrating already-used cache pages (detailed in Figure~\ref{fig:kv_scale_time}).
Moreover, frequent operations like model loading/unloading coexist with these resizes. 
When multiple instances co-reside, arbitrary operations can lead to OOM and compromise system stability.
\textit{Thus, the system should balance memory utilization and operational cost, constructing a global-orchestrated memory scaling mechanism.}

\textbf{Challenge-3: Maintaining resource efficiency in shared environments.}
LLM inference relies on batching to improve compute efficiency, as larger batches yield sub-linear growth in compute cost (see Figure~\ref{fig:cpu_tpot_profile_7b}). 
To increase the batch size, an instance needs to scale up its compute and memory resources. However, in a shared setup, these resources may already be occupied by co-located models, forcing the instance to scale out by launching a fragmented replica on another node.
This not only leads to scattered batches, but also incurs redundant memory overhead from duplicated model weights. 
\textit{Therefore, it is essential to proactively identify potential fragmentation issues and assist instances in scaling up.}

\section{Design Overview of \oursys}

To address the above challenges,
we present \textit{\oursys}, a \underline{S}erverless \underline{L}LM \underline{Infer}ence scheme designed for small- to mid-sized LLMs in heterogeneous data centers. 
It transparently leverages diverse hardware and elastically shares resources on demand to maximize serving capacity.

\begin{figure}
	\centering
	\includegraphics[width=.82\linewidth]{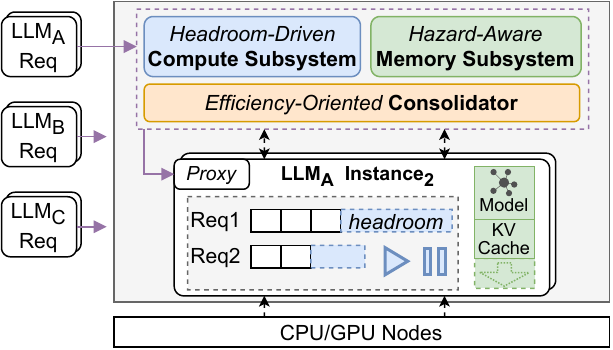}
	\caption{The design architecture of \oursys.}
        \vspace{-2mm}
	\label{fig:overview}
\end{figure}

Specifically, \oursys coordinates multiple LLMs on both CPU and GPU nodes through the compute and memory subsystems, alongside a consolidation module, as shown in Figure~\ref{fig:overview}. 
\oursys follows an event-driven approach to deploy multiple instances, where instances are placed using a bin-packing strategy to minimize resource usage. 
It handles a request’s prefill and decode within the same instance: the prefill runs independently, while the decode joins the instance’s existing batch.
Assuming an LLM already has several instances, we illustrate the components and workflow of SLINFER through a request lifecycle.

When a new request arrives, \oursys first attempts to assign it to existing instances, prioritizing those on CPU nodes.
Since CPU generations differ substantially in performance (see Table~\ref{tab:cpu-comparison}), \oursys excludes CPUs that lack dedicated matrix-acceleration (e.g., AMX) support. Moreover, as Section~\ref{sec:cpu_capability} shows that CPUs can only serve a limited range of models and workloads, \oursys profiles CPUs in advance and transparently falls back to GPU instances whenever a CPU cannot meet the request’s SLO requirements.

Specifically, to schedule a new request, the compute subsystem performs shadow validation, checking whether a candidate instance can absorb the request without violating the SLOs of other requests on the same node by calculating per-request headroom.
Simultaneously, the memory subsystem verifies whether the node has enough available memory to accommodate the request. 
If both checks succeed, the request is dispatched to the selected instance.

Subsequently, the compute subsystem orchestrates execution at token-level, focusing on request headroom (Challenge 1, see §\ref{sec:compute_subsystem}). The memory subsystem employs a watermark-based scaling and a hazard-aware out-of-order operation strategy, ensuring efficient and safe sharing (Challenge 2, see §\ref{sec:memory_subsystem}).

If no instance passes the validation, \oursys introduces a consolidator, which attempts to proactively preempt resources from neighboring instances to avoid launching a new fragmented one, thereby improving overall efficiency (Challenge 3, see §\ref{sec:defragementation}). If all attempts fail, it falls back to creating a new instance, using the same validation procedure.

Upon request completion, \oursys scales down the instance’s KV-cache via the memory subsystem and reclaims the instance if it stays idle beyond a keep-alive threshold.

\section{Headroom-Driven Compute Subsystem}
\label{sec:compute_subsystem}
\subsection{Headroom-based Token-level Scheduling}
\label{sec:headroom_schedule}

\begin{figure}
	\centering
	\includegraphics[width=.96\linewidth]{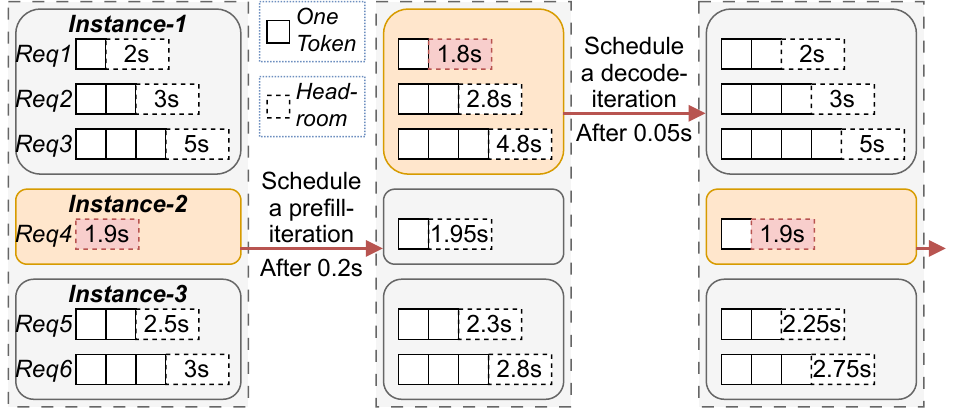}
	\caption{Procedure of token-level scheduling. At each cycle, \oursys schedules the instance with the shortest $headroom$.}
	\label{fig:token_level_schedule}
\end{figure}

To schedule compute resources at token-level, \oursys dynamically orchestrates the iterations of multiple instances, as each new token results from a prefill or decode iteration.
Specifically, as illustrated in Figure \ref{fig:token_level_schedule}, it selects one instance at a time to compute one iteration. 
Once complete, it moves on to the next instance for another iteration cycle and repeats. 

By continuously assigning token-level tasks to instances, the node is full-time utilized without idle periods. 
However, it is still uncertain which instance should be selected for each scheduling cycle. 
To minimize SLO violations, \oursys prioritizes the instance handling the most urgent request. 

\oursys introduces $headroom$ to characterize the degree of urgency. 
Let TTFT$_{\text{SLO}}$ and TPOT$_{\text{SLO}}$ denote the SLO for TTFT and TPOT. 
Suppose a request started at time $ST$, has generated $O$ tokens, and the current time is $CT$. 
The headroom of this request, which represents the maximal delay for generating the next token within the SLO, is given by:
\begin{equation}
	headroom = ST+\text{TTFT}_{\text{SLO}}+\text{TPOT}_{\text{SLO}}\cdot O-CT \label{eq:headroom_def}
\end{equation}

Therefore, at each scheduling cycle, \oursys selects the instance with the shortest request headroom and assigns it an iteration. 
In Figure \ref{fig:token_level_schedule}, it first selects instance-2. 
Suppose the TPOT$_{\text{SLO}}$ is 0.25 s and the iteration takes 0.2 s, the headroom then updates to $1.9-0.2+0.25=1.95$ s. 
\oursys then re-compares the headroom and repeats the process. 

\subsection{Performance Quantification}
\label{sec:quantify_compute}

Since headroom represents the time a request can delay its output, a negative headroom indicates that an SLO violation has occurred. 
To make sure this does not happen, 
it is essential to quantify the performance of each model instance, specifically the computation time per iteration under varying loads.
Since a prefill iteration is significantly different from a decode iteration, \oursys characterizes them separately.

\textbf{Quantify Prefill Time.}
As shown in Figure \ref{fig:cpu_ttft_profile}, the prefill time is approximately linearly correlated with the input token length. 
Therefore, \oursys uses linear interpolation. 
For a given model, \oursys collects the TTFT results for an input length samples $S_L$. 
Then, for a new request of length $L$, it finds the two closest known points and applies the interpolation.

\textbf{Quantify Decode Time.}
As evaluated in Figure \ref{fig:cpu_tpot_profile_7b} and Figure \ref{fig:cpu_tpot_profile_13b}, the time of decode iteration is correlated with both length and batch size. 
This is because the computation involves both the attention and the feed-forward network: the former scales with the total token length in the batch, while the latter scales with the batch size.
Thus, \oursys uses these two factors as two dimensions and applies 2D linear interpolation.  
For a given model, \oursys generates the batch size samples $S_B$ and the average token length samples $S_L$. 
For each $B'\in S_B$ and $L'\in S_L$, \oursys collects the corresponding TPOT results. 
Then, for a batch size $B$ and average token length $L$, it finds the four closest points and applies the interpolation.

Considering the hardware heterogeneity, \oursys quantifies for each hardware type.
To reduce sampling overhead, it uses $2^X$ to generate $S_L$ and $S_B$. 
If a model's maximum token length is $L_{\max}$ (e.g., 4096) and the maximum batch size is $B_{\max}$ (e.g., 256), \oursys only needs to collect $O(\log_{L_{\max}}\cdot \log_{B_{\max}})$ cases, which amounts to only a few hundred samples that can be completed within minutes, enabling it to quickly adapt to diverse platforms.
Lastly, to evaluate the accuracy, we randomly generated 100 workloads with various batch sizes and token lengths. 
The average relative deviations between the actual TTFT/TPOT and the estimated values were only 5.9\% and 3.9\%, respectively. 

\subsection{Adding Request via Shadow Validation}

\begin{figure}
	\centering
	\includegraphics[width=.99\linewidth]{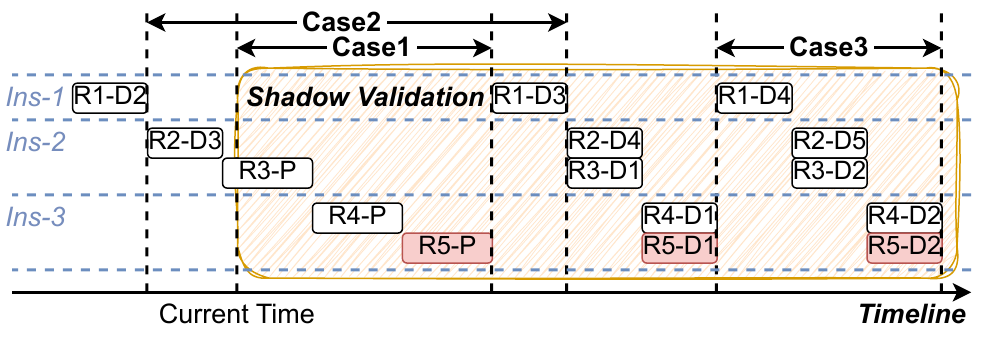}
	\caption{A shadow validation example with three cases.}
	\label{fig:shadow_validation}
\end{figure}

\label{sec:shadow_validation}
Based on quantification, \oursys can estimate the time of each iteration under various loads, we next focus on how the SLO violation could occur and how to avoid it. 
Specifically, as shown in Figure \ref{fig:shadow_validation}, when request-5 tries to join instance-3, there are three possible cases: (1)
The prefill of the new request (R5-P) is finished too late, making R5's headroom negative with TTFT SLO violation. (2) Existing request (R1-D3) is delayed too late due to the prefill of new request, making R1's headroom negative with TPOT SLO violation. (3) After the new request, the target instance takes longer time to decode (R5-D1, D2), causing the aggregate time for a single decode iteration across all instances in the node to exceed TPOT SLO.

Therefore, when trying to add a new request to a target instance, \oursys performs a shadow validation to virtually add and simulate the future compute procedure. 
This is particularly important because \oursys prioritizes scheduling requests to compute-bound CPU instances.
Considering the runtime fluctuations and the ever-growing token length during decode, \oursys overestimates each iteration by 10\%. 
Finally, the instance will only accept the request if none of the above cases occur in the simulation. 
Otherwise, \oursys will retry the validation on other instances, including creating a new instance to serve the new request.

\section{Hazard-Aware Memory Subsystem}
\label{sec:memory_subsystem}
\subsection{Characterizing Memory Demands}
\label{sec:quantify_memory}

The memory demand of one instance consists of model weights and KV-cache of ongoing requests, while the latter is dynamic and hard to determine since the final output length is hard to know in advance.
To avoid memory over-provisioning, \oursys estimates that each request's final output length is \textit{at least} the average output length $\bar{O}$ obtained from the historical logs. Additionally, to improve robustness, it introduces a lower bound $L_{min}$, which is set to the maximum context length in practice.

Consider a model instance where each token's KV-cache occupies $C$ bytes.
If $R$ requests are currently running, with the $r$-th request having an input length of $I_r$ and having generated $O_r$ tokens, assuming requests peak at the same time, the memory requirement of KV-cache is:
\begin{equation}
	M_{require} = C \cdot \max(\sum_{r=1}^{R} (I_r+\max(O_r, \bar{O})),L_{min})
\end{equation}

\subsection{KV-Cache Scaling via Watermark}
\label{sec:watermark_scale}

\begin{figure}
	\begin{minipage}{0.55\linewidth}
		\centering
		\includegraphics[width=.85\linewidth]{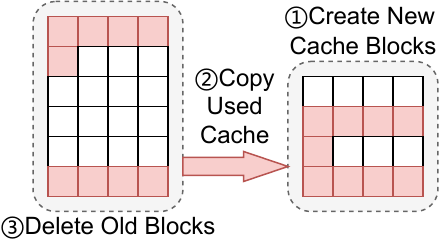}
		\caption{KV-cache scaling.}
		\label{fig:kv_cache_scale_procedure}
	\end{minipage}
	\hspace{1mm}
	\begin{minipage}{0.42\linewidth}
		\centering
		\includegraphics[width=.6\linewidth]{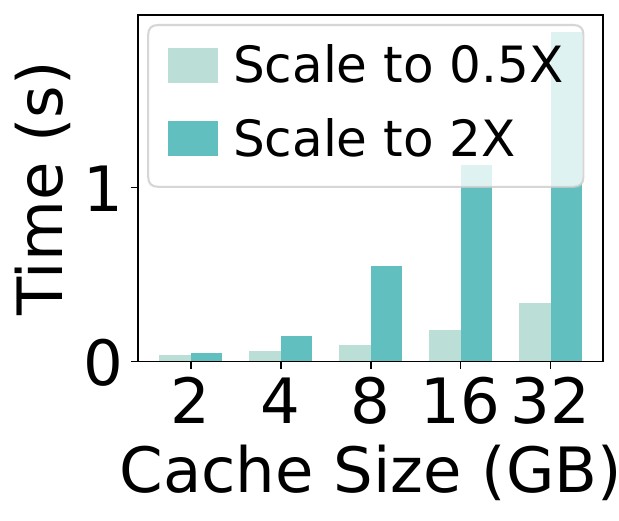}
		\caption{KV-cache scaling overhead on the GPU.}
		\label{fig:kv_scale_time}
	\end{minipage}
\end{figure}

As the KV-cache demand fluctuates with each received or completed request, \oursys should dynamically respond by adjusting the allocated memory resource accordingly, rather than statically assigning the entire node memory to a single instance.
However, we find that the adjusting procedure incurs non-negligible overhead. 
As demonstrated in Figure \ref{fig:kv_cache_scale_procedure}, based on the widely adopted paged-attention mechanism~\cite{DBLP:conf/sosp/KwonLZ0ZY0ZS23}, scaling it requires re-allocating cache blocks, and copy the original KV-cache from old blocks to new blocks. 
As evaluated in Figure \ref{fig:kv_scale_time}, scaling the original 32GB KV-cache blocks down to 16GB or up to 64GB requires 0.3 s and 1.9 s, respectively. 

Given the scaling overhead and the memory underestimation risk, \oursys adopts an early scale-up and lazy scale-down strategy.
Specifically, it utilizes a watermark hyperparameter $w$, which is used to calculate the recommended size of KV-cache $M_{recommend} \gets M_{require} \cdot (1 + w\%)$. 
Suppose the current KV-cache size is $M_{cur}$.
When adding a new request and the current cache is insufficient ($M_{cur} < M_{require}$), \oursys scales up directly to $M_{recommend}$. 
This reserves space for upcoming requests and the bursty long outputs, as one long-output request can steal reserved memory from others.
When a request completes, \oursys defers scaling down the KV-cache unless the recommended size falls below the watermark ($M_{recommend} \cdot (1+w\%) < M_{cur}$). 
This helps mitigate the ping-pong effect caused by load fluctuations. 
We set the watermark to 25\% and detail its sensitivity in \S\ref{sec:eval_watermark}.

\subsection{Inter-Instance Scaling Orchestration}
\label{sec:memory_orchestration}
\begin{figure}
	\begin{minipage}{0.51\linewidth}
		\centering
		\includegraphics[width=\linewidth]{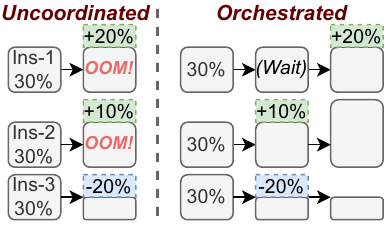}
		\caption{For example, uncoordinated memory scaling can spike usage to 120\% (OOM).}
		\label{fig:OOM}
	\end{minipage}
	\hspace{1mm}
	\begin{minipage}{0.46\linewidth}
		\centering
		\includegraphics[width=\linewidth]{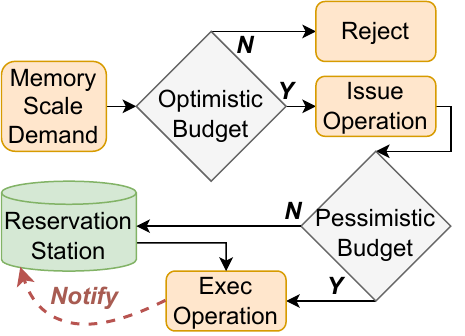}
		\caption{Flowchart of memory scaling operation.}
		\label{fig:memory_scale_orchestration}
	\end{minipage}
\end{figure}

Since each instance dynamically scales its KV-cache while also handling model loading/unloading, multiple instances within a node would simultaneously undergo multiple memory scaling operations, all of which are inherently asynchronous due to their execution latency. 
To efficiently manage memory adjustments and respond to fluctuations in real time, as illustrated in Figure \ref{fig:memory_scale_orchestration}, \oursys combines optimistic budgeting with pessimistic scheduling, enabling parallel execution of operations while avoiding OOM errors (e.g., Figure~\ref{fig:OOM}).

\oursys maintains an optimistic total memory budget within a node. 
When handling a scale-down demand, it directly reduces the budget and issues a corresponding operation. 
This budget update is optimistic because the actual memory release only takes effect once the operation completes. 
Conversely, for a scale-up demand, it first checks whether the current budget can be increased to fit the needs. 
If it does, the budget is updated, and an operation is issued.

However, parallel execution introduces hazards, such as a scale-up immediately following a scale-down, which could lead to OOM errors. 
To avoid such risks, \oursys employs a pessimistic global memory tracking mechanism to determine when to execute each issued operation. 
In this scheme, instances undergoing scale-down are accounted for based on their previous memory size.
An issued scale-down operation will be executed directly. 
For a scale-up operation, if pessimistic tracking suggests a risk of OOM, the operation is placed in a reservation station rather than executing immediately. 
When a scale-down operation completes, it notifies the reservation station, which then reevaluates the risk and attempts to execute any pending operations accordingly.

\subsection{Intra-Instance Scaling Compromise}
\label{sec:memory_check}
The orchestration mechanism may reject a scale-up demand if there is not enough memory (recall Figure \ref{fig:memory_scale_orchestration}). 
When trying to add a new request to a instance, \oursys also performs a shadow check on whether the potential scale-up demand can be approved.
To fully utilize all available memory, if the shadow check fails, \oursys will attempt to compromise the scale-up demand, allowing the request to be accepted as long as it can scale up to $M_{require}$ rather than $M_{recommend}$. 

Additionally, although we have strengthened the robustness of the estimates for the KV-cache, there is still a possibility of underestimation. 
In this rare case, \oursys will attempt to scale up the cache again. 
If the attempt fails due to the node memory shortage, \oursys will evict and re-schedule the request with the longest $headroom$. 

\section{Efficiency-Oriented Consolidation}
\label{sec:defragementation}
\begin{figure}
	\centering
	\subfloat[\label{fig:fragmented_case_a}Fragmented]
	{\includegraphics[width=.25\linewidth]{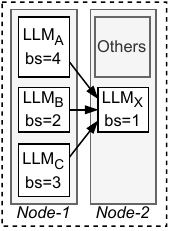}}
	\hspace{0.2mm}
	\subfloat[\label{fig:fragmented_case_b}Proactive]
	{\includegraphics[width=.25\linewidth]{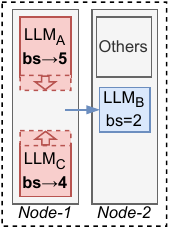}}
	\hspace{0.2mm}
	\subfloat[\label{fig:fragmented_case_c}Reactive]
	{\includegraphics[width=.47\linewidth]{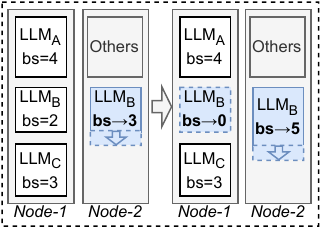}}
\caption{\label{fig:fragmented_case}(a) By default, A's, B's, or C's new request will create a fragmented instance. (b) To avoid fragment, A's or C's new request can trigger in-place scale-up by \textit{proactively} preempting B's instance. (c) When B holds multiple instances, its small-bs instance is \textit{reactively} reclaimed by prioritizing new requests to large-bs instance. ``bs'' represents ``batch size''.}
	\end{figure}
	
	In a shared environment, the presence of neighboring instances may block an instance from scaling up to accommodate a new request.
	Instead, the system is forced to scale out a new, fragmented instance (Figure \ref{fig:fragmented_case_a}), leading to degraded compute and memory efficiency.
	\oursys performs consolidation to reduce the fragmentation with two strategies.
	
	\subsection{Proactive Consolidation with Preemption}
	\label{sec:proactive_defragementation}
	
When the scale-up is hindered by neighboring instances, \oursys allows an instance to preempt them to make room for the new request, as shown in Figure \ref{fig:fragmented_case_b}. 
The requests of the preempted instances are then rescheduled to other nodes.

However, such preemption risks increasing fragmentation by disintegrating already-enlarged neighboring instances.
To avoid this, \oursys only allows an instance to preempt those with smaller batch sizes than itself and prioritizes the smallest one.
Additionally, \oursys also performs shadow validation to ensure preempted requests can still meet their SLOs after rescheduling, allowing preemption only when passed.
As a result, even in a crowded environment, a small instance can still hold promise for growing into a larger one without affecting the existing large instances, thereby minimizing fragmentation.

\subsection{Reactive Consolidation with Bin-Packing}
\label{sec:Reactive_defragementation}
While fragmentation can be minimized proactively, scaling out may still be necessary. 
To reduce its impact, when multiple instances of the same model exist, \oursys adopts a bin-packing strategy that preferentially routes new requests to the instance with the largest batch size.
On one hand, large-batch instances have more opportunities to grow larger through preemption. 
On the other hand, small-batch instances are more likely to finish their remaining requests sooner, so avoiding them increases the chances of reclaiming them earlier.

Figure \ref{fig:fragmented_case_c} illustrates this behavior.
Suppose $LLM_B$ needs to scale and creates a new instance on Node-2 with batch size $bs=3$, while an existing instance on Node-1 has $bs=2$. 
The Node-1 instance is now considered fragmented.
Subsequent requests are preferentially scheduled to Node 2, allowing \oursys to reclaim the Node-1 instance once its current requests are finished.

\section{Evaluation}

\subsection{Experimental Setup}
\label{sec:experimental_setup}
\textbf{Testbed.} 
We use 4 32-core Intel Xeon 6462C @3.3 GHz CPU nodes and 4 NVIDIA A100-80GB GPU nodes, which are logically separated from two physical machines with 2 GPUs each.

\textbf{Models.} 
We use popular LLMs with 16-bit precision of different sizes: Llama-3.2-3B, Llama-2-7B, and Llama-2-13B. 
As the resource requirement is primarily determined by the model size, same scale models exhibit similar performance. 
For instance, the TTFT and TPOT (1-batch and 1K-length) of DeepSeek-R1-Distill-Qwen-7B (7.6B) on CPU is 650 ms and 74 ms, while Llama-2-7B (6.7B) is 567 ms and 71 ms. 

\begin{figure}
	\centering
	\includegraphics[width=.9\linewidth]{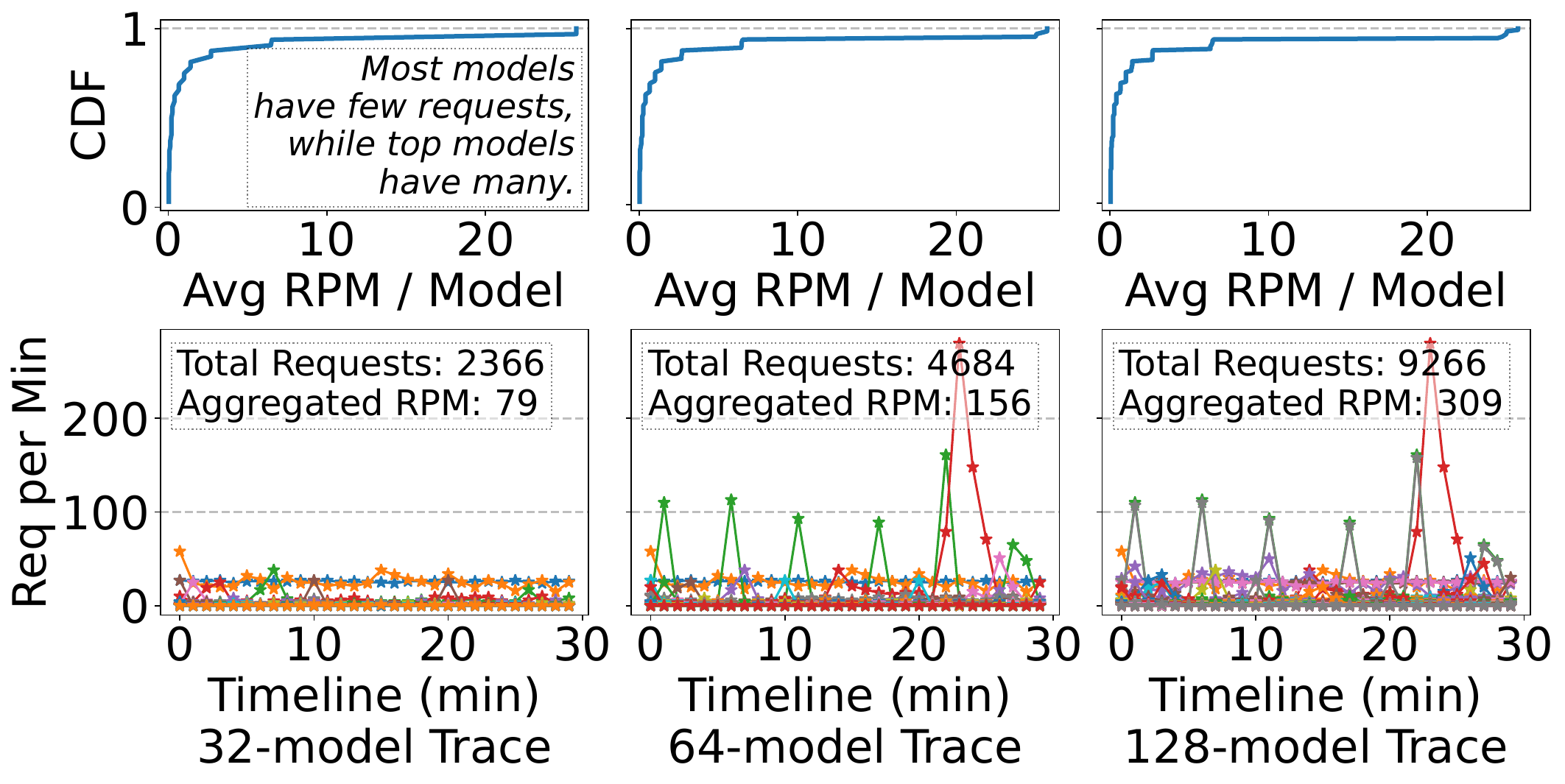}
	\caption{Azure Trace under different number of models.}
	\vspace{-1mm}
	\label{fig:trace_timeline}
\end{figure}

\textbf{Workloads and SLOs.} 
The input and output length of each request are sampled from Azure LLM Conversation dataset~\cite{DBLP:conf/isca/PatelCZSGMB24} (depicted in Figure \ref{fig:dataset_info}). 
In \S \ref{sec:eval_dataset}, we test four other datasets.
Since LLM traces contain only a single model and lack the multi-model hot–cold characteristics,
following ServerlessLLM~\cite{DBLP:conf/osdi/FuXHBUPM24},we use Azure Serverless Trace~\cite{DBLP:conf/usenix/ShahradFGCBCLTR20} and map each LLM to a function. 
We extracted the first 30-minute segment of the trace and uniformly sampled 32, 64, and 128 functions from it (depicted in Figure \ref{fig:trace_timeline}). 
For a request of input length $L$, following previous works~\cite{DBLP:conf/osdi/ZhongLCHZL0024,DBLP:conf/osdi/AgrawalKPMKGTR24}, we set TTFT and TPOT SLO to $\min(\max(0.5, L/512),8)$ s and 0.25 s.

\textbf{Baselines.} 
(1) We treat ServerlessLLM~\cite{DBLP:conf/osdi/FuXHBUPM24} as the baseline, denoted as \texttt{sllm}, which only supports GPUs. 
(2) \texttt{sllm+c} is modified to also support the CPUs. 
(3) Based on \texttt{sllm+c}, we further extend it to support time-sharing on both CPU and GPU nodes, denoted as \texttt{sllm+c+s}.
In this setting, each model instance (except for 13B-sized models on CPU) is allocated only half of the per-node resources.

\textbf{Systems Behavior and Fairness.}
\texttt{sllm+c}, \texttt{sllm+c+s}, and \texttt{\oursys} all prioritize the CPU nodes. 
All models are cached in CPU memory, and the cold-start procedure is similar across all systems, as \texttt{\oursys} utilizes \texttt{sllm}'s loader to enable fast loading. 
Although \texttt{sllm}'s loader has reduced the cold start latency to a few seconds (e.g., 1 second to load a 7B model in our environment), requests that experience cold-start may still violate the TTFT SLO. To address this, we relax the TTFT requirement for such requests by allowing a grace window equal to the cold-start duration. The keep-alive threshold is set to 1 s and all systems use same inference engines: vLLM 0.5.2 and OpenVINO 2024.6.0.

Unlike \texttt{\oursys}'s dynamic decision-making, \texttt{sllm} triggers instance scale-out based on a fixed concurrency limit of 2, which leads to extreme inefficiency. 
Based on the profiling, we tried our best to conservatively tailor a set of higher concurrency limits for \texttt{sllm} and \texttt{sllm+c}, which are (59, 15, 6) and (160, 32, 16) for the 3B, 7B, and 13B models on CPU and GPU, respectively. 
As for \texttt{sllm+c+s}, since the compute and memory shortages can easily occur when each instance is provisioned with constrained resources, the corresponding limits are (23, 4, 6) and (71, 12, 4).

\begin{figure*}
	\centering
	\subfloat[\label{fig:e2e_3b}3B-sized cases.]
	{\includegraphics[width=.31\linewidth]{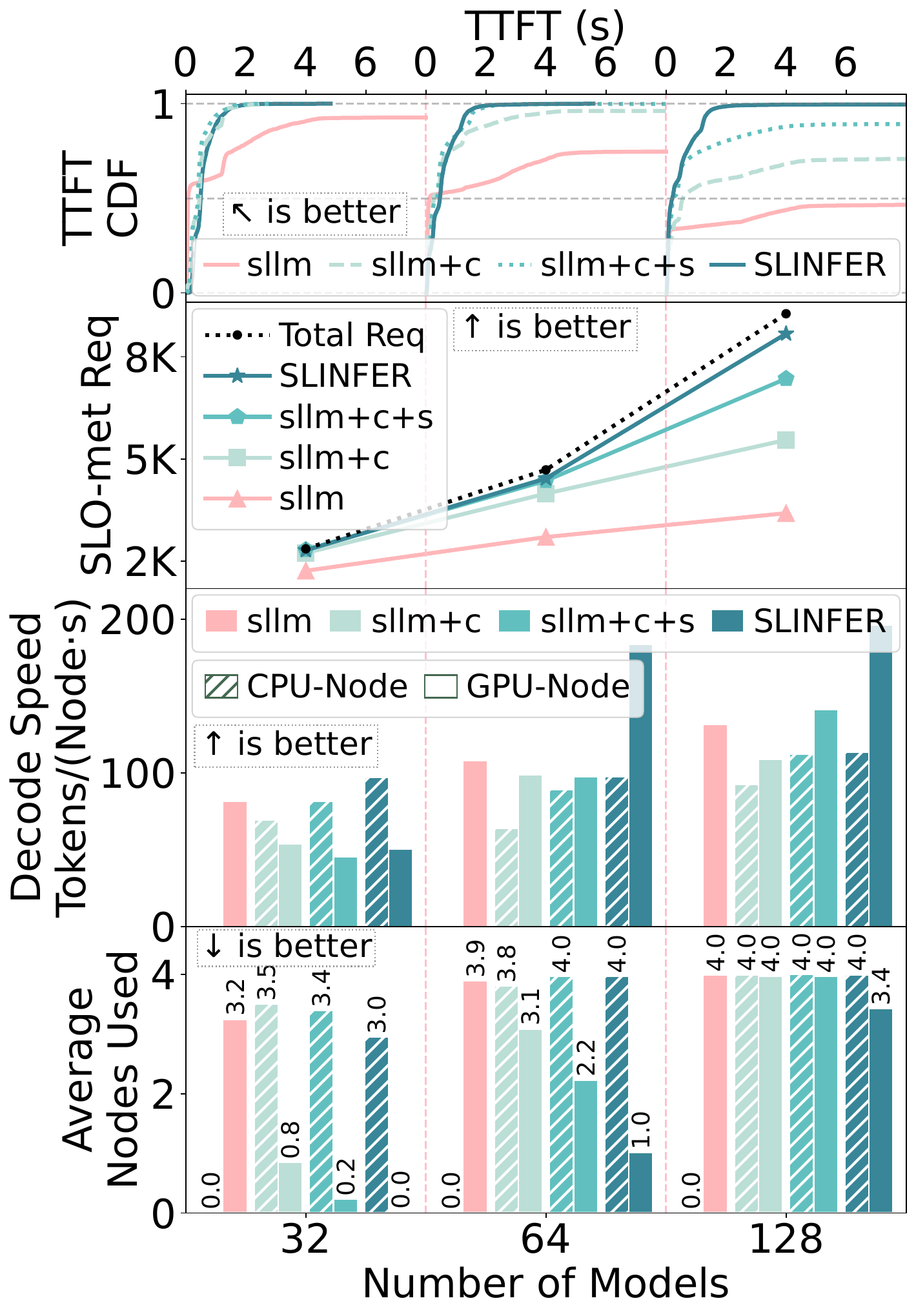}
		\vspace{-1mm}}
	\hspace{0.1mm}
	\subfloat[\label{fig:e2e_7b}7B-sized cases.]
	{\includegraphics[width=.31\linewidth]{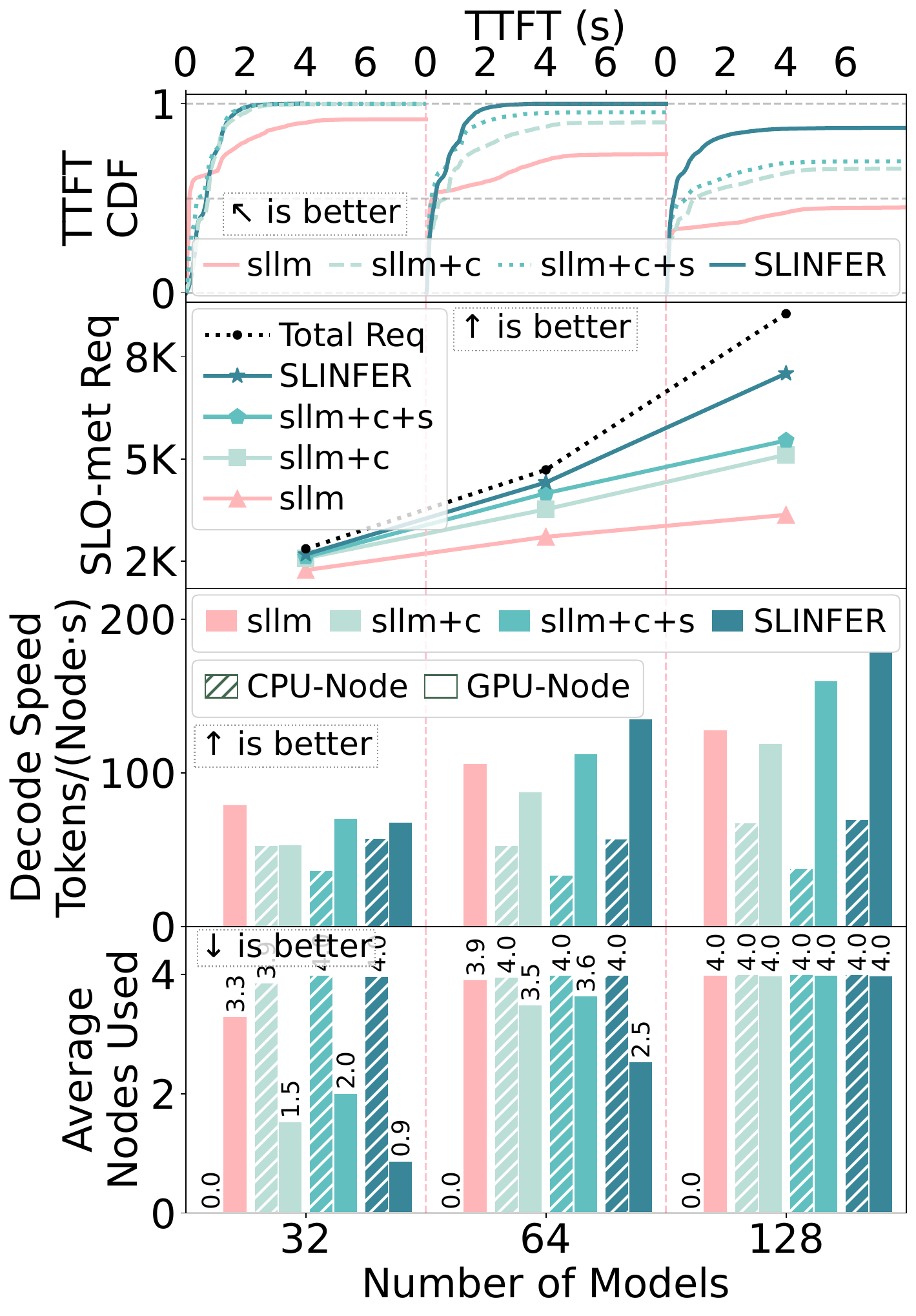}
		\vspace{-1mm}}
	\hspace{0.1mm}
	\subfloat[\label{fig:e2e_13b}13B-sized cases.]
	{\includegraphics[width=.31\linewidth]{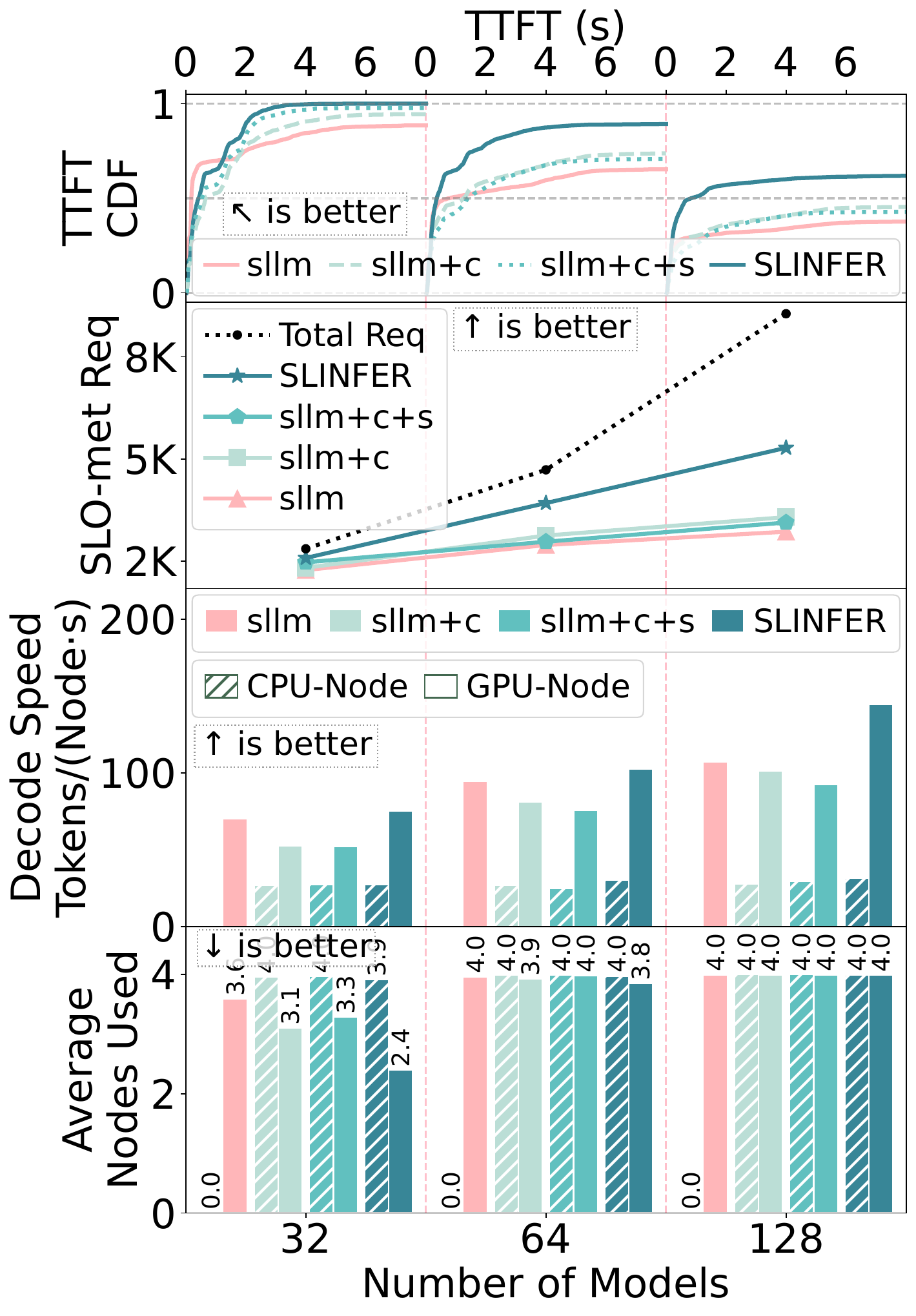}
		\vspace{-1mm}}
	\caption{Diverse performance metrics of each system under different model sizes and quantities.}
	\vspace{-1mm}
\end{figure*}

\subsection{End-to-end Experiments}
\label{sec:eval_e2e}
In this section, we present diverse performance metrics of \texttt{\oursys} under different model sizes and quantities, comparing it with \texttt{sllm} and its variants. 
Figure \ref{fig:e2e_3b} shows the results for the 3B-sized cases, where 32, 64, and 128 replica models are generated from Llama-3.2-3B and mapped to the Azure Trace (Figure \ref{fig:trace_timeline} details the trace). 
Figure \ref{fig:e2e_7b} and \ref{fig:e2e_13b} depict the scenarios for the 7B-sized and the 13B-sized, respectively. 

\textbf{\oursys uses less resources (Nodes Used) with higher per-node throughput (Decode Speed).} 
When serving 32 3B-sized models in Figure \ref{fig:e2e_3b}, \texttt{\oursys} consumes only 3.0 CPUs with 0 GPU, whereas \texttt{sllm} requires 3.2 exclusive GPUs. 
\texttt{sllm+c} and \texttt{sllm+c+s} can also reduce GPU usage by leveraging CPUs and sharing resources, but it is less effective than \texttt{\oursys}. 
Moreover, \texttt{sllm+c+s} can result in negative optimization effects due to the fixed resource partitioning (detailed in \S\ref{sec:eval_mixed_deployment}). 
For example, when serving 32 7B-sized models in Figure \ref{fig:e2e_7b}, \texttt{sllm+c} consumes 1.5 GPUs, while \texttt{sllm+c+s} consumes even more (2.0 GPUs), whereas \texttt{\oursys} uses only 0.9 GPUs.

To further investigate the reasons behind \texttt{\oursys}'s resource savings, we measured the average decode throughput per node. 
Compared to \texttt{sllm+c+s}, \texttt{\oursys} achieves higher throughput by 0\% - 84\% on CPUs and by (-4)\% - 88\% on GPUs. 
Note that the improvements can be negative since \texttt{\oursys} uses little GPUs when serving 32 models. 
The reasons for the improvements are twofold: 
First, \texttt{\oursys} can achieve higher batch size (detailed in \S \ref{sec:gpu_efficiency}). 
Second, \texttt{sllm} and its variants waste the allocated compute resources during instance cold-start and keep-alive, while \texttt{\oursys} can immediately reassign resources to other instances instead.

Finally, as the number of models increases or model size grows, the resource usage gap among four systems gradually narrows. 
For example, when serving 128 13B-sized models (Figure \ref{fig:e2e_13b}), each system exhausts all nodes.  
This is because, on one hand, the excessive load begins to saturate each system, and on the other hand, larger models diminish the sharing potential of \texttt{\oursys} (detailed in \S \ref{sec:eval_mixed_deployment}).

\textbf{\oursys achieves superior serving capacity (SLO-met Req) with quick response (TTFT CDF).}
When serving 128 models, it improves the number of SLO-met requests by 86\% - 154\% compared to \texttt{sllm}, by 47\% - 62\% compared to \texttt{sllm+c}, and by 18\% - 70\% compared to \texttt{sllm+c+s}. 
Meanwhile, it maintains sub-second TTFT for most requests.
This demonstrates the effectiveness of \texttt{\oursys}'s shadow validation and memory scaling mechanisms, ensuring that resource sharing does not compromise request SLOs.
\texttt{sllm+c+s} does not exhibit significant improvement, as the fixed resource partitioning leads to resource inefficiency (detailed in \S \ref{sec:gpu_efficiency}).

Note that \texttt{sllm} instead achieves a lower median TTFT when serving 32 models, since it only utilizes GPUs, whereas \texttt{\oursys} prioritizes CPUs.
Meanwhile, the CDFs of all systems do not always reach 1, as they proactively drop requests whose queuing delays exceed the TTFT SLO under heavy load.
When serving 128 models, \texttt{\oursys}'s CDF curves flatten at much higher percentiles compared to \texttt{sllm} and its variants, indicating \texttt{\oursys}'s superior serving capacity. 

\subsection{Ablation Study}
\begin{figure}
\centering
\includegraphics[width=.9\linewidth]{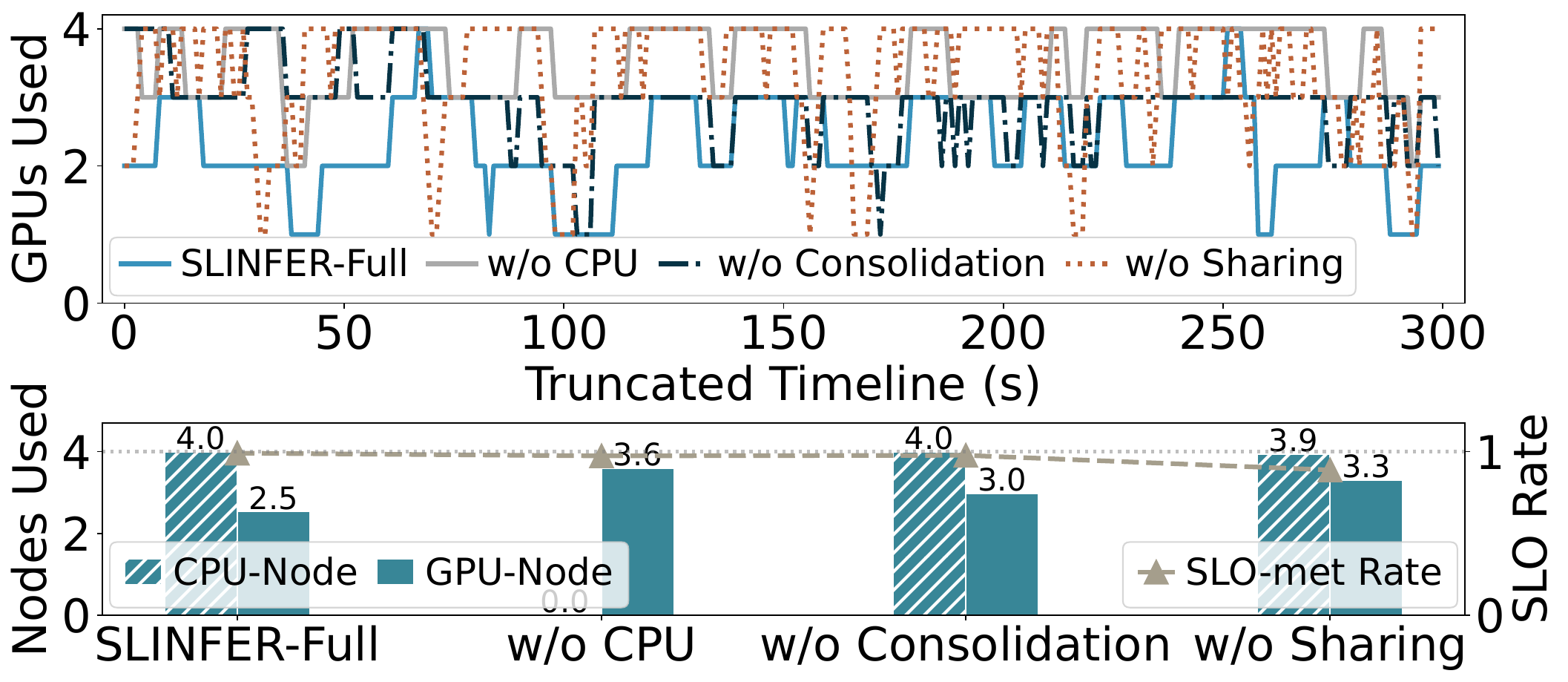}
\caption{The resource usage and SLO compliance rate when disabling each component of \oursys.}
\label{fig:ablation_study}
\end{figure}

We further study the effectiveness of \texttt{\oursys}'s design. 
Figure \ref{fig:ablation_study} shows the results when disabling each component when serving 64 7B-sized models. 
Disabling any component results in a increase in GPU resource usage. Notably, after disabling sharing, the SLO compliance rate drops substantially to 89\%. 
This is because sharing is a key factor in increasing deployment density; without it, \texttt{\oursys} struggles to handle such a large number of models simultaneously. 

From the truncated timeline of GPU usage, we observe that after disabling the CPU, GPU usage is consistently high, whereas \texttt{\oursys}-full rarely exhausts all four GPUs. 
Additionally, after disabling consolidation, when handling fluctuating loads (at 50 s and 250 s) and after load spikes, the GPU usage is notably higher compared to \texttt{\oursys}-full. 
This is because it creates fragmented instances to handle the surge, which cannot be reclaimed promptly. 

\begin{figure}
	\begin{minipage}{0.5\linewidth}
		\centering
		\includegraphics[width=.88\linewidth]{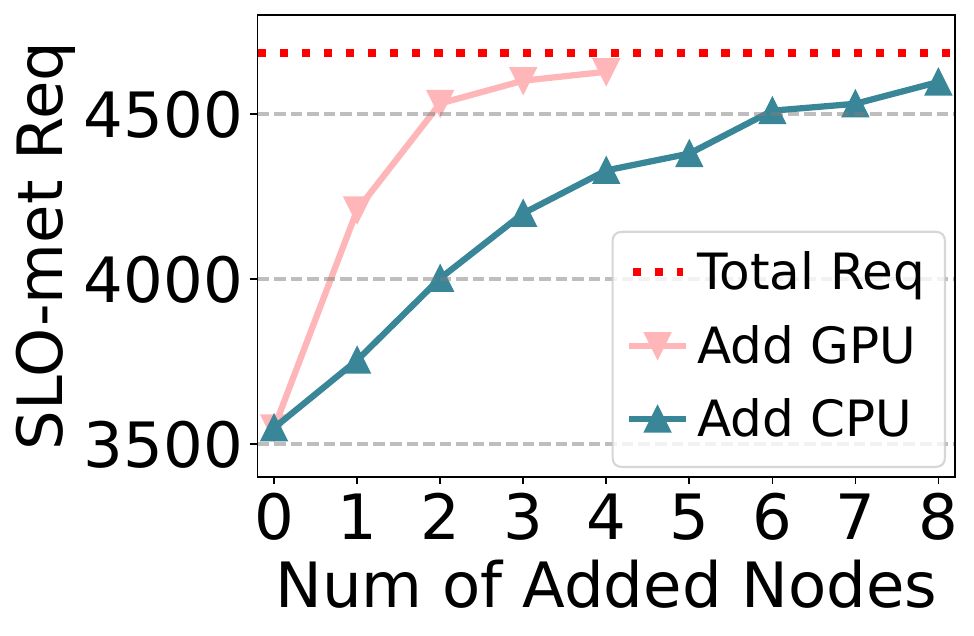}
		\caption{CPU scalability.}
		\label{fig:cpu_scalability}
	\end{minipage}
	\begin{minipage}{0.48\linewidth}
		\centering
		\includegraphics[width=.91\linewidth]{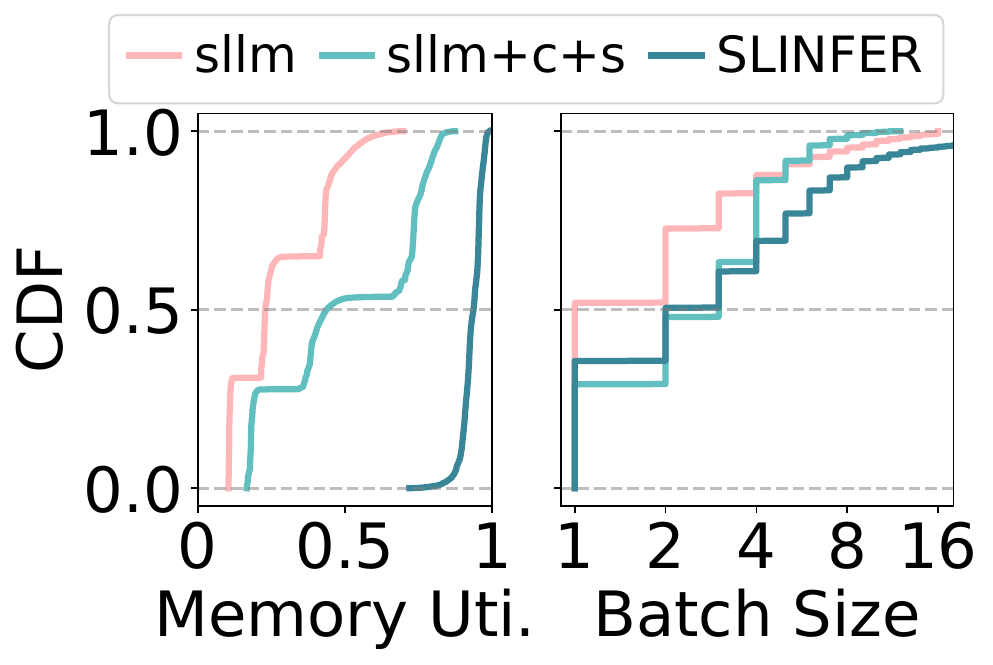}
		\caption{GPU efficiency.}
		\label{fig:gpu_efficiency}
	\end{minipage}
\end{figure}

\subsection{Evaluate CPU Scalability}
\label{sec:eval_cpu_scalability}

Having validated the effectiveness of the CPU, we further examine its scalability. 
In Figure \ref{fig:cpu_scalability}, we assume that \texttt{\oursys} initially has only two GPU nodes and zero CPU nodes, which are insufficient to handle all requests for 64 7B-sized models. 
As observed, continuously adding CPU nodes gradually increases the system’s serving capacity to accommodate all requests. 
However, the scaling efficiency is lower compared to adding GPU nodes—roughly 3 to 4 CPU nodes are required to match the capacity of a single GPU node. 
This aligns with expectations, as CPUs have relatively lower compute power. 

\begin{figure}
	\centering
	\includegraphics[width=.82\linewidth]{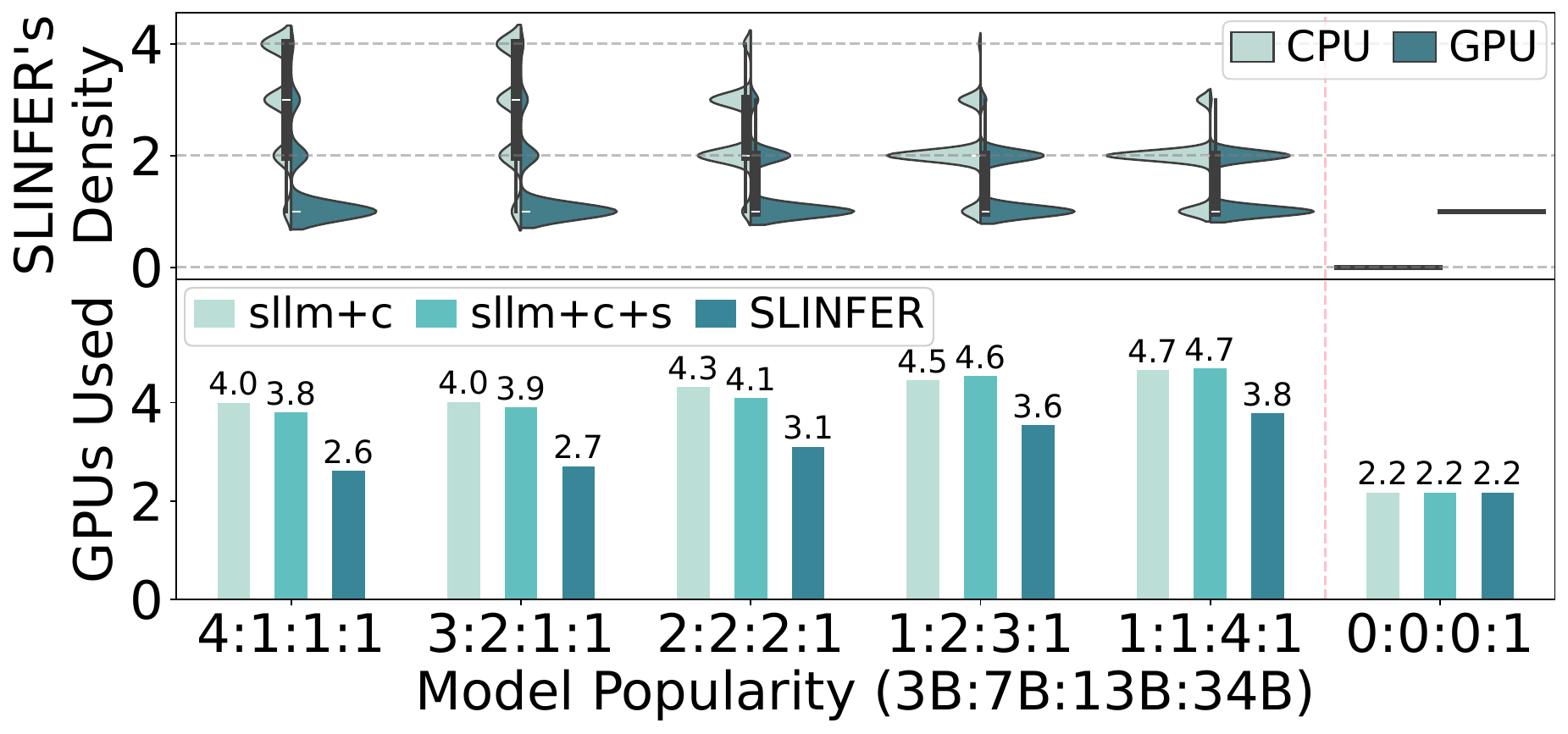}
	\caption{Performance when various sized models co-exist.}
	\label{fig:mixed_deployment}
\end{figure}

\subsection{Mixed Deployment}
\label{sec:eval_mixed_deployment}

To reflect real-world scenarios with mixed model sizes, we evaluate \texttt{\oursys} under mixed-sized workloads, including CodeLlama-34B deployed with tensor parallelism (2 GPUs/instance). 
To accommodate the increased workload scale, this experiment runs on 4 CPUs and 6 GPUs. 
The CPU results are omitted as all systems saturate CPU usage.

Figure~\ref{fig:mixed_deployment} shows that \texttt{\oursys} consistently uses fewer GPUs than both \texttt{sllm+c} and \texttt{sllm+c+s}, but its efficiency varies with model popularity.
When small models dominate (4:1:1:1), \texttt{\oursys} can deploy up to four instances per CPU, while reserving GPUs primarily for large models. 
In contrast, when large models dominate (1:1:4:1), the deployment density drops due to higher resource demands, reducing sharing efficiency.
In the extreme case (0:0:0:1), \texttt{\oursys} falls back to exclusive GPU allocation, similar to \texttt{sllm+c} and \texttt{sllm+c+s}.

We also observe that \texttt{sllm+c+s} performs worse under large models due to static partitioning that severely limits concurrency under high demands.
Overall, since most popular models are relatively small~\cite{huggingface-popular}, \texttt{\oursys} can achieve significant resource savings in practice.

\subsection{Investigate GPU Efficiency}
\label{sec:gpu_efficiency}

Figure \ref{fig:gpu_efficiency} presents an analysis of GPU efficiency when serving mixed models (3B, 7B, and 13B) of 2:2:2 ratio. 
As discussed in \S\ref{sec:eval_mixed_deployment}, \texttt{\oursys}'s behavior for larger models aligns with baselines and is omitted for brevity.

\texttt{\oursys} achieves near-optimal memory utilization with close to 1. 
In contrast, \texttt{sllm} and \texttt{sllm+c+s} both exhibit a three-tier memory utilization pattern corresponding to the three model sizes, with most instances using less than half of their allocated memory. 
This suggests significant over-provision, since they allocate all available memory in a node (or half of the node) to each instance for KV-cache space.

Despite the sparsity of serverless workloads, \texttt{\oursys} achieves a 74\% higher average batch size than \texttt{sllm}, as instance sharing prolongs execution intervals and accumulates more requests.
\texttt{sllm+c+s} suffers from lower peak batch sizes due to fixed resource partitioning that limits concurrency.

\begin{table}[htbp]
	\centering
	\caption{Performance under prefill–decode disaggregation. Each cell shows results for aggregated PD / disaggregated PD.}
	\label{tab:PD}
	\begin{tabular}{llcc}
		\toprule
		System & Load (models) & Avg. GPU Usage & SLO Rate (\%) \\
		\midrule
		\multirow{3}{*}{\texttt{sllm+c+s}}
		& 32  & 2.0 / 3.0  & 99 / 93 \\
		& 64  & 3.6 / 3.9  & 93 / 70 \\
		& 128 & 4.0 / 4.0  & 65 / 35 \\
		\midrule
		\multirow{3}{*}{\texttt{\oursys}}
		& 32  & 0.9 / 1.0  & 99 / 99 \\
		& 64  & 2.5 / 2.9  & 99 / 98 \\
		& 128 & 4.0 / 4.0  & 86 / 69 \\
		\bottomrule
	\end{tabular}
\end{table}

\subsection{Exploring Prefill-Decode Disaggregated Architecture}
To minimize resource usage, we co-locate the prefill and decode stages of each request within the same instance. An alternative design, known as prefill-decode (PD) disaggregation~\cite{DBLP:conf/isca/PatelCZSGMB24}, launches dedicated instances for each stage per model.
Table~\ref{tab:PD} shows the performance impact of this approach. 
The cross-node communication bandwidth is 100 Gbps.
We observe that PD disaggregation instead leads to increased resource usage and reduced serving capacity. 
This is because the prefill stage is short-lived, and infrequent requests result in prefill instances spending 93\% of their lifetime on average in cold starts or idle.
This finding aligns with DistServe~\cite{DBLP:conf/osdi/ZhongLCHZL0024}, which also argues that PD disaggregation is ill-suited for resource-constrained scenarios.

\subsection{Scalability and Scheduling Overhead}

\begin{figure*}
	\centering
	\begin{minipage}{0.191\textwidth}
		\centering
		\includegraphics[width=\linewidth]{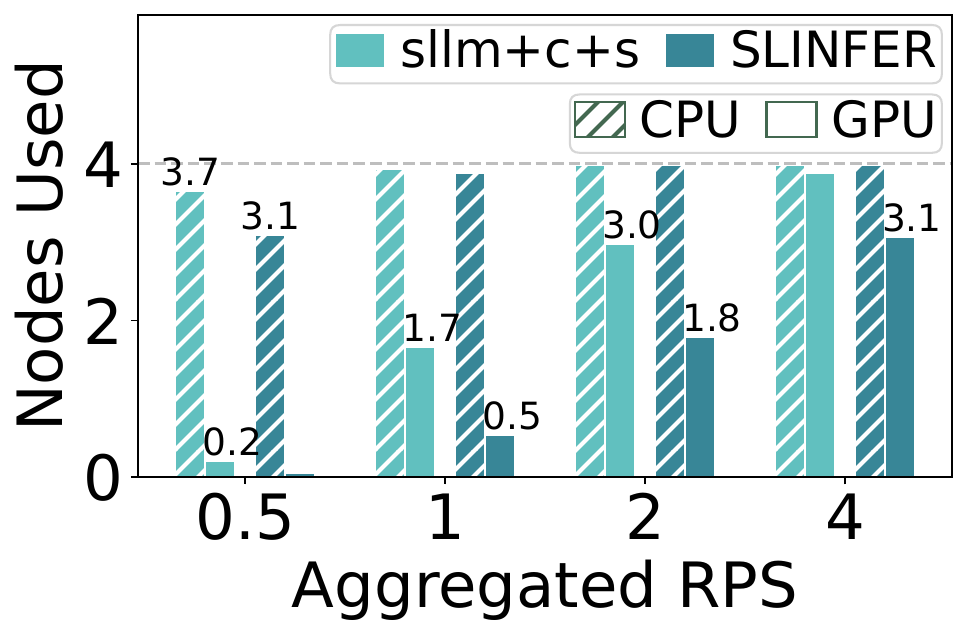}
		\caption{The resource usage of BurstGPT under different load-levels.}
		\label{fig:sensitivity_burstgpt_rps}
	\end{minipage}
	\hspace{1mm}
	\begin{minipage}{0.14\textwidth}
		\centering
		\includegraphics[width=.875\linewidth]{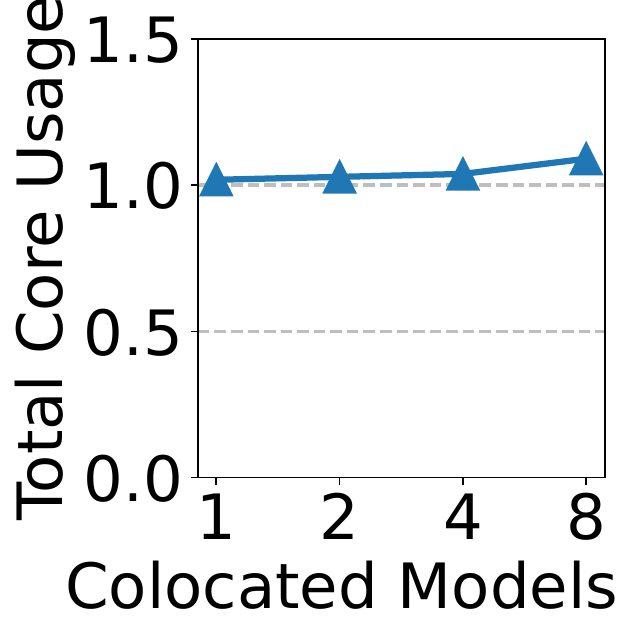}
		\caption{CPU usage during multi-model colocation.}
		\label{fig:multi_model_cpu_usage}
	\end{minipage}
	\hspace{1mm}
	\begin{minipage}{0.168\textwidth}
		\centering
		\includegraphics[width=\linewidth]{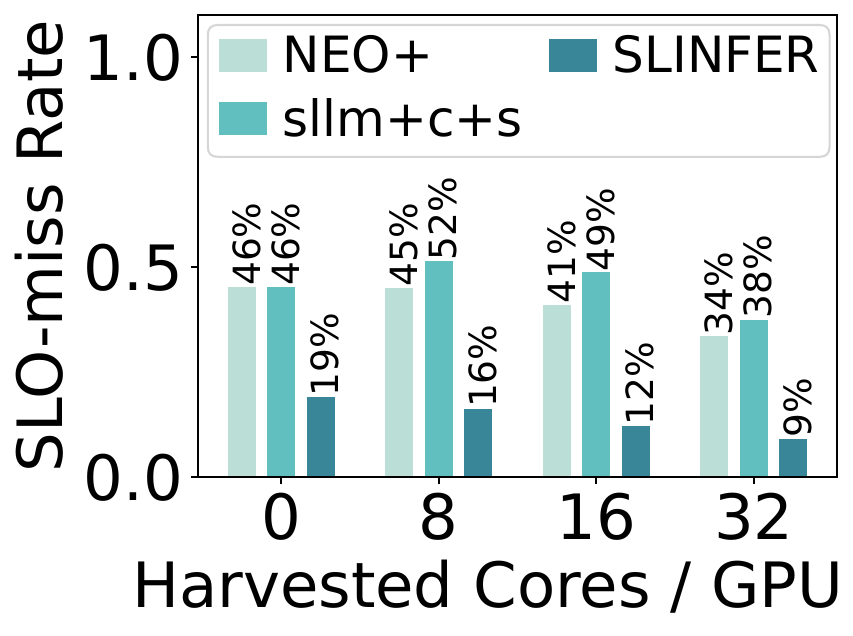}
		\caption{Performance under varying numbers of CPU cores.}
		\label{fig:sensitivity_harvested_cores}
	\end{minipage}
	\hspace{1mm}
	\begin{minipage}{0.199\textwidth}
		\centering
		\includegraphics[width=\linewidth]{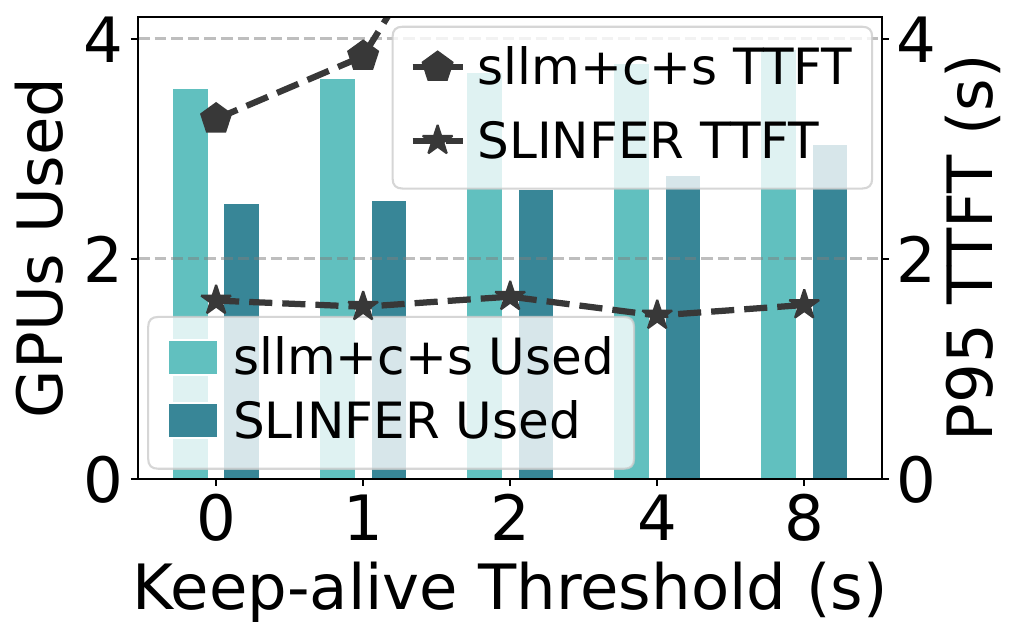}
		\caption{Performance under different keep-alive thresholds.}
		\label{fig:sensitivity_keep_alive}
	\end{minipage}
	\hspace{1mm}
	\begin{minipage}{0.23\textwidth}
		\centering
		\includegraphics[width=\linewidth]{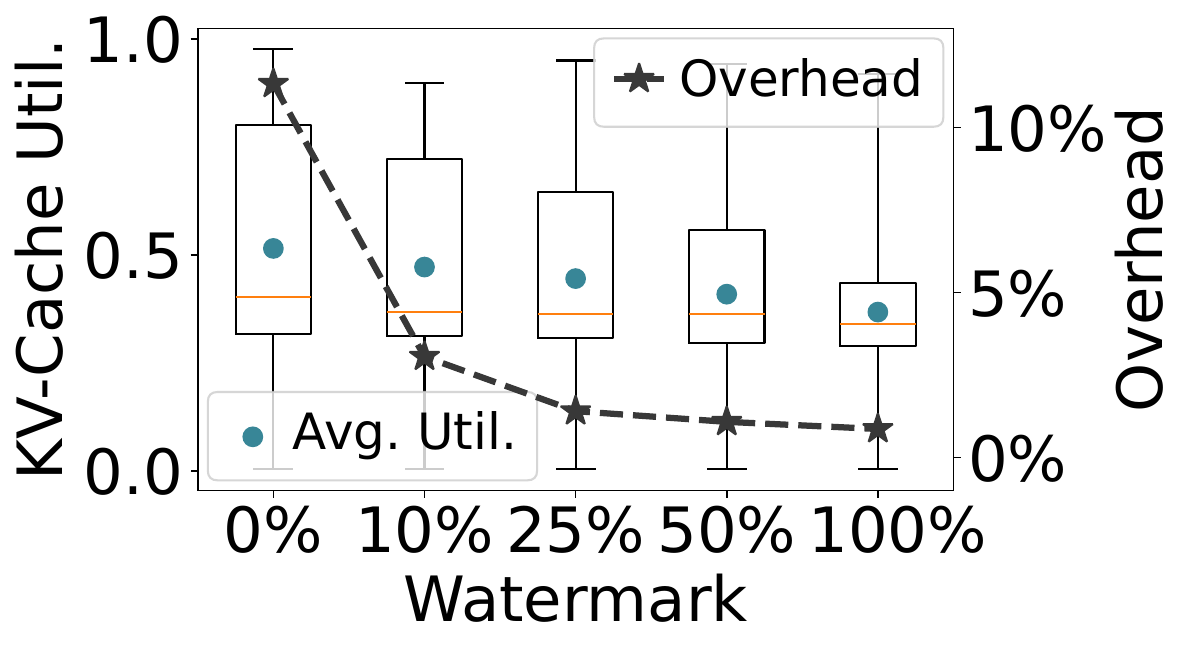}
		\caption{KV-cache utilization and scaling overhead under different watermarks.}
		\label{fig:sensitivity_watermark}
	\end{minipage}
\end{figure*}

\begin{figure}
	\begin{minipage}{0.50\linewidth}
		\centering
		\includegraphics[width=.9\linewidth]{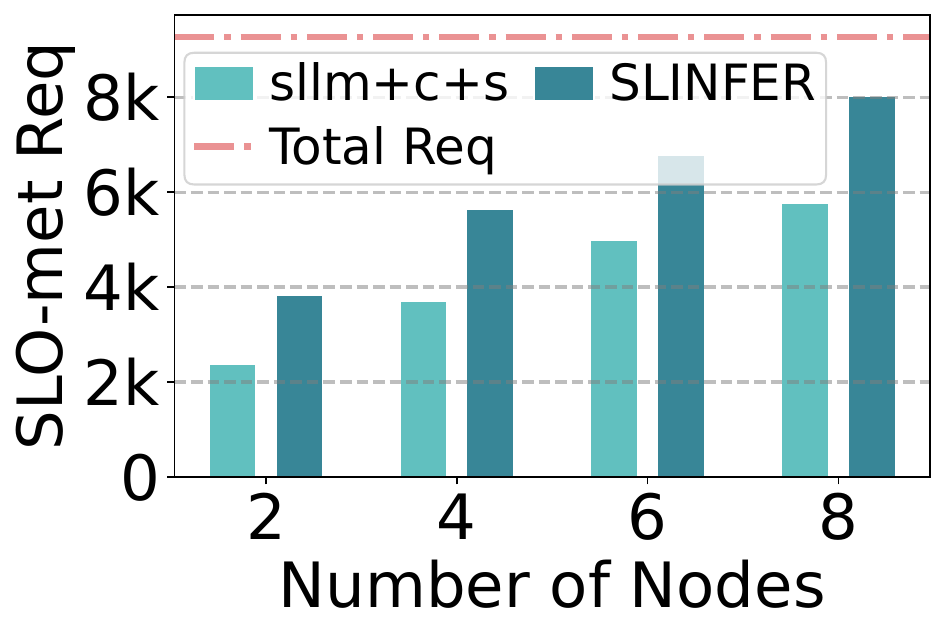}
		\caption{Performance under different node counts.}
		\label{fig:sensitivity_node_scalability}
	\end{minipage}
	\hspace{1mm}
	\begin{minipage}{0.45\linewidth}
		\centering
		\includegraphics[width=.9\linewidth]{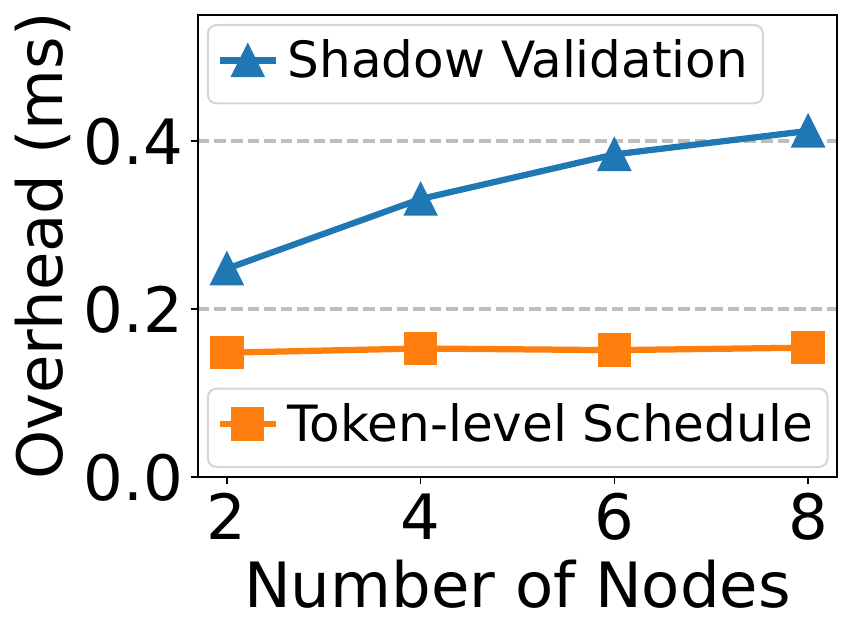}
		\caption{The scheduling overhead of \oursys.}
		\label{fig:sensitivity_multinode_overhead}
	\end{minipage}
\end{figure}

As shown in Figure~\ref{fig:sensitivity_node_scalability}, we compare the serving capacity under the same workload while varying the number of nodes—from 1 CPU + 1 GPU to 4 CPU + 4 GPU. 
Across all configurations, \texttt{\oursys} achieves a higher number of SLO-met requests. 
With four nodes, \texttt{\oursys} delivers equivalent performance to \texttt{sllm+c+s} running on eight nodes.
Note that performance gains show diminishing returns, as we evaluate SLO-met requests under fixed load with many infrequent and concurrent model invocations.
For instance, a single node can serve ten requests from one model, but handling one request each from ten different models requires much more nodes.

We further analyze the scheduling overhead of \texttt{\oursys} and find that it remains low, as shown in Figure~\ref{fig:sensitivity_multinode_overhead}.
First, when a request arrives, it undergoes shadow validation to select an instance. 
The time cost slightly increases with the number of nodes, because a heavily loaded model tends to have more instances as the cluster scales, leading shadow validation to probe more candidates.
Second, \texttt{\oursys} dynamically schedules instances at token-level (recall Figure~\ref{fig:token_level_schedule}). 
This overhead remains stable regardless of the scales, since this scheduling decision is performed independently on each node.

\subsection{Sensitivity Analyses}

Previously, we used Azure Conversation dataset and Azure Serverless Trace as workloads. 
Each CPU node was provisioned with 32 cores, the keep-alive threshold was set to 1 second, and \texttt{\oursys}'s KV-cache scaling watermark was set to 25\%.
In this section, we conduct a series of sensitivity analyses to evaluate how these settings affect system performance.

\begin{figure}
	\begin{minipage}{0.42\linewidth}
		\centering
		\includegraphics[width=.91\linewidth]{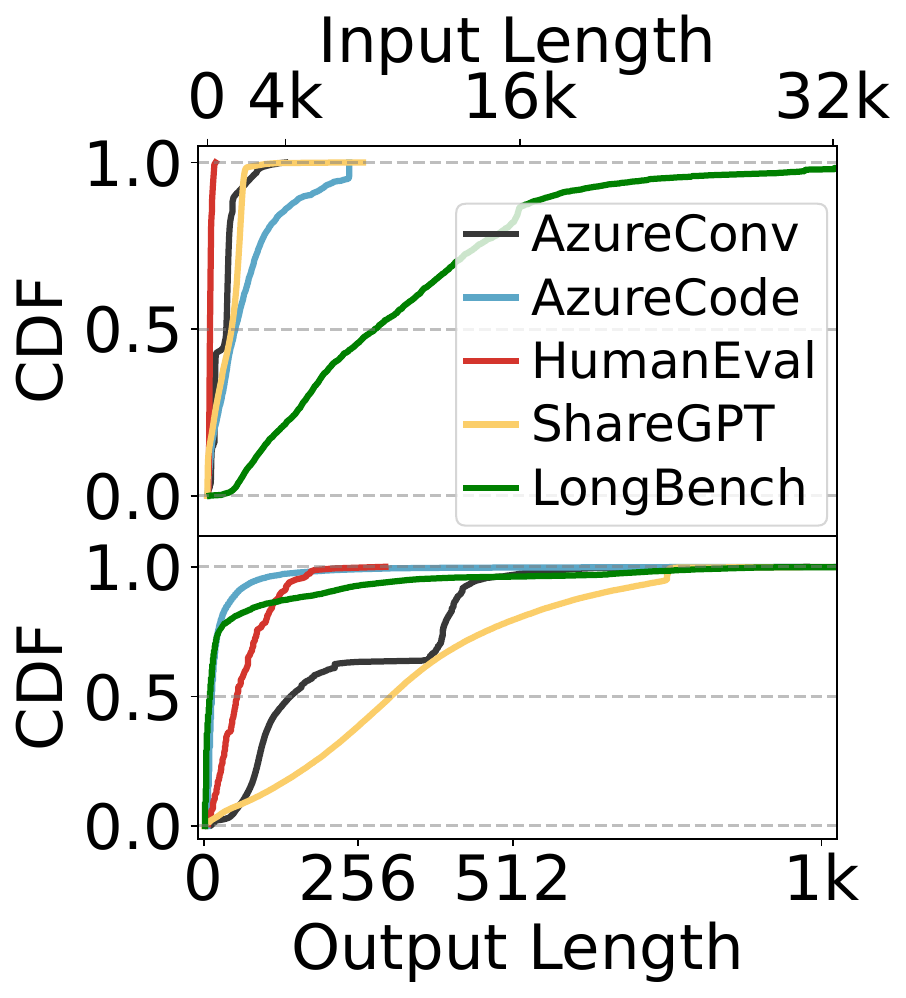}
		\caption{Characterization of different LLM datasets.}
		\label{fig:dataset_info}
	\end{minipage}
	\hspace{1mm}
	\begin{minipage}{0.56\linewidth}
		\centering
		\includegraphics[width=\linewidth]{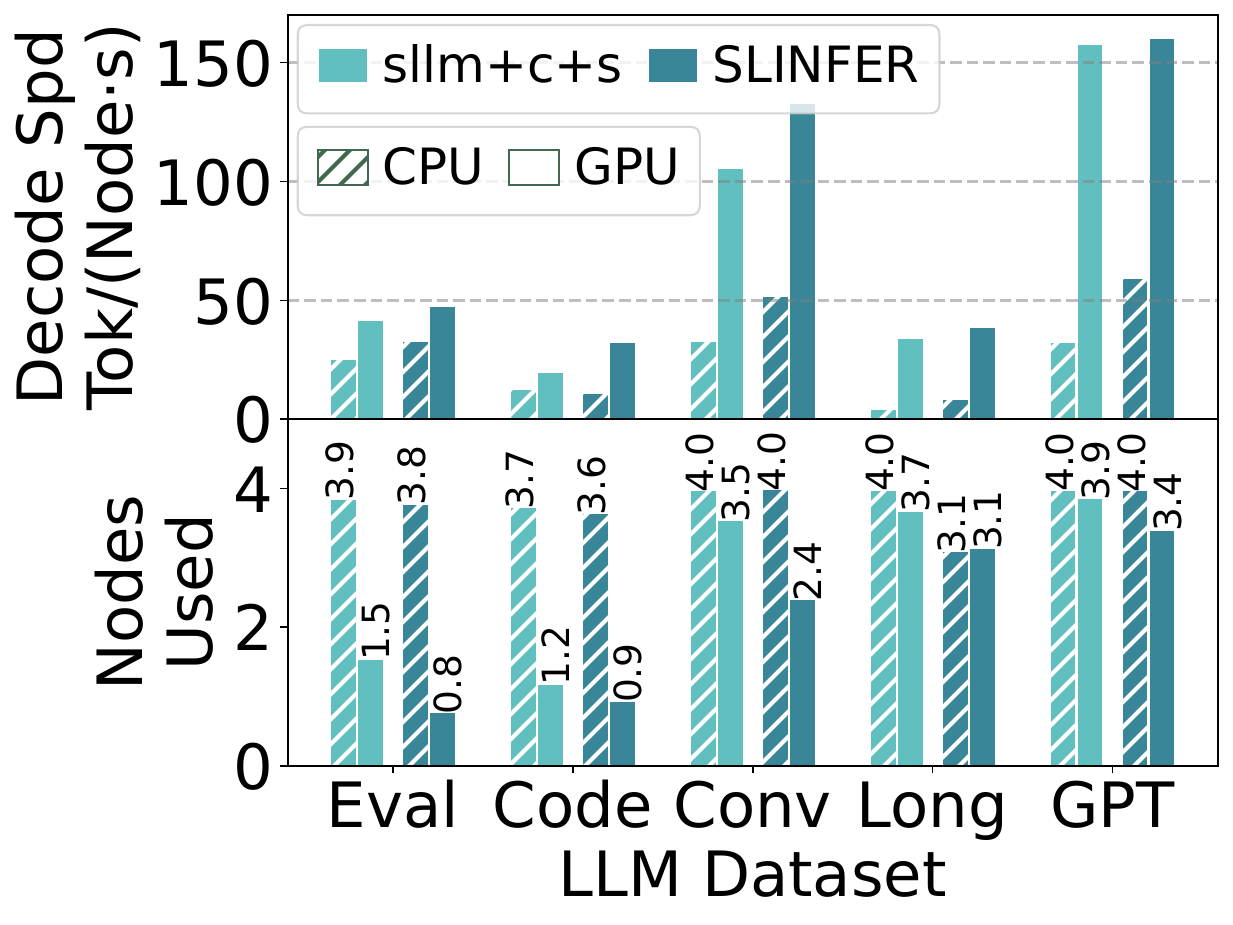}
		\caption{Eval of different datasets when serving 64 8B-sized models.}
		\label{fig:sensitivity_dataset}
	\end{minipage}
\end{figure}
\subsubsection{Length Patterns}
\label{sec:eval_dataset}
We further evaluate on the Azure Code~\cite{DBLP:conf/isca/PatelCZSGMB24}, HumanEval~\cite{chen2021evaluatinglargelanguagemodels}, ShareGPT~\cite{sharegpt}, and Longbench~\cite{DBLP:conf/acl/BaiLZL0HDLZHDTL24} datasets. 
They are characterized in Figure \ref{fig:dataset_info}. 
To support Longbench with up to 32k tokens, we use Llama-3.1-8B models across all datasets.
As shown in Figure~\ref{fig:sensitivity_dataset}, \texttt{\oursys} consistently consumes fewer resources than \texttt{sllm+c+s}. 
We observe that datasets with longer outputs, such as ShareGPT, consume more resources but achieve higher decode throughput. 
This is because longer generations provides more batching opportunities.
For LongBench, however, CPUs cannot satisfy the long-sequence TTFT SLO, so \texttt{\oursys} does not prefer CPUs. In comparison, \texttt{sllm+c+s} fully utilizes CPUs but violates 63.4\% of SLOs. Overall, CPUs can handle inputs up to 8.4k tokens within the 8s TTFT SLO.

\subsubsection{Invocation Traces}
In addition to serverless trace, we also experimented with an LLM trace, BurstGPT~\cite{10.1145/3711896.3737413}. 
However, the LLM trace represents a centralized single-model invocation pattern, which does not match the multi-model scenarios. 
To emulate the serverless environments, we distributed all invocations across 64 models following a Pareto distribution. 
Figure~\ref{fig:sensitivity_burstgpt_rps} shows system resource utilization under different load levels by sampling various time segments from BurstGPT. 
\texttt{\oursys} consistently consumes fewer resources. 
When the RPS increases to 4, \texttt{sllm+c+s} incurs 7.7\% SLO violations, whereas \texttt{\oursys} maintains only 1.0\%.

\subsubsection{CPU Resources}
CPU resources may become constrained in shared environments. 
However, as shown in Figure~\ref{fig:multi_model_cpu_usage}, even when eight model instances are deployed on a single GPU, their total average CPU usage only slightly exceeds one core. 
This is because each instance takes turns to use the GPU and only keeps the CPU busy-waiting during GPU interactions. 
Apart from that, tasks such as data preprocessing consume negligible CPU resources ($<$ 0.1 core).

Nevertheless, Figure~\ref{fig:sensitivity_harvested_cores} compares system performance under varying harvested CPU cores. 
In addition to being used independently, CPU resources can also assist GPU instances, as proposed in NEO~\cite{jiang2025neo}. 
We also compare this approach. 
Results show that SLINFER consistently achieves the lowest SLO-miss rate across all resource conditions. 
In contrast, NEO lags behind, as it is primarily optimized for single-instance, high-load scenarios, whereas in serverless multi-model settings, elastic and independent utilization of heterogeneous resources to increase deployment density is the top priority.

\subsubsection{Keep-alive Threshold} 
Counterintuitively, extending the threshold can even worsen the TTFT, due to: (1) cold-start latency is already low, and (2) prolonged idle instances exacerbates resource contention, leading to requests queuing—particularly for \texttt{sllm+c+s}. 
We therefore recommend a short threshold (e.g., 1 s) to balance resource efficiency and user experience.

\subsubsection{KV-cache Scaling Watermark}
\label{sec:eval_watermark}
As shown in Figure \ref{fig:sensitivity_watermark}, setting a watermark is essential since disabling it (set to 0\%) causes each instance to spend 11.3\% of its lifetime on scaling due to frequent adjustments. 
Besides, even a low watermark can significantly reduce this overhead, as \texttt{\oursys} leverages early scale-up to accommodate upcoming requests in a single event and delays scale-down to mitigate short-term fluctuations. 
Thus, we recommend using a low watermark (e.g., 25\%), where scaling overhead is already minimal (1.4\%), and the request migration rate due to underestimations is only 0–0.3\%. 
Raising the watermark further provides negligible benefit but lowering KV-cache utilization, leading to memory inefficiency.

\section{Discussion}

\textbf{Impact of Hardware Advancements.}
\oursys currently targets small- to mid-sized LLMs. 
For large models, \oursys falls back to ServerlessLLM~\cite{DBLP:conf/osdi/FuXHBUPM24}’s exclusive allocation approach (recall \S\ref{sec:eval_mixed_deployment}). 
Besides, current CPUs are still slow for tight SLOs and long inputs—decoding of Llama-3.1-8B takes at least 74 ms, and processing 32k inputs takes 84 s.
However, CPU's capabilities are rapidly evolving: the 32-core 4th Gen Xeon we use delivers 105 TFLOPS (BF16) compared to 13 TFLOPS on a 32-core 3rd Gen Xeon, and the latest 96-core 6th Gen~\cite{6th-xeon} reaches 297 TFLOPS.
Meanwhile, GPU memory capacity is also increasing.
These advancements offer further performance gains and greater model-sharing potential. 

\textbf{Serving Quantized Models.}
Applying quantization further enhances \oursys's sharing capacity by reducing the memory footprint of each instance. 
When serving 32 22B-sized models~\cite{Codestral-22B}, applying INT4 quantization~\cite{DBLP:conf/mlsys/0002TTYCWXDG024} reduced GPU usage from 3.8 to 2.6. 
This improvement stems from the fact that the model weights alone consume 44GB, making quantization essential for sharing on a 80GB GPU.

\section{Conclusion}
We propose \oursys, a resource-efficient serverless LLM inference scheme.
Motivated by evolving hardware architectures and real-world workload characteristics, \oursys brings a new solution in the face of GPU scarcity. 
We consider \oursys as a first step in applying the serverless paradigm to explore transparent sharing of heterogeneous resources for LLM inference. 
As hardware continues to advance, more opportunities will emerge.


\bibliographystyle{IEEEtranS}
\bibliography{refs}

@Misc{Phi-2,
title  = {Phi-2: The surprising power of small language models - Microsoft Research},
howpublished={\url{https://www.microsoft.com/en-us/research/blog/phi-2-the-surprising-power-of-small-language-models/}},
year = 2023
}

@Misc{llama-2-7b,
title  = {meta-llama/Llama-2-7b-hf · Hugging Face},
howpublished={\url{https://huggingface.co/meta-llama/Llama-2-7b-hf}},
year = 2023
}

@Misc{sharegpt,
title  = {ShareGPT · Datasets at Hugging Face},
howpublished={\url{https://huggingface.co/datasets/anon8231489123/ShareGPT_Vicuna_unfiltered}},
year = 2023
}

@Misc{6th-xeon,
title  = {Intel® Xeon® 6966P-C Processor},
howpublished={\url{https://www.intel.com/content/www/us/en/products/sku/240782/intel-xeon-6966pc-processor-432m-cache-3-00-ghz/specifications.html}},
year = 2025
}

@Misc{Codestral-22B,
	title  = {mistralai/Codestral-22B-v0.1 · Hugging Face},
	howpublished={\url{https://huggingface.co/mistralai/Codestral-22B-v0.1}},
	year = 2024
}

@Misc{Claude,
title  = {Meet Claude \ Anthropic},
howpublished={\url{https://www.anthropic.com/claude}},
year = 2025
}

@Misc{chatgpt,
title  = {ChatGPT | OpenAI},
howpublished={\url{https://openai.com/chatgpt/overview/}},
year = 2025
}

@Misc{aws-sagemaker,
	title  = {Amazon SageMaker},
	howpublished={\url{https://cloud.google.com/blog/products/application-development/run-your-ai-inference-applications-on-cloud-run-with-nvidia-gpus}},
	year = 2025
}

@Misc{google-serverless,
	title  = {Host your LLMs on Cloud Run | Google Cloud Blog},
	howpublished={\url{https://aws.amazon.com/sagemaker/}},
	year = 2025
}

@Misc{huggingface-serverless,
title  = {Inference API (serverless) - Hugging Face},
howpublished={\url{https://huggingface.co/inference-api/serverless}},
year = 2025
}

@Misc{azure-serverless,
title  = {Deploy models as serverless APIs - Azure Machine Learning | Microsoft Learn},
howpublished={\url{https://learn.microsoft.com/en-us/azure/machine-learning/how-to-deploy-models-serverless}},
year = 2025
}

@Misc{TensorRT,
title  = {TensorRT-LLM},
howpublished={\url{https://github.com/NVIDIA/TensorRT-LLM}},
year = 2025
}

@Misc{AMX,
title  = {What Is Intel® Advanced Matrix Extensions (Intel® AMX)? – Intel},
howpublished={\url{https://www.intel.com/content/www/us/en/products/docs/accelerator-engines/what-is-intel-amx.html}},
year = 2025
}

@Misc{openvino,
title  = {Intel® Distribution of OpenVINO™ Toolkit},
howpublished={\url{https://www.intel.com/content/www/us/en/developer/tools/openvino-toolkit/overview.html}},
year = 2025
}

@Misc{huggingface-popular,
title  = {Open Source Ai Year In Review 2024 - a Hugging Face Space by huggingface},
howpublished={\url{https://huggingface.co/spaces/huggingface/open-source-ai-year-in-review-2024?day=2}},
year = 2024
}

@inproceedings{DBLP:conf/osdi/FuXHBUPM24,
  author       = {Yao Fu and
                  Leyang Xue and
                  Yeqi Huang and
                  Andrei{-}Octavian Brabete and
                  Dmitrii Ustiugov and
                  Yuvraj Patel and
                  Luo Mai},
  editor       = {Ada Gavrilovska and
                  Douglas B. Terry},
  title        = {ServerlessLLM: Low-Latency Serverless Inference for Large Language
                  Models},
  booktitle    = {18th {USENIX} Symposium on Operating Systems Design and Implementation,
                  {OSDI} 2024, Santa Clara, CA, USA, July 10-12, 2024},
  pages        = {135--153},
  publisher    = {{USENIX} Association},
  year         = {2024},
  url          = {https://www.usenix.org/conference/osdi24/presentation/fu},
  bibsource    = {dblp computer science bibliography, https://dblp.org}
}

@inproceedings{DBLP:conf/nsdi/CrankshawWZFGS17,
  author       = {Daniel Crankshaw and
                  Xin Wang and
                  Giulio Zhou and
                  Michael J. Franklin and
                  Joseph E. Gonzalez and
                  Ion Stoica},
  editor       = {Aditya Akella and
                  Jon Howell},
  title        = {Clipper: {A} Low-Latency Online Prediction Serving System},
  booktitle    = {14th {USENIX} Symposium on Networked Systems Design and Implementation,
                  {NSDI} 2017, Boston, MA, USA, March 27-29, 2017},
  pages        = {613--627},
  publisher    = {{USENIX} Association},
  year         = {2017},
  url          = {https://www.usenix.org/conference/nsdi17/technical-sessions/presentation/crankshaw},
  bibsource    = {dblp computer science bibliography, https://dblp.org}
}

@inproceedings{DBLP:conf/osdi/GujaratiKAHKVM20,
  author       = {Arpan Gujarati and
                  Reza Karimi and
                  Safya Alzayat and
                  Wei Hao and
                  Antoine Kaufmann and
                  Ymir Vigfusson and
                  Jonathan Mace},
  title        = {Serving DNNs like Clockwork: Performance Predictability from the Bottom
                  Up},
  booktitle    = {14th {USENIX} Symposium on Operating Systems Design and Implementation,
                  {OSDI} 2020, Virtual Event, November 4-6, 2020},
  pages        = {443--462},
  publisher    = {{USENIX} Association},
  year         = {2020},
  url          = {https://www.usenix.org/conference/osdi20/presentation/gujarati},
  bibsource    = {dblp computer science bibliography, https://dblp.org}
}

@inproceedings{DBLP:conf/nsdi/GunasekaranMTSK22,
  author       = {Jashwant Raj Gunasekaran and
                  Cyan Subhra Mishra and
                  Prashanth Thinakaran and
                  Bikash Sharma and
                  Mahmut Taylan Kandemir and
                  Chita R. Das},
  editor       = {Amar Phanishayee and
                  Vyas Sekar},
  title        = {Cocktail: {A} Multidimensional Optimization for Model Serving in Cloud},
  booktitle    = {19th {USENIX} Symposium on Networked Systems Design and Implementation,
                  {NSDI} 2022, Renton, WA, USA, April 4-6, 2022},
  pages        = {1041--1057},
  publisher    = {{USENIX} Association},
  year         = {2022},
  url          = {https://www.usenix.org/conference/nsdi22/presentation/gunasekaran},
  bibsource    = {dblp computer science bibliography, https://dblp.org}
}

@inproceedings{DBLP:conf/nsdi/0025TKS23,
  author       = {Hong Zhang and
                  Yupeng Tang and
                  Anurag Khandelwal and
                  Ion Stoica},
  editor       = {Mahesh Balakrishnan and
                  Manya Ghobadi},
  title        = {{SHEPHERD:} Serving DNNs in the Wild},
  booktitle    = {20th {USENIX} Symposium on Networked Systems Design and Implementation,
                  {NSDI} 2023, Boston, MA, April 17-19, 2023},
  pages        = {787--808},
  publisher    = {{USENIX} Association},
  year         = {2023},
  url          = {https://www.usenix.org/conference/nsdi23/presentation/zhang-hong},
  bibsource    = {dblp computer science bibliography, https://dblp.org}
}

@inproceedings{DBLP:conf/asplos/YangZLZLZCL22,
  author       = {Yanan Yang and
                  Laiping Zhao and
                  Yiming Li and
                  Huanyu Zhang and
                  Jie Li and
                  Mingyang Zhao and
                  Xingzhen Chen and
                  Keqiu Li},
  editor       = {Babak Falsafi and
                  Michael Ferdman and
                  Shan Lu and
                  Thomas F. Wenisch},
  title        = {INFless: a native serverless system for low-latency, high-throughput
                  inference},
  booktitle    = {{ASPLOS} '22: 27th {ACM} International Conference on Architectural
                  Support for Programming Languages and Operating Systems, Lausanne,
                  Switzerland, 28 February 2022 - 4 March 2022},
  pages        = {768--781},
  publisher    = {{ACM}},
  year         = {2022},
  url          = {https://doi.org/10.1145/3503222.3507709},
  doi          = {10.1145/3503222.3507709},
  bibsource    = {dblp computer science bibliography, https://dblp.org}
}

@inproceedings{DBLP:conf/usenix/HuXLHCXLMZWDDRL25,
  author       = {Junhao Hu and
                  Jiang Xu and
                  Zhixia Liu and
                  Yulong He and
                  Yuetao Chen and
                  Hao Xu and
                  Jiang Liu and
                  Jie Meng and
                  Baoquan Zhang and
                  Shining Wan and
                  Gengyuan Dan and
                  Zhiyu Dong and
                  Zhihao Ren and
                  Changhong Liu and
                  Tao Xie and
                  Dayun Lin and
                  Qin Zhang and
                  Yue Yu and
                  Hao Feng and
                  Xusheng Chen and
                  Yizhou Shan},
  editor       = {Deniz Altinb{\"{u}}ken and
                  Ryan Stutsman},
  title        = {{DEEPSERVE:} Serverless Large Language Model Serving at Scale},
  booktitle    = {Proceedings of the 2025 {USENIX} Annual Technical Conference, {USENIX}
                  {ATC} 2025, Boston, MA, USA, July 7-9, 2025},
  pages        = {57--72},
  publisher    = {{USENIX} Association},
  year         = {2025},
  url          = {https://www.usenix.org/conference/atc25/presentation/hu-junhao},
  bibsource    = {dblp computer science bibliography, https://dblp.org}
}

@inproceedings{DBLP:conf/sc/AliP0S20,
  author       = {Ahsan Ali and
                  Riccardo Pinciroli and
                  Feng Yan and
                  Evgenia Smirni},
  editor       = {Christine Cuicchi and
                  Irene Qualters and
                  William T. Kramer},
  title        = {Batch: machine learning inference serving on serverless platforms
                  with adaptive batching},
  booktitle    = {Proceedings of the International Conference for High Performance Computing,
                  Networking, Storage and Analysis, {SC} 2020, Virtual Event / Atlanta,
                  Georgia, USA, November 9-19, 2020},
  pages        = {69},
  publisher    = {{IEEE/ACM}},
  year         = {2020},
  url          = {https://doi.org/10.1109/SC41405.2020.00073},
  doi          = {10.1109/SC41405.2020.00073},
  bibsource    = {dblp computer science bibliography, https://dblp.org}
}

@article{DBLP:journals/corr/abs-2408-00118,
  author       = {Morgane Rivi{\`{e}}re and
                  Shreya Pathak and
                  Pier Giuseppe Sessa and
                  Cassidy Hardin and
                  Surya Bhupatiraju and
                  L{\'{e}}onard Hussenot and
                  Thomas Mesnard and
                  Bobak Shahriari and
                  Alexandre Ram{\'{e}} and
                  Johan Ferret and
                  Peter Liu and
                  Pouya Tafti and
                  Abe Friesen and
                  Michelle Casbon and
                  Sabela Ramos and
                  Ravin Kumar and
                  Charline Le Lan and
                  Sammy Jerome and
                  Anton Tsitsulin and
                  Nino Vieillard and
                  Piotr Stanczyk and
                  Sertan Girgin and
                  Nikola Momchev and
                  Matt Hoffman and
                  Shantanu Thakoor and
                  Jean{-}Bastien Grill and
                  Behnam Neyshabur and
                  Olivier Bachem and
                  Alanna Walton and
                  Aliaksei Severyn and
                  Alicia Parrish and
                  Aliya Ahmad and
                  Allen Hutchison and
                  Alvin Abdagic and
                  Amanda Carl and
                  Amy Shen and
                  Andy Brock and
                  Andy Coenen and
                  Anthony Laforge and
                  Antonia Paterson and
                  Ben Bastian and
                  Bilal Piot and
                  Bo Wu and
                  Brandon Royal and
                  Charlie Chen and
                  Chintu Kumar and
                  Chris Perry and
                  Chris Welty and
                  Christopher A. Choquette{-}Choo and
                  Danila Sinopalnikov and
                  David Weinberger and
                  Dimple Vijaykumar and
                  Dominika Rogozinska and
                  Dustin Herbison and
                  Elisa Bandy and
                  Emma Wang and
                  Eric Noland and
                  Erica Moreira and
                  Evan Senter and
                  Evgenii Eltyshev and
                  Francesco Visin and
                  Gabriel Rasskin and
                  Gary Wei and
                  Glenn Cameron and
                  Gus Martins and
                  Hadi Hashemi and
                  Hanna Klimczak{-}Plucinska and
                  Harleen Batra and
                  Harsh Dhand and
                  Ivan Nardini and
                  Jacinda Mein and
                  Jack Zhou and
                  James Svensson and
                  Jeff Stanway and
                  Jetha Chan and
                  Jin Peng Zhou and
                  Joana Carrasqueira and
                  Joana Iljazi and
                  Jocelyn Becker and
                  Joe Fernandez and
                  Joost van Amersfoort and
                  Josh Gordon and
                  Josh Lipschultz and
                  Josh Newlan and
                  Ju{-}yeong Ji and
                  Kareem Mohamed and
                  Kartikeya Badola and
                  Kat Black and
                  Katie Millican and
                  Keelin McDonell and
                  Kelvin Nguyen and
                  Kiranbir Sodhia and
                  Kish Greene and
                  Lars Lowe Sj{\"{o}}sund and
                  Lauren Usui and
                  Laurent Sifre and
                  Lena Heuermann and
                  Leticia Lago and
                  Lilly McNealus},
  title        = {Gemma 2: Improving Open Language Models at a Practical Size},
  journal      = {CoRR},
  volume       = {abs/2408.00118},
  year         = {2024},
  url          = {https://doi.org/10.48550/arXiv.2408.00118},
  doi          = {10.48550/ARXIV.2408.00118},
  eprinttype    = {arXiv},
  eprint       = {2408.00118},
  bibsource    = {dblp computer science bibliography, https://dblp.org}
}

@inproceedings{DBLP:conf/usenix/Romero0YK21,
  author       = {Francisco Romero and
                  Qian Li and
                  Neeraja J. Yadwadkar and
                  Christos Kozyrakis},
  editor       = {Irina Calciu and
                  Geoff Kuenning},
  title        = {INFaaS: Automated Model-less Inference Serving},
  booktitle    = {Proceedings of the 2021 {USENIX} Annual Technical Conference, {USENIX}
                  {ATC} 2021, July 14-16, 2021},
  pages        = {397--411},
  publisher    = {{USENIX} Association},
  year         = {2021},
  url          = {https://www.usenix.org/conference/atc21/presentation/romero},
  bibsource    = {dblp computer science bibliography, https://dblp.org}
}

@inproceedings{DBLP:conf/asplos/DuLJXZC22,
  author       = {Dong Du and
                  Qingyuan Liu and
                  Xueqiang Jiang and
                  Yubin Xia and
                  Binyu Zang and
                  Haibo Chen},
  editor       = {Babak Falsafi and
                  Michael Ferdman and
                  Shan Lu and
                  Thomas F. Wenisch},
  title        = {Serverless computing on heterogeneous computers},
  booktitle    = {{ASPLOS} '22: 27th {ACM} International Conference on Architectural
                  Support for Programming Languages and Operating Systems, Lausanne,
                  Switzerland, 28 February 2022 - 4 March 2022},
  pages        = {797--813},
  publisher    = {{ACM}},
  year         = {2022},
  url          = {https://doi.org/10.1145/3503222.3507732},
  doi          = {10.1145/3503222.3507732},
  bibsource    = {dblp computer science bibliography, https://dblp.org}
}

@inproceedings{DBLP:conf/asplos/MahapatraGAKWXK24,
  author       = {Rohan Mahapatra and
                  Soroush Ghodrati and
                  Byung Hoon Ahn and
                  Sean Kinzer and
                  Shu{-}Ting Wang and
                  Hanyang Xu and
                  Lavanya Karthikeyan and
                  Hardik Sharma and
                  Amir Yazdanbakhsh and
                  Mohammad Alian and
                  Hadi Esmaeilzadeh},
  editor       = {Rajiv Gupta and
                  Nael B. Abu{-}Ghazaleh and
                  Madan Musuvathi and
                  Dan Tsafrir},
  title        = {In-Storage Domain-Specific Acceleration for Serverless Computing},
  booktitle    = {Proceedings of the 29th {ACM} International Conference on Architectural
                  Support for Programming Languages and Operating Systems, Volume 2,
                  {ASPLOS} 2024, La Jolla, CA, USA, 27 April 2024- 1 May 2024},
  pages        = {530--548},
  publisher    = {{ACM}},
  year         = {2024},
  url          = {https://doi.org/10.1145/3620665.3640413},
  doi          = {10.1145/3620665.3640413},
  bibsource    = {dblp computer science bibliography, https://dblp.org}
}

@inproceedings{DBLP:conf/asplos/RoyPT22,
  author       = {Rohan Basu Roy and
                  Tirthak Patel and
                  Devesh Tiwari},
  editor       = {Babak Falsafi and
                  Michael Ferdman and
                  Shan Lu and
                  Thomas F. Wenisch},
  title        = {IceBreaker: warming serverless functions better with heterogeneity},
  booktitle    = {{ASPLOS} '22: 27th {ACM} International Conference on Architectural
                  Support for Programming Languages and Operating Systems, Lausanne,
                  Switzerland, 28 February 2022 - 4 March 2022},
  pages        = {753--767},
  publisher    = {{ACM}},
  year         = {2022},
  url          = {https://doi.org/10.1145/3503222.3507750},
  doi          = {10.1145/3503222.3507750},
  bibsource    = {dblp computer science bibliography, https://dblp.org}
}

@inproceedings{DBLP:conf/middleware/PfandzelterDFCE23,
  author       = {Tobias Pfandzelter and
                  Aditya Dhakal and
                  Eitan Frachtenberg and
                  Sai Rahul Chalamalasetti and
                  Darel Emmot and
                  Ninad Hogade and
                  Rolando Pablo Hong Enriquez and
                  Gourav Rattihalli and
                  David Bermbach and
                  Dejan S. Milojicic},
  title        = {Kernel-as-a-Service: {A} Serverless Programming Model for Heterogeneous
                  Hardware Accelerators},
  booktitle    = {Proceedings of the 24th International Middleware Conference, Middleware
                  2023, Bologna, Italy, December 11-15, 2023},
  pages        = {192--206},
  publisher    = {{ACM}},
  year         = {2023},
  url          = {https://doi.org/10.1145/3590140.3629115},
  doi          = {10.1145/3590140.3629115},
  bibsource    = {dblp computer science bibliography, https://dblp.org}
}

@inproceedings{DBLP:conf/ppopp/FangYZZ21,
  author       = {Jiarui Fang and
                  Yang Yu and
                  Chengduo Zhao and
                  Jie Zhou},
  editor       = {Jaejin Lee and
                  Erez Petrank},
  title        = {TurboTransformers: an efficient {GPU} serving system for transformer
                  models},
  booktitle    = {PPoPP '21: 26th {ACM} {SIGPLAN} Symposium on Principles and Practice
                  of Parallel Programming, Virtual Event, Republic of Korea, February
                  27- March 3, 2021},
  pages        = {389--402},
  publisher    = {{ACM}},
  year         = {2021},
  url          = {https://doi.org/10.1145/3437801.3441578},
  doi          = {10.1145/3437801.3441578},
  bibsource    = {dblp computer science bibliography, https://dblp.org}
}

@inproceedings{DBLP:conf/osdi/LeeSCSWI18,
  author       = {Yunseong Lee and
                  Alberto Scolari and
                  Byung{-}Gon Chun and
                  Marco Domenico Santambrogio and
                  Markus Weimer and
                  Matteo Interlandi},
  editor       = {Andrea C. Arpaci{-}Dusseau and
                  Geoff Voelker},
  title        = {{PRETZEL:} Opening the Black Box of Machine Learning Prediction Serving
                  Systems},
  booktitle    = {13th {USENIX} Symposium on Operating Systems Design and Implementation,
                  {OSDI} 2018, Carlsbad, CA, USA, October 8-10, 2018},
  pages        = {611--626},
  publisher    = {{USENIX} Association},
  year         = {2018},
  url          = {https://www.usenix.org/conference/osdi18/presentation/lee},
  bibsource    = {dblp computer science bibliography, https://dblp.org}
}

@inproceedings{DBLP:conf/osdi/LiZZL00HCZGS23,
  author       = {Zhuohan Li and
                  Lianmin Zheng and
                  Yinmin Zhong and
                  Vincent Liu and
                  Ying Sheng and
                  Xin Jin and
                  Yanping Huang and
                  Zhifeng Chen and
                  Hao Zhang and
                  Joseph E. Gonzalez and
                  Ion Stoica},
  editor       = {Roxana Geambasu and
                  Ed Nightingale},
  title        = {AlpaServe: Statistical Multiplexing with Model Parallelism for Deep
                  Learning Serving},
  booktitle    = {17th {USENIX} Symposium on Operating Systems Design and Implementation,
                  {OSDI} 2023, Boston, MA, USA, July 10-12, 2023},
  pages        = {663--679},
  publisher    = {{USENIX} Association},
  year         = {2023},
  url          = {https://www.usenix.org/conference/osdi23/presentation/li-zhouhan},
  bibsource    = {dblp computer science bibliography, https://dblp.org}
}

@inproceedings{DBLP:conf/cloud/RomeroZYK21,
  author       = {Francisco Romero and
                  Mark Zhao and
                  Neeraja J. Yadwadkar and
                  Christos Kozyrakis},
  editor       = {Carlo Curino and
                  Georgia Koutrika and
                  Ravi Netravali},
  title        = {Llama: {A} Heterogeneous {\&} Serverless Framework for Auto-Tuning
                  Video Analytics Pipelines},
  booktitle    = {SoCC '21: {ACM} Symposium on Cloud Computing, Seattle, WA, USA, November
                  1 - 4, 2021},
  pages        = {1--17},
  publisher    = {{ACM}},
  year         = {2021},
  url          = {https://doi.org/10.1145/3472883.3486972},
  doi          = {10.1145/3472883.3486972},
  bibsource    = {dblp computer science bibliography, https://dblp.org}
}

@inproceedings{DBLP:conf/asplos/MiaoOZCWZWZYSSC24,
  author       = {Xupeng Miao and
                  Gabriele Oliaro and
                  Zhihao Zhang and
                  Xinhao Cheng and
                  Zeyu Wang and
                  Zhengxin Zhang and
                  Rae Ying Yee Wong and
                  Alan Zhu and
                  Lijie Yang and
                  Xiaoxiang Shi and
                  Chunan Shi and
                  Zhuoming Chen and
                  Daiyaan Arfeen and
                  Reyna Abhyankar and
                  Zhihao Jia},
  editor       = {Rajiv Gupta and
                  Nael B. Abu{-}Ghazaleh and
                  Madan Musuvathi and
                  Dan Tsafrir},
  title        = {SpecInfer: Accelerating Large Language Model Serving with Tree-based
                  Speculative Inference and Verification},
  booktitle    = {Proceedings of the 29th {ACM} International Conference on Architectural
                  Support for Programming Languages and Operating Systems, Volume 3,
                  {ASPLOS} 2024, La Jolla, CA, USA, 27 April 2024- 1 May 2024},
  pages        = {932--949},
  publisher    = {{ACM}},
  year         = {2024},
  url          = {https://doi.org/10.1145/3620666.3651335},
  doi          = {10.1145/3620666.3651335},
  bibsource    = {dblp computer science bibliography, https://dblp.org}
}

@inproceedings{DBLP:conf/osdi/YuJKKC22,
  author       = {Gyeong{-}In Yu and
                  Joo Seong Jeong and
                  Geon{-}Woo Kim and
                  Soojeong Kim and
                  Byung{-}Gon Chun},
  editor       = {Marcos K. Aguilera and
                  Hakim Weatherspoon},
  title        = {Orca: {A} Distributed Serving System for Transformer-Based Generative
                  Models},
  booktitle    = {16th {USENIX} Symposium on Operating Systems Design and Implementation,
                  {OSDI} 2022, Carlsbad, CA, USA, July 11-13, 2022},
  pages        = {521--538},
  publisher    = {{USENIX} Association},
  year         = {2022},
  url          = {https://www.usenix.org/conference/osdi22/presentation/yu},
  bibsource    = {dblp computer science bibliography, https://dblp.org}
}

@inproceedings{DBLP:conf/asplos/OhKKKLC024,
  author       = {Hyungjun Oh and
                  Kihong Kim and
                  Jaemin Kim and
                  Sungkyun Kim and
                  Junyeol Lee and
                  Du{-}Seong Chang and
                  Jiwon Seo},
  editor       = {Rajiv Gupta and
                  Nael B. Abu{-}Ghazaleh and
                  Madan Musuvathi and
                  Dan Tsafrir},
  title        = {ExeGPT: Constraint-Aware Resource Scheduling for {LLM} Inference},
  booktitle    = {Proceedings of the 29th {ACM} International Conference on Architectural
                  Support for Programming Languages and Operating Systems, Volume 2,
                  {ASPLOS} 2024, La Jolla, CA, USA, 27 April 2024- 1 May 2024},
  pages        = {369--384},
  publisher    = {{ACM}},
  year         = {2024},
  url          = {https://doi.org/10.1145/3620665.3640383},
  doi          = {10.1145/3620665.3640383},
  bibsource    = {dblp computer science bibliography, https://dblp.org}
}

@inproceedings{DBLP:conf/asplos/MiaoSDXL0J24,
  author       = {Xupeng Miao and
                  Chunan Shi and
                  Jiangfei Duan and
                  Xiaoli Xi and
                  Dahua Lin and
                  Bin Cui and
                  Zhihao Jia},
  editor       = {Rajiv Gupta and
                  Nael B. Abu{-}Ghazaleh and
                  Madan Musuvathi and
                  Dan Tsafrir},
  title        = {SpotServe: Serving Generative Large Language Models on Preemptible
                  Instances},
  booktitle    = {Proceedings of the 29th {ACM} International Conference on Architectural
                  Support for Programming Languages and Operating Systems, Volume 2,
                  {ASPLOS} 2024, La Jolla, CA, USA, 27 April 2024- 1 May 2024},
  pages        = {1112--1127},
  publisher    = {{ACM}},
  year         = {2024},
  url          = {https://doi.org/10.1145/3620665.3640411},
  doi          = {10.1145/3620665.3640411},
  bibsource    = {dblp computer science bibliography, https://dblp.org}
}

@inproceedings{DBLP:conf/asplos/PatelCZGWMB24,
  author       = {Pratyush Patel and
                  Esha Choukse and
                  Chaojie Zhang and
                  {\'{I}}{\~{n}}igo Goiri and
                  Brijesh Warrier and
                  Nithish Mahalingam and
                  Ricardo Bianchini},
  editor       = {Rajiv Gupta and
                  Nael B. Abu{-}Ghazaleh and
                  Madan Musuvathi and
                  Dan Tsafrir},
  title        = {Characterizing Power Management Opportunities for LLMs in the Cloud},
  booktitle    = {Proceedings of the 29th {ACM} International Conference on Architectural
                  Support for Programming Languages and Operating Systems, Volume 3,
                  {ASPLOS} 2024, La Jolla, CA, USA, 27 April 2024- 1 May 2024},
  pages        = {207--222},
  publisher    = {{ACM}},
  year         = {2024},
  url          = {https://doi.org/10.1145/3620666.3651329},
  doi          = {10.1145/3620666.3651329},
  bibsource    = {dblp computer science bibliography, https://dblp.org}
}

@inproceedings{DBLP:conf/sosp/KwonLZ0ZY0ZS23,
  author       = {Woosuk Kwon and
                  Zhuohan Li and
                  Siyuan Zhuang and
                  Ying Sheng and
                  Lianmin Zheng and
                  Cody Hao Yu and
                  Joseph Gonzalez and
                  Hao Zhang and
                  Ion Stoica},
  editor       = {Jason Flinn and
                  Margo I. Seltzer and
                  Peter Druschel and
                  Antoine Kaufmann and
                  Jonathan Mace},
  title        = {Efficient Memory Management for Large Language Model Serving with
                  PagedAttention},
  booktitle    = {Proceedings of the 29th Symposium on Operating Systems Principles,
                  {SOSP} 2023, Koblenz, Germany, October 23-26, 2023},
  pages        = {611--626},
  publisher    = {{ACM}},
  year         = {2023},
  url          = {https://doi.org/10.1145/3600006.3613165},
  doi          = {10.1145/3600006.3613165},
  bibsource    = {dblp computer science bibliography, https://dblp.org}
}

@inproceedings{DBLP:conf/osdi/SunHZXZL024,
  author       = {Biao Sun and
                  Ziming Huang and
                  Hanyu Zhao and
                  Wencong Xiao and
                  Xinyi Zhang and
                  Yong Li and
                  Wei Lin},
  editor       = {Ada Gavrilovska and
                  Douglas B. Terry},
  title        = {Llumnix: Dynamic Scheduling for Large Language Model Serving},
  booktitle    = {18th {USENIX} Symposium on Operating Systems Design and Implementation,
                  {OSDI} 2024, Santa Clara, CA, USA, July 10-12, 2024},
  pages        = {173--191},
  publisher    = {{USENIX} Association},
  year         = {2024},
  url          = {https://www.usenix.org/conference/osdi24/presentation/sun-biao},
  bibsource    = {dblp computer science bibliography, https://dblp.org}
}

@inproceedings{DBLP:conf/sosp/WuLZ0L024,
  author       = {Bingyang Wu and
                  Shengyu Liu and
                  Yinmin Zhong and
                  Peng Sun and
                  Xuanzhe Liu and
                  Xin Jin},
  editor       = {Emmett Witchel and
                  Christopher J. Rossbach and
                  Andrea C. Arpaci{-}Dusseau and
                  Kimberly Keeton},
  title        = {LoongServe: Efficiently Serving Long-Context Large Language Models
                  with Elastic Sequence Parallelism},
  booktitle    = {Proceedings of the {ACM} {SIGOPS} 30th Symposium on Operating Systems
                  Principles, {SOSP} 2024, Austin, TX, USA, November 4-6, 2024},
  pages        = {640--654},
  publisher    = {{ACM}},
  year         = {2024},
  url          = {https://doi.org/10.1145/3694715.3695948},
  doi          = {10.1145/3694715.3695948},
  bibsource    = {dblp computer science bibliography, https://dblp.org}
}

@inproceedings{DBLP:conf/isca/PatelCZSGMB24,
  author       = {Pratyush Patel and
                  Esha Choukse and
                  Chaojie Zhang and
                  Aashaka Shah and
                  {\'{I}}{\~{n}}igo Goiri and
                  Saeed Maleki and
                  Ricardo Bianchini},
  title        = {Splitwise: Efficient Generative {LLM} Inference Using Phase Splitting},
  booktitle    = {51st {ACM/IEEE} Annual International Symposium on Computer Architecture,
                  {ISCA} 2024, Buenos Aires, Argentina, June 29 - July 3, 2024},
  pages        = {118--132},
  publisher    = {{IEEE}},
  year         = {2024},
  url          = {https://doi.org/10.1109/ISCA59077.2024.00019},
  doi          = {10.1109/ISCA59077.2024.00019},
  bibsource    = {dblp computer science bibliography, https://dblp.org}
}

@inproceedings{DBLP:conf/osdi/ZhongLCHZL0024,
  author       = {Yinmin Zhong and
                  Shengyu Liu and
                  Junda Chen and
                  Jianbo Hu and
                  Yibo Zhu and
                  Xuanzhe Liu and
                  Xin Jin and
                  Hao Zhang},
  editor       = {Ada Gavrilovska and
                  Douglas B. Terry},
  title        = {DistServe: Disaggregating Prefill and Decoding for Goodput-optimized
                  Large Language Model Serving},
  booktitle    = {18th {USENIX} Symposium on Operating Systems Design and Implementation,
                  {OSDI} 2024, Santa Clara, CA, USA, July 10-12, 2024},
  pages        = {193--210},
  publisher    = {{USENIX} Association},
  year         = {2024},
  url          = {https://www.usenix.org/conference/osdi24/presentation/zhong-yinmin},
  bibsource    = {dblp computer science bibliography, https://dblp.org}
}

@inproceedings{DBLP:conf/osdi/AgrawalKPMKGTR24,
  author       = {Amey Agrawal and
                  Nitin Kedia and
                  Ashish Panwar and
                  Jayashree Mohan and
                  Nipun Kwatra and
                  Bhargav S. Gulavani and
                  Alexey Tumanov and
                  Ramachandran Ramjee},
  editor       = {Ada Gavrilovska and
                  Douglas B. Terry},
  title        = {Taming Throughput-Latency Tradeoff in {LLM} Inference with Sarathi-Serve},
  booktitle    = {18th {USENIX} Symposium on Operating Systems Design and Implementation,
                  {OSDI} 2024, Santa Clara, CA, USA, July 10-12, 2024},
  pages        = {117--134},
  publisher    = {{USENIX} Association},
  year         = {2024},
  url          = {https://www.usenix.org/conference/osdi24/presentation/agrawal},
  bibsource    = {dblp computer science bibliography, https://dblp.org}
}

@inproceedings{DBLP:conf/usenix/ChoiLKPKH22,
  author       = {Seungbeom Choi and
                  Sunho Lee and
                  Yeonjae Kim and
                  Jongse Park and
                  Youngjin Kwon and
                  Jaehyuk Huh},
  editor       = {Jiri Schindler and
                  Noa Zilberman},
  title        = {Serving Heterogeneous Machine Learning Models on Multi-GPU Servers
                  with Spatio-Temporal Sharing},
  booktitle    = {Proceedings of the 2022 {USENIX} Annual Technical Conference, {USENIX}
                  {ATC} 2022, Carlsbad, CA, USA, July 11-13, 2022},
  pages        = {199--216},
  publisher    = {{USENIX} Association},
  year         = {2022},
  url          = {https://www.usenix.org/conference/atc22/presentation/choi-seungbeom},
  bibsource    = {dblp computer science bibliography, https://dblp.org}
}

@misc{lou2025swiftserverlessllmcold,
      title={Towards Swift Serverless LLM Cold Starts with ParaServe}, 
      author={Chiheng Lou and Sheng Qi and Chao Jin and Dapeng Nie and Haoran Yang and Xuanzhe Liu and Xin Jin},
      year={2025},
      eprint={2502.15524},
      archivePrefix={arXiv},
      primaryClass={cs.DC},
      url={https://arxiv.org/abs/2502.15524}, 
}

@inproceedings{DBLP:conf/sosp/SongMX024,
  author       = {Yixin Song and
                  Zeyu Mi and
                  Haotong Xie and
                  Haibo Chen},
  editor       = {Emmett Witchel and
                  Christopher J. Rossbach and
                  Andrea C. Arpaci{-}Dusseau and
                  Kimberly Keeton},
  title        = {PowerInfer: Fast Large Language Model Serving with a Consumer-grade
                  {GPU}},
  booktitle    = {Proceedings of the {ACM} {SIGOPS} 30th Symposium on Operating Systems
                  Principles, {SOSP} 2024, Austin, TX, USA, November 4-6, 2024},
  pages        = {590--606},
  publisher    = {{ACM}},
  year         = {2024},
  url          = {https://doi.org/10.1145/3694715.3695964},
  doi          = {10.1145/3694715.3695964},
  bibsource    = {dblp computer science bibliography, https://dblp.org}
}

@article{DBLP:journals/corr/abs-2403-11421,
  author       = {Jiaao He and
                  Jidong Zhai},
  title        = {FastDecode: High-Throughput GPU-Efficient {LLM} Serving using Heterogeneous
                  Pipelines},
  journal      = {CoRR},
  volume       = {abs/2403.11421},
  year         = {2024},
  url          = {https://doi.org/10.48550/arXiv.2403.11421},
  doi          = {10.48550/ARXIV.2403.11421},
  eprinttype    = {arXiv},
  eprint       = {2403.11421},
  bibsource    = {dblp computer science bibliography, https://dblp.org}
}

@article{DBLP:journals/corr/abs-2401-11240,
  author       = {Suyi Li and
                  Hanfeng Lu and
                  Tianyuan Wu and
                  Minchen Yu and
                  Qizhen Weng and
                  Xusheng Chen and
                  Yizhou Shan and
                  Binhang Yuan and
                  Wei Wang},
  title        = {CaraServe: CPU-Assisted and Rank-Aware LoRA Serving for Generative
                  {LLM} Inference},
  journal      = {CoRR},
  volume       = {abs/2401.11240},
  year         = {2024},
  url          = {https://doi.org/10.48550/arXiv.2401.11240},
  doi          = {10.48550/ARXIV.2401.11240},
  eprinttype    = {arXiv},
  eprint       = {2401.11240},
  bibsource    = {dblp computer science bibliography, https://dblp.org}
}

@inproceedings{DBLP:conf/IEEEpact/Park024,
  author       = {Daon Park and
                  Bernhard Egger},
  title        = {Improving Throughput-oriented {LLM} Inference with {CPU} Computations},
  booktitle    = {Proceedings of the 2024 International Conference on Parallel Architectures
                  and Compilation Techniques, {PACT} 2024, Long Beach, CA, USA, October
                  14-16, 2024},
  pages        = {233--245},
  publisher    = {{ACM}},
  year         = {2024},
  url          = {https://doi.org/10.1145/3656019.3676949},
  doi          = {10.1145/3656019.3676949},
  bibsource    = {dblp computer science bibliography, https://dblp.org}
}

@article{DBLP:journals/corr/abs-2411-09317,
  author       = {Yi Xu and
                  Ziming Mao and
                  Xiangxi Mo and
                  Shu Liu and
                  Ion Stoica},
  title        = {Pie: Pooling {CPU} Memory for {LLM} Inference},
  journal      = {CoRR},
  volume       = {abs/2411.09317},
  year         = {2024},
  url          = {https://doi.org/10.48550/arXiv.2411.09317},
  doi          = {10.48550/ARXIV.2411.09317},
  eprinttype    = {arXiv},
  eprint       = {2411.09317},
  bibsource    = {dblp computer science bibliography, https://dblp.org}
}

@inproceedings{DBLP:conf/nsdi/WengXYWWHLZLD22,
  author       = {Qizhen Weng and
                  Wencong Xiao and
                  Yinghao Yu and
                  Wei Wang and
                  Cheng Wang and
                  Jian He and
                  Yong Li and
                  Liping Zhang and
                  Wei Lin and
                  Yu Ding},
  editor       = {Amar Phanishayee and
                  Vyas Sekar},
  title        = {MLaaS in the Wild: Workload Analysis and Scheduling in Large-Scale
                  Heterogeneous {GPU} Clusters},
  booktitle    = {19th {USENIX} Symposium on Networked Systems Design and Implementation,
                  {NSDI} 2022, Renton, WA, USA, April 4-6, 2022},
  pages        = {945--960},
  publisher    = {{USENIX} Association},
  year         = {2022},
  url          = {https://www.usenix.org/conference/nsdi22/presentation/weng},
  bibsource    = {dblp computer science bibliography, https://dblp.org}
}

@inproceedings{DBLP:conf/osdi/NarayananSKPZ20,
  author       = {Deepak Narayanan and
                  Keshav Santhanam and
                  Fiodar Kazhamiaka and
                  Amar Phanishayee and
                  Matei Zaharia},
  title        = {Heterogeneity-Aware Cluster Scheduling Policies for Deep Learning
                  Workloads},
  booktitle    = {14th {USENIX} Symposium on Operating Systems Design and Implementation,
                  {OSDI} 2020, Virtual Event, November 4-6, 2020},
  pages        = {481--498},
  publisher    = {{USENIX} Association},
  year         = {2020},
  url          = {https://www.usenix.org/conference/osdi20/presentation/narayanan-deepak},
  bibsource    = {dblp computer science bibliography, https://dblp.org}
}

@article{DBLP:journals/tpds/ChenLOZL18,
  author       = {Cen Chen and
                  Kenli Li and
                  Aijia Ouyang and
                  Zeng Zeng and
                  Keqin Li},
  title        = {GFlink: An In-Memory Computing Architecture on Heterogeneous {CPU-GPU}
                  Clusters for Big Data},
  journal      = {{IEEE} Trans. Parallel Distributed Syst.},
  volume       = {29},
  number       = {6},
  pages        = {1275--1288},
  year         = {2018},
  url          = {https://doi.org/10.1109/TPDS.2018.2794343},
  doi          = {10.1109/TPDS.2018.2794343},
  bibsource    = {dblp computer science bibliography, https://dblp.org}
}

@inproceedings{DBLP:conf/osdi/JiangZLYCG20,
  author       = {Yimin Jiang and
                  Yibo Zhu and
                  Chang Lan and
                  Bairen Yi and
                  Yong Cui and
                  Chuanxiong Guo},
  title        = {A Unified Architecture for Accelerating Distributed {DNN} Training
                  in Heterogeneous {GPU/CPU} Clusters},
  booktitle    = {14th {USENIX} Symposium on Operating Systems Design and Implementation,
                  {OSDI} 2020, Virtual Event, November 4-6, 2020},
  pages        = {463--479},
  publisher    = {{USENIX} Association},
  year         = {2020},
  url          = {https://www.usenix.org/conference/osdi20/presentation/jiang},
  bibsource    = {dblp computer science bibliography, https://dblp.org}
}

@inproceedings{DBLP:conf/usenix/JeonVPQXY19,
  author       = {Myeongjae Jeon and
                  Shivaram Venkataraman and
                  Amar Phanishayee and
                  Junjie Qian and
                  Wencong Xiao and
                  Fan Yang},
  editor       = {Dahlia Malkhi and
                  Dan Tsafrir},
  title        = {Analysis of Large-Scale Multi-Tenant {GPU} Clusters for {DNN} Training
                  Workloads},
  booktitle    = {Proceedings of the 2019 {USENIX} Annual Technical Conference, {USENIX}
                  {ATC} 2019, Renton, WA, USA, July 10-12, 2019},
  pages        = {947--960},
  publisher    = {{USENIX} Association},
  year         = {2019},
  url          = {https://www.usenix.org/conference/atc19/presentation/jeon},
  bibsource    = {dblp computer science bibliography, https://dblp.org}
}

@inproceedings{DBLP:conf/isscc/NassifMMPLYMHVK22,
  author       = {Nevine Nassif and
                  Ashley O. Munch and
                  Carleton L. Molnar and
                  Gerald Pasdast and
                  Sitaraman V. Lyer and
                  Zibing Yang and
                  Oscar Mendoza and
                  Mark Huddart and
                  Srikrishnan Venkataraman and
                  Sireesha Kandula and
                  Rafi Marom and
                  Alexandra M. Kern and
                  William J. Bowhill and
                  David R. Mulvihill and
                  Srikanth Nimmagadda and
                  Varma Kalidindi and
                  Jonathan Krause and
                  Mohammad M. Haq and
                  Roopali Sharma and
                  Kevin Duda},
  title        = {Sapphire Rapids: The Next-Generation Intel Xeon Scalable Processor},
  booktitle    = {{IEEE} International Solid-State Circuits Conference, {ISSCC} 2022,
                  San Francisco, CA, USA, February 20-26, 2022},
  pages        = {44--46},
  publisher    = {{IEEE}},
  year         = {2022},
  url          = {https://doi.org/10.1109/ISSCC42614.2022.9731107},
  doi          = {10.1109/ISSCC42614.2022.9731107},
  bibsource    = {dblp computer science bibliography, https://dblp.org}
}

@inproceedings{DBLP:conf/usenix/ShahradFGCBCLTR20,
  author       = {Mohammad Shahrad and
                  Rodrigo Fonseca and
                  I{\~{n}}igo Goiri and
                  Gohar Irfan Chaudhry and
                  Paul Batum and
                  Jason Cooke and
                  Eduardo Laureano and
                  Colby Tresness and
                  Mark Russinovich and
                  Ricardo Bianchini},
  editor       = {Ada Gavrilovska and
                  Erez Zadok},
  title        = {Serverless in the Wild: Characterizing and Optimizing the Serverless
                  Workload at a Large Cloud Provider},
  booktitle    = {Proceedings of the 2020 {USENIX} Annual Technical Conference, {USENIX}
                  {ATC} 2020, July 15-17, 2020},
  pages        = {205--218},
  publisher    = {{USENIX} Association},
  year         = {2020},
  url          = {https://www.usenix.org/conference/atc20/presentation/shahrad},
  bibsource    = {dblp computer science bibliography, https://dblp.org}
}

@inproceedings{DBLP:conf/mlsys/0002TTYCWXDG024,
  author       = {Ji Lin and
                  Jiaming Tang and
                  Haotian Tang and
                  Shang Yang and
                  Wei{-}Ming Chen and
                  Wei{-}Chen Wang and
                  Guangxuan Xiao and
                  Xingyu Dang and
                  Chuang Gan and
                  Song Han},
  editor       = {Phillip B. Gibbons and
                  Gennady Pekhimenko and
                  Christopher De Sa},
  title        = {{AWQ:} Activation-aware Weight Quantization for On-Device {LLM} Compression
                  and Acceleration},
  booktitle    = {Proceedings of the Seventh Annual Conference on Machine Learning and
                  Systems, MLSys 2024, Santa Clara, CA, USA, May 13-16, 2024},
  publisher    = {mlsys.org},
  year         = {2024},
  url          = {https://proceedings.mlsys.org/paper\_files/paper/2024/hash/42a452cbafa9dd64e9ba4aa95cc1ef21-Abstract-Conference.html},
  bibsource    = {dblp computer science bibliography, https://dblp.org}
}

@inproceedings{DBLP:conf/asplos/ZengXGCL25,
	author       = {Shaoxun Zeng and
	Minhui Xie and
	Shiwei Gao and
	Youmin Chen and
	Youyou Lu},
	editor       = {Lieven Eeckhout and
	Georgios Smaragdakis and
	Kaitai Liang and
	Adrian Sampson and
	Martha A. Kim and
	Christopher J. Rossbach},
	title        = {Medusa: Accelerating Serverless {LLM} Inference with Materialization},
	booktitle    = {Proceedings of the 30th {ACM} International Conference on Architectural
	Support for Programming Languages and Operating Systems, Volume 1,
	{ASPLOS} 2025, Rotterdam, The Netherlands, 30 March 2025 - 3 April
	2025},
	pages        = {653--668},
	publisher    = {{ACM}},
	year         = {2025},
	url          = {https://doi.org/10.1145/3669940.3707285},
	doi          = {10.1145/3669940.3707285},
	bibsource    = {dblp computer science bibliography, https://dblp.org}
}

@article{DBLP:journals/corr/abs-2410-18038,
  author       = {Aditya K. Kamath and
                  Ramya Prabhu and
                  Jayashree Mohan and
                  Simon Peter and
                  Ramachandran Ramjee and
                  Ashish Panwar},
  title        = {POD-Attention: Unlocking Full Prefill-Decode Overlap for Faster {LLM}
                  Inference},
  journal      = {CoRR},
  volume       = {abs/2410.18038},
  year         = {2024},
  url          = {https://doi.org/10.48550/arXiv.2410.18038},
  doi          = {10.48550/ARXIV.2410.18038},
  eprinttype    = {arXiv},
  eprint       = {2410.18038},
  bibsource    = {dblp computer science bibliography, https://dblp.org}
}

@misc{touvron2023llama2openfoundation,
      title={Llama 2: Open Foundation and Fine-Tuned Chat Models}, 
      author={Hugo Touvron and Louis Martin and Kevin Stone and Peter Albert and Amjad Almahairi and Yasmine Babaei and Nikolay Bashlykov and Soumya Batra and Prajjwal Bhargava and Shruti Bhosale and Dan Bikel and Lukas Blecher and Cristian Canton Ferrer and Moya Chen and Guillem Cucurull and David Esiobu and Jude Fernandes and Jeremy Fu and Wenyin Fu and Brian Fuller and Cynthia Gao and Vedanuj Goswami and Naman Goyal and Anthony Hartshorn and Saghar Hosseini and Rui Hou and Hakan Inan and Marcin Kardas and Viktor Kerkez and Madian Khabsa and Isabel Kloumann and Artem Korenev and Punit Singh Koura and Marie-Anne Lachaux and Thibaut Lavril and Jenya Lee and Diana Liskovich and Yinghai Lu and Yuning Mao and Xavier Martinet and Todor Mihaylov and Pushkar Mishra and Igor Molybog and Yixin Nie and Andrew Poulton and Jeremy Reizenstein and Rashi Rungta and Kalyan Saladi and Alan Schelten and Ruan Silva and Eric Michael Smith and Ranjan Subramanian and Xiaoqing Ellen Tan and Binh Tang and Ross Taylor and Adina Williams and Jian Xiang Kuan and Puxin Xu and Zheng Yan and Iliyan Zarov and Yuchen Zhang and Angela Fan and Melanie Kambadur and Sharan Narang and Aurelien Rodriguez and Robert Stojnic and Sergey Edunov and Thomas Scialom},
      year={2023},
      eprint={2307.09288},
      archivePrefix={arXiv},
      primaryClass={cs.CL},
      url={https://arxiv.org/abs/2307.09288}, 
}

@misc{gemmateam2024gemmaopenmodelsbased,
      title={Gemma: Open Models Based on Gemini Research and Technology}, 
      author={Gemma Team and Thomas Mesnard and Cassidy Hardin and Robert Dadashi and Surya Bhupatiraju and Shreya Pathak and Laurent Sifre and Morgane Rivière and Mihir Sanjay Kale and Juliette Love and Pouya Tafti and Léonard Hussenot and Pier Giuseppe Sessa and Aakanksha Chowdhery and Adam Roberts and Aditya Barua and Alex Botev and Alex Castro-Ros and Ambrose Slone and Amélie Héliou and Andrea Tacchetti and Anna Bulanova and Antonia Paterson and Beth Tsai and Bobak Shahriari and Charline Le Lan and Christopher A. Choquette-Choo and Clément Crepy and Daniel Cer and Daphne Ippolito and David Reid and Elena Buchatskaya and Eric Ni and Eric Noland and Geng Yan and George Tucker and George-Christian Muraru and Grigory Rozhdestvenskiy and Henryk Michalewski and Ian Tenney and Ivan Grishchenko and Jacob Austin and James Keeling and Jane Labanowski and Jean-Baptiste Lespiau and Jeff Stanway and Jenny Brennan and Jeremy Chen and Johan Ferret and Justin Chiu and Justin Mao-Jones and Katherine Lee and Kathy Yu and Katie Millican and Lars Lowe Sjoesund and Lisa Lee and Lucas Dixon and Machel Reid and Maciej Mikuła and Mateo Wirth and Michael Sharman and Nikolai Chinaev and Nithum Thain and Olivier Bachem and Oscar Chang and Oscar Wahltinez and Paige Bailey and Paul Michel and Petko Yotov and Rahma Chaabouni and Ramona Comanescu and Reena Jana and Rohan Anil and Ross McIlroy and Ruibo Liu and Ryan Mullins and Samuel L Smith and Sebastian Borgeaud and Sertan Girgin and Sholto Douglas and Shree Pandya and Siamak Shakeri and Soham De and Ted Klimenko and Tom Hennigan and Vlad Feinberg and Wojciech Stokowiec and Yu-hui Chen and Zafarali Ahmed and Zhitao Gong and Tris Warkentin and Ludovic Peran and Minh Giang and Clément Farabet and Oriol Vinyals and Jeff Dean and Koray Kavukcuoglu and Demis Hassabis and Zoubin Ghahramani and Douglas Eck and Joelle Barral and Fernando Pereira and Eli Collins and Armand Joulin and Noah Fiedel and Evan Senter and Alek Andreev and Kathleen Kenealy},
      year={2024},
      eprint={2403.08295},
      archivePrefix={arXiv},
      primaryClass={cs.CL},
      url={https://arxiv.org/abs/2403.08295}, 
}

@misc{chen2021evaluatinglargelanguagemodels,
      title={Evaluating Large Language Models Trained on Code}, 
      author={Mark Chen and Jerry Tworek and Heewoo Jun and Qiming Yuan and Henrique Ponde de Oliveira Pinto and Jared Kaplan and Harri Edwards and Yuri Burda and Nicholas Joseph and Greg Brockman and Alex Ray and Raul Puri and Gretchen Krueger and Michael Petrov and Heidy Khlaaf and Girish Sastry and Pamela Mishkin and Brooke Chan and Scott Gray and Nick Ryder and Mikhail Pavlov and Alethea Power and Lukasz Kaiser and Mohammad Bavarian and Clemens Winter and Philippe Tillet and Felipe Petroski Such and Dave Cummings and Matthias Plappert and Fotios Chantzis and Elizabeth Barnes and Ariel Herbert-Voss and William Hebgen Guss and Alex Nichol and Alex Paino and Nikolas Tezak and Jie Tang and Igor Babuschkin and Suchir Balaji and Shantanu Jain and William Saunders and Christopher Hesse and Andrew N. Carr and Jan Leike and Josh Achiam and Vedant Misra and Evan Morikawa and Alec Radford and Matthew Knight and Miles Brundage and Mira Murati and Katie Mayer and Peter Welinder and Bob McGrew and Dario Amodei and Sam McCandlish and Ilya Sutskever and Wojciech Zaremba},
      year={2021},
      eprint={2107.03374},
      archivePrefix={arXiv},
      primaryClass={cs.LG},
      url={https://arxiv.org/abs/2107.03374}, 
}

@inproceedings{DBLP:conf/iclr/ZhengC0LZW00LXG24,
  author       = {Lianmin Zheng and
                  Wei{-}Lin Chiang and
                  Ying Sheng and
                  Tianle Li and
                  Siyuan Zhuang and
                  Zhanghao Wu and
                  Yonghao Zhuang and
                  Zhuohan Li and
                  Zi Lin and
                  Eric P. Xing and
                  Joseph E. Gonzalez and
                  Ion Stoica and
                  Hao Zhang},
  title        = {LMSYS-Chat-1M: {A} Large-Scale Real-World {LLM} Conversation Dataset},
  booktitle    = {The Twelfth International Conference on Learning Representations,
                  {ICLR} 2024, Vienna, Austria, May 7-11, 2024},
  publisher    = {OpenReview.net},
  year         = {2024},
  url          = {https://openreview.net/forum?id=BOfDKxfwt0},
  bibsource    = {dblp computer science bibliography, https://dblp.org}
}

@inproceedings{DBLP:conf/cloud/JoosenHASDWB23,
  author       = {Artjom Joosen and
                  Ahmed Hassan and
                  Martin Asenov and
                  Rajkarn Singh and
                  Luke Nicholas Darlow and
                  Jianfeng Wang and
                  Adam Barker},
  title        = {How Does It Function?: Characterizing Long-term Trends in Production
                  Serverless Workloads},
  booktitle    = {Proceedings of the 2023 {ACM} Symposium on Cloud Computing, SoCC 2023,
                  Santa Cruz, CA, USA, 30 October 2023 - 1 November 2023},
  pages        = {443--458},
  publisher    = {{ACM}},
  year         = {2023},
  url          = {https://doi.org/10.1145/3620678.3624783},
  doi          = {10.1145/3620678.3624783},
  bibsource    = {dblp computer science bibliography, https://dblp.org}
}

@inproceedings{DBLP:conf/isca/KimWX0YK25,
  author       = {Hyungyo Kim and
                  Nachuan Wang and
                  Qirong Xia and
                  Jinghan Huang and
                  Amir Yazdanbakhsh and
                  Nam Sung Kim},
  title        = {{LIA:} {A} Single-GPU {LLM} Inference Acceleration with Cooperative
                  AMX-Enabled {CPU-GPU} Computation and {CXL} Offloading},
  booktitle    = {Proceedings of the 52nd Annual International Symposium on Computer
                  Architecture, {ISCA} 2025, Tokyo, Japan, June 21-25, 2025},
  pages        = {544--558},
  publisher    = {{ACM}},
  year         = {2025},
  url          = {https://doi.org/10.1145/3695053.3731092},
  doi          = {10.1145/3695053.3731092},
  bibsource    = {dblp computer science bibliography, https://dblp.org}
}

@inproceedings{DBLP:conf/asplos/Lv0LHTZZ25,
  author       = {Cunchi Lv and
                  Xiao Shi and
                  Zhengyu Lei and
                  Jinyue Huang and
                  Wenting Tan and
                  Xiaohui Zheng and
                  Xiaofang Zhao},
  editor       = {Lieven Eeckhout and
                  Georgios Smaragdakis and
                  Kaitai Liang and
                  Adrian Sampson and
                  Martha A. Kim and
                  Christopher J. Rossbach},
  title        = {Dilu: Enabling {GPU} Resourcing-on-Demand for Serverless {DL} Serving
                  via Introspective Elasticity},
  booktitle    = {Proceedings of the 30th {ACM} International Conference on Architectural
                  Support for Programming Languages and Operating Systems, Volume 1,
                  {ASPLOS} 2025, Rotterdam, The Netherlands, 30 March 2025 - 3 April
                  2025},
  pages        = {311--325},
  publisher    = {{ACM}},
  year         = {2025},
  url          = {https://doi.org/10.1145/3669940.3707251},
  doi          = {10.1145/3669940.3707251},
  bibsource    = {dblp computer science bibliography, https://dblp.org}
}

@inproceedings{DBLP:conf/acl/BaiLZL0HDLZHDTL24,
  author       = {Yushi Bai and
                  Xin Lv and
                  Jiajie Zhang and
                  Hongchang Lyu and
                  Jiankai Tang and
                  Zhidian Huang and
                  Zhengxiao Du and
                  Xiao Liu and
                  Aohan Zeng and
                  Lei Hou and
                  Yuxiao Dong and
                  Jie Tang and
                  Juanzi Li},
  editor       = {Lun{-}Wei Ku and
                  Andre Martins and
                  Vivek Srikumar},
  title        = {LongBench: {A} Bilingual, Multitask Benchmark for Long Context Understanding},
  booktitle    = {Proceedings of the 62nd Annual Meeting of the Association for Computational
                  Linguistics (Volume 1: Long Papers), {ACL} 2024, Bangkok, Thailand,
                  August 11-16, 2024},
  pages        = {3119--3137},
  publisher    = {Association for Computational Linguistics},
  year         = {2024},
  url          = {https://doi.org/10.18653/v1/2024.acl-long.172},
  doi          = {10.18653/V1/2024.ACL-LONG.172},
  bibsource    = {dblp computer science bibliography, https://dblp.org}
}

@inproceedings{10.1145/3711896.3737413,
author = {Wang, Yuxin and Chen, Yuhan and Li, Zeyu and Kang, Xueze and Fang, Yuchu and Zhou, Yeju and Zheng, Yang and Tang, Zhenheng and He, Xin and Guo, Rui and Wang, Xin and Wang, Qiang and Zhou, Amelie Chi and Chu, Xiaowen},
title = {BurstGPT: A Real-World Workload Dataset to Optimize LLM Serving Systems},
year = {2025},
isbn = {9798400714542},
publisher = {Association for Computing Machinery},
address = {New York, NY, USA},
url = {https://doi.org/10.1145/3711896.3737413},
doi = {10.1145/3711896.3737413},
booktitle = {Proceedings of the 31st ACM SIGKDD Conference on Knowledge Discovery and Data Mining V.2},
pages = {5831–5841},
numpages = {11},
location = {Toronto ON, Canada},
series = {KDD '25}
}

@inproceedings{
jiang2025neo,
title={{NEO}: Saving {GPU} Memory Crisis with {CPU} Offloading for Online {LLM} Inference},
author={Xuanlin Jiang and Yang Zhou and Shiyi Cao and Ion Stoica and Minlan Yu},
booktitle={Eighth Conference on Machine Learning and Systems},
year={2025},
url={https://openreview.net/forum?id=umgy9tWBLA}
}

@inproceedings{DBLP:conf/icml/DuanLDLZLS024,
  author       = {Jiangfei Duan and
                  Runyu Lu and
                  Haojie Duanmu and
                  Xiuhong Li and
                  Xingcheng Zhang and
                  Dahua Lin and
                  Ion Stoica and
                  Hao Zhang},
  title        = {MuxServe: Flexible Spatial-Temporal Multiplexing for Multiple {LLM}
                  Serving},
  booktitle    = {Forty-first International Conference on Machine Learning, {ICML} 2024,
                  Vienna, Austria, July 21-27, 2024},
  publisher    = {OpenReview.net},
  year         = {2024},
  url          = {https://openreview.net/forum?id=R0SoZvqXyQ},
  bibsource    = {dblp computer science bibliography, https://dblp.org}
}

@inproceedings{
na2025flexinfer,
title={FlexInfer: Flexible {LLM} Inference with {CPU} Computations},
author={Seonjin Na and Geonhwa Jeong and Byung Hoon Ahn and Aaron Jezghani and Jeffrey Young and Christopher J. Hughes and Tushar Krishna and Hyesoon Kim},
booktitle={Eighth Conference on Machine Learning and Systems},
year={2025},
url={https://openreview.net/forum?id=sFNRNTduKO}
}

@inproceedings{DBLP:conf/iiswc/NaJA0KK24,
  author       = {Seonjin Na and
                  Geonhwa Jeong and
                  Byung Hoon Ahn and
                  Jeffrey Young and
                  Tushar Krishna and
                  Hyesoon Kim},
  title        = {Understanding Performance Implications of {LLM} Inference on CPUs},
  booktitle    = {{IEEE} International Symposium on Workload Characterization, {IISWC}
                  2024, Vancouver, BC, Canada, September 15-17, 2024},
  pages        = {169--180},
  publisher    = {{IEEE}},
  year         = {2024},
  url          = {https://doi.org/10.1109/IISWC63097.2024.00024},
  doi          = {10.1109/IISWC63097.2024.00024},
  bibsource    = {dblp computer science bibliography, https://dblp.org}
}

@inproceedings{DBLP:conf/asplos/PrabhuNMRP25,
  author       = {Ramya Prabhu and
                  Ajay Nayak and
                  Jayashree Mohan and
                  Ramachandran Ramjee and
                  Ashish Panwar},
  editor       = {Lieven Eeckhout and
                  Georgios Smaragdakis and
                  Kaitai Liang and
                  Adrian Sampson and
                  Martha A. Kim and
                  Christopher J. Rossbach},
  title        = {vAttention: Dynamic Memory Management for Serving LLMs without PagedAttention},
  booktitle    = {Proceedings of the 30th {ACM} International Conference on Architectural
                  Support for Programming Languages and Operating Systems, Volume 1,
                  {ASPLOS} 2025, Rotterdam, The Netherlands, 30 March 2025 - 3 April
                  2025},
  pages        = {1133--1150},
  publisher    = {{ACM}},
  year         = {2025},
  url          = {https://doi.org/10.1145/3669940.3707256},
  doi          = {10.1145/3669940.3707256},
  bibsource    = {dblp computer science bibliography, https://dblp.org}
}

\end{document}